\documentclass[12pt]{article}

\usepackage{amsmath}

\renewcommand{\theequation}{\arabic{equation}}
\newcommand{\EQ}{\begin{equation}}
\newcommand{\EN}{\end{equation}}

\newcommand{\ket}[1]{\left|#1\right\rangle}      % Ket-Zustand
\newcommand{\bear}{\begin{eqnarray}}
\newcommand{\ear}{\end{eqnarray}}
\newcommand{\bt} { \begin{tabular} }
\newcommand{\et}{ \end{tabular} }
\newcommand{\bc} { \begin{center} }
\newcommand{\ec}{ \end{center} }

\newcommand{\btb} { \begin{table} }
\newcommand{\etb}{ \end{table} }
\begin{document}

\topmargin 0pt
\oddsidemargin 5mm
\newcommand{\NP}[1]{Nucl.\ Phys.\ {\bf #1}}
\newcommand{\PL}[1]{Phys.\ Lett.\ {\bf #1}}
\newcommand{\NC}[1]{Nuovo Cimento {\bf #1}}
\newcommand{\CMP}[1]{Comm.\ Math.\ Phys.\ {\bf #1}}
\newcommand{\PR}[1]{Phys.\ Rev.\ {\bf #1}}
\newcommand{\PRL}[1]{Phys.\ Rev.\ Lett.\ {\bf #1}}
\newcommand{\MPL}[1]{Mod.\ Phys.\ Lett.\ {\bf #1}}
\newcommand{\JETP}[1]{Sov.\ Phys.\ JETP {\bf #1}}
\newcommand{\TMP}[1]{Teor.\ Mat.\ Fiz.\ {\bf #1}}

\renewcommand{\thefootnote}{\fnsymbol{footnote}}

\newpage
\setcounter{page}{0}
\begin{titlepage}
\begin{flushright}
UFSCARF-TH-08-10
\end{flushright}
\vspace{0.5cm}
\begin{center}
{\large Algebraic Bethe Ansatz for $U(1)$ Invariant Integrable Models: \\
       The Method and General Results }\\
\vspace{1cm}
{\large C.S. Melo and M.J. Martins} \\
\vspace{0.5cm}
{\em Universidade Federal de S\~ao Carlos\\
Departamento de F\'{\i}sica \\
C.P. 676, 13565-905~~S\~ao Carlos(SP), Brasil}\\
\end{center}
\vspace{0.5cm}

\begin{abstract}
In this work we have developed the essential tools for the algebraic Bethe
ansatz solution of integrable vertex models
invariant by a unique $U(1)$ charge
symmetry.
The formulation is valid for
arbitrary statistical weights and respective
number $N$ of edge states.
We show that the fundamental commutation
rules between the  monodromy matrix elements are derived  by solving
linear systems of equations. This makes possible the construction
of the transfer matrix eigenstates
by means of a new recurrence relation depending on $N-1$ distinct types of creation fields.
The necessary identities to solve the eigenvalue problem  are obtained exploring the
unitarity property and the Yang-Baxter equation satisfied by the $R$-matrix.
The on-shell and off-shell properties of the algebraic Bethe ansatz are explicitly presented
in terms of the arbitrary $R$-matrix elements. This includes the transfer matrix eigenvalues,
the Bethe ansatz equations and the structure of the vectors not parallel
to the eigenstates.
\end{abstract}

\vspace{.15cm}
\centerline{PACS numbers:  05.50+q, 02.30.IK}
\vspace{.1cm}
\centerline{Keywords: Algebraic Bethe Ansatz, Lattice Integrable Models}
\vspace{.15cm}
\centerline{June 2008}

\end{titlepage}

%\tableofcontents

\pagestyle{empty}

\newpage

\pagestyle{plain}
\pagenumbering{arabic}

\renewcommand{\thefootnote}{\arabic{footnote}}

\section{Introduction}

The quantum inverse scattering method has been playing
a major role in the development of the theory of two-dimensional integrable models \cite{FA,TA,KO}.
This method  paved the way for the discovery of
important generalizations of the six-vertex model \cite{KS} as well as it
has helped to prompt the notion of quantum group symmetry \cite{QG,JI}. This approach also offers us the
basic tools to solve exactly quantum integrable systems including the computation
of lattice correlation functions
\cite{KO1,KI}.

The central object in the quantum inverse scattering method is the
monodromy matrix ${\cal T}_{\cal A}(\lambda)$ which depends on the continuous spectral parameter $\lambda$.
This operator is frequently viewed as a matrix on the auxiliary
space ${\cal A}$ and for an arbitrary $N$-dimensional space one can write,
\EQ
{\cal T}_{\cal A}(\lambda)=\sum_{a,b=1}^{N} {\cal T}_{a,b}(\lambda) e_{a,b}~,
\label{eq1}
\EN
where $e_{a,b}$ are the standard $N \times N$ Weyl matrices.

The matrix elements ${\cal T}_{a,b}(\lambda)$ act on the space of states of some quantum
physical system and they are
the generators of a quadratic algebra denominated Yang-Baxter algebra. The set of relations
defining this algebra are,
\EQ
\sum_{f,g=1}^{N} R(\lambda,\mu)_{a,b}^{f,g} {\cal T}_{f,c}(\lambda)
{\cal T}_{g,d}(\mu) =
\sum_{f,g=1}^{N} {\cal T}_{b,f}(\mu) {\cal T}_{a,g}(\lambda) R(\lambda,\mu)_{g,f}^{c,d}~,
\label{fundre}
\EN

In analogy to Lie algebras, functions
$ R(\lambda,\mu)_{a,b}^{c,d}$
can be interpreted as the structure constants of the Yang-Baxter algebra (\ref{fundre}). They are often
viewed as the elements of a
$N^2 \times N^2$ $R$-matrix acting on the tensor product of two auxiliary spaces. This matrix can be defined as,
\EQ
R_{12}(\lambda_1,\lambda_2)= \sum_{a,b,c,d=1}^{N} R_{a,b}^{c,d}(\lambda_1,\lambda_2) e_{a,c} \otimes e_{b,d} .
\label{rma}
\EN

The associativity of the Yang-Baxter algebra requires that the $R$-matrix satisfies the
celebrated Yang-Baxter equation \cite{BA1},
\EQ
R_{12}(\lambda_{1},\lambda_{2}) R_{13}(\lambda_{1},\lambda_{3}) R_{23}(\lambda_{2},\lambda_{3}) = R_{23}(\lambda_{2},\lambda_{3})
R_{13}(\lambda_{1},\lambda_{3})
R_{12}(\lambda_{1},\lambda_{2}) ,
\label{ybr}
\EN
where $R_{ab}(\lambda_a,\lambda_b)$ denotes the $R$-matrix
acting on the tensor product of the
spaces ${\cal{A}}_a \otimes {\cal{A}}_b$.

In this paper we shall consider the solutions of the
Yang-Baxter equation (\ref{ybr}) that can be normalized in
order to satisfy the unitarity property,
\EQ
R_{21}(\lambda,\mu)
R_{12}(\mu,\lambda)
= I_N \otimes I_N,
\label{uni}
\EN
where $I_N$ is the $N \times N$ identity matrix.

At this point we remark that the unitarity property (\ref{uni}) follows
from the Yang-Baxter equation (\ref{ybr}) under the extra assumption
that the $R$-matrix is regular. This hypothesis is equivalent to say that
$R_{ab}(\lambda,\lambda)$ is proportional to the permutator on $C_a^{N} \otimes C_b^{N}$. The
unitarity assures us that the $R$-matrix is invertible and this property is important to show
that the Yang-Baxter algebra leads us to mutually commuting operators. This family
of commuting operators is then obtained by taking the trace of the monodromy matrix
on the auxiliary space,
\EQ T(\lambda) =\sum_{a=1}^{N} {\cal T}_{a, a}(\lambda) ,
\label{tran0}
\EN

In the theory of integrable models the operator
$T(\lambda)$ is regarded as the generating function
of the corresponding quantum integrals of motion such as the underlying one-dimensional Hamiltonian.
The understanding of the physical properties of these quantum systems should therefore include at least the
knowledge of the eigenvalues and the eigenvectors of $T(\lambda)$.
In principle, the diagonalization of $T(\lambda)$ can be accomplished through the formulation of
the  algebraic Bethe ansatz. For comprehensive reviews  on this subject see for instance the references \cite{KO,TATA}.
The basic idea of this method is to exploit commutation relations between the monodromy matrix elements
coming from the
Yang-Baxter algebra (\ref{fundre}). In particular,
the eigenstates of
$T(\lambda)$ are constructed
by applying appropriate off-diagonal monodromy matrix elements,
usually named creation operators,
on a previously chosen  reference state. The main expected feature of this state is that the action
of the monodromy matrix on it gives us as result a triangular matrix. We emphasize, however,
that the existence of such reference state does not immediately guarantee the
success of an algebraic Bethe ansatz solution.  In fact, for general values of $N$,
we are not aware of any recipe to implement the algebraic Bethe ansatz
even when a possible reference state  is the trivial ferromagnetic highest vector.  For this class of models,
our current knowledge on algebraic Bethe ansatz formulations remains restrict to very specific
integrable systems. A representative category of such models are those
directly related to $R$-matrices based on the vector
representation of some special Lie algebras \cite{RE,TA1,MA} and
$Z_2$ graded superalgebras \cite{KU,ES,MA1}.

In order to bring new insights into the algebraic Bethe ansatz approach we have to consider
the diagonalization of
$T(\lambda)$ without referring
to any particular dependence of the $R$-matrix on the spectral parameters. This point of view is
rather evident when the $R$-matrix commutes with at least one $U(1)$ charge symmetry. This feature
assures us that the ferromagnetic pseudovacuum will play the role of a
suitable reference state for the most fundamental integrable model associated to such
$R$-matrix. The diagonalization of the corresponding $T(\lambda)$
should then be completed solely on basis of the commutation
relations derived from the Yang-Baxter algebra and the constraints
imposed by both the Yang-Baxter equation and the
unitarity property. Despite of its relevance, this strategy
of solving integrable models has so far remained largely ignored in the
literature. This is particularly the case when the dimension of the monodromy
matrix is $ N \ge 3$ since we have to consider
the presence of different
types of creation fields.
The basic problem is to unveil the role played by each creation operator on the structure
of the eigenvectors. This understanding is certainly more difficult when the $R$-matrices
elements are not specified. It is expected that identities among the weights will be crucial
to solve this problem.

In this paper we report on some progress towards the algebraic Bethe ansatz solution
of integrable models with arbitrary $R$-matrix
\footnote{ A small part of our results has been
briefly announced in reference \cite{CSB}.}. We will consider the simplest possible
family of models whose Hilbert space description requires us to consider
many independent quasi-particle excitations.
This turns out to be the systems  that their
$R$-matrices commute with a single $U(1)$ symmetry for arbitrary
values of $N$,
\EQ
[ R_{12}(\lambda,\mu), S^z \otimes I_N +I_N \otimes S^{z} ]=0
\label{u1}
\EN
where $S^{z}$ denotes the azimuthal component of an operator with spin
$s=(N-1)/2$.
Note that this invariance implies that
$R(\lambda,\mu)_{a,b}^{c,d} \ne 0$  only when the ice
rule $a+b=c+d$ is satisfied.

We can use property (\ref{u1}) to  express
the $R$-matrix in terms of the sectors labeling the $2N-1$ eigenvalues
of the $U(1)$ operator. In the Weyl basis indices these
sectors are easily parameterized by the charge $q=a+b-1$ and the
$R$-matrix can be written as,
\EQ
R_{12}(\lambda,\mu)=\sum_{q=1}^{2N-1}~~\sum_{a,c=M \lbrace 1,q+1-N
\rbrace }^{m \lbrace q, N \rbrace }
R(\lambda,\mu)_{a,q+1-a}^{c,q+1-c} e_{a,c}\otimes e_{q+1-a,q+1-c} .
\label{RMA}
\EN
where $M \lbrace x,y \rbrace$ and $m \lbrace x,y
\rbrace $ denotes the maximum and minimum integer of the pair
$\lbrace x,y \rbrace$, respectively.

These integrable systems can be seen as  multistate extensions of the totally asymmetric
six-vertex model  preserving a single conserved quantum number. The classical example
is the model whose $R$-matrix is based on the higher spin representation of the
quantum $U_q[SU(2)]$ algebra \cite{RES}.
Here we stress that a systematic classification
of $U(1)$ invariant solutions of the non-linear
Yang-Baxter equation is beyond the reach at present.
It is therefore rather probable that a variety of $U(1)$ invariant $R$-matrices
are still  waiting to be discovered.
This should include solutions with non-additive $R$-matrices as well as those with infinity
number of degrees of freedom typical of non-compact models.  An example of the latter system
could be the vertex model based on the discrete infinite
dimensional representation of the $SL(2,R)$ algebra. Note that such representation has a
highest weight which plays the role of possible reference state. The
framework developed
here has the nice feature of being able to
accommodate the algebraic solution of all
such different systems
in a rather unified way.

We have organized this paper as follows. In next section we describe the basic properties
of the vertex model associated to the fundamental representation of the Yang-Baxter
algebra based on the $R$-matrices satisfying the ice rule (\ref{u1}). In section 3 we describe the essential
tools that are necessary to obtain the appropriate commutation rules between the monodromy
matrix elements. It is argued that these relations are obtained by solving
coupled  systems of
linear equations. To fix up the main idea of our method we presented the structure
of specific sets of commutation
rules.  In section 4 we discuss a procedure to obtain suitable identities between
the elements of the $R$-matrix from that Yang-Baxter and unitarity relations. These
identities are decisive for the solution of the transfer matrix eigenvalue problem.
In section 5 we diagonalize the transfer matrix by means of an algebraic Bethe ansatz.
The respective eigenstates are
constructed similarly to that of a bosonic Fock space with $N-1$ distinct creation fields. We
present the explicit expressions for both the on-shell and off-shell parts of the action of
the transfer matrix on the eigenstates in terms of the statistical weights.
Our conclusions are presented in section 6.
In appendices $A$ and $B$ we summarize the technical details
entering the solution of the two and three particle problems, respectively.

\section{The vertex model representation}
\label{sec2}

Each solution of the Yang-Baxter equation (\ref{ybr}) gives rise
to representations of the Yang-Baxter algebra (\ref{fundre}).
The representations depending on the spectral parameter are named
Lax operators denoted here by ${\cal L}_{{\cal A} i}(\lambda)$. The
simplest such representation can be obtained directly from the $R$-matrix amplitudes by the expression,
\EQ
\label{lr}
{\cal L}_{{\cal A} i}(\lambda)=\sum_{a,b,c,d=1}^{N} R(\lambda,\mu_i)_{a,b}^{c,d}  e_{a,c} \otimes e_{b,d}^{(i)} ,
\EN
where $e_{b,d}^{(i)}$ are $N\times N$ Weyl matrices acting on the
tensor product $\prod_{i=1}^{L}\otimes _{i}^{N}$ space of an one-dimensional lattice of length L.
The variables $\mu_i$ play the role of free continuous parameters.

The Yang-Baxter algebra (\ref{fundre}) has the co-multiplication property that the
tensor product of two representations is still another possible representation.
Hence, the following ordered product of ${\cal L}$-operators
\EQ
\label{mono}
{\cal T}_{\cal A}(\lambda) =
{\cal L}_{{\cal A}L}(\lambda,\mu_{L}) {\cal L}_{{\cal A}L-1}(\lambda,\mu_{L-1}) \dots
{\cal L}_{{\cal A}1}(\lambda,\mu_{1}) ,
\EN
is indeed a representation of the quadratic algebra (\ref{fundre}).

In the context of classical vertex models of statistical mechanics
${\cal L}_{{\cal A} i}(\lambda)$ encoded the structure of the local Boltzmann weights at
the $i$-th site of a square lattice of size $L$. The possible states of such  statistical system
are then associated to the possible bond configurations
$a,b,c,d =1, \dots, N $  of each vertex on the $L \times L$ lattice.
The energy of each $i$-th vertex configuration is then associated
to the statistical weight $R(\lambda,\mu_i)_{a,b}^{c,d}$.
The row-to-row transfer matrix of these vertex models $T(\lambda)$
can be written in a compact form with the help of
the monodromy matrix (\ref{mono}).  It is given by the trace of
${\cal T}_{\cal A}(\lambda)$ over the auxiliary space,
\EQ
T(\lambda)=Tr_{\cal A} [
{\cal L}_{{\cal A}L}(\lambda,\mu_{L}) {\cal L}_{{\cal A}L-1}(\lambda,\mu_{L-1}) \dots
{\cal L}_{{\cal A}1}(\lambda,\mu_{1})] .
\label{tran}
\EN

In order to diagonalize
the operator $T(\lambda)$ within the quantum inverse
scattering framework one needs to know an exact eigenstate $\ket{0}$ of $T(\lambda)$.
This vector works a reference state in the construction of the Hilbert space of $T(\lambda)$ by an algebraic Bethe
ansatz.
In general, such state is searched by asking that the action of the lower left monodromy matrix elements on
$\ket{0}$ are annihilated for arbitrary $\lambda$.
This means that the monodromy matrix acts as an upper triangular matrix on $\ket{0}$, namely
\EQ
{\cal T}_{a,b}(\lambda) \ket{0}=\begin{cases}
\displaystyle  w_{a}(\lambda) \ket{0},
~~~~\mbox{for} ~~ a=b \cr \displaystyle 0, ~~~~ \mbox{for} ~~ a>b \cr \displaystyle
\ket{ab}, ~~~~\mbox{for} ~~ a<b ,
\label{mono1}
\end{cases}
\EN
where $\ket{ab}$ denotes a general non-null vector.

The action of the upper and the lower elements
${\cal T}_{a,b}(\lambda)$ on $\ket{0}$ has therefore distinct meanings.  For $a <b$ they play
the role of creation fields while for $a >b$ they can be thought as annihilators.
The presence of the $U(1)$ invariance (\ref{u1}) makes it possible to build up such reference state
in terms of the tensor product of local ferromagnetic vectors,
\EQ
\ket{0}=\prod_{i=1}^{L} \otimes \ket{s}_{i}, ~~~~ \ket{s}_{i}=\left(\begin{array}{c}
1 \\ 0 \\ \vdots \\ 0 \end{array}\right)_{N}.
\EN

The state $\ket{s}_i$ can be viewed as the highest eigenstate
of the spin operators  $S_i^{z}$ acting on the $i$-th site of a chain of length $L$. The total spin of the reference state
is therefore $\frac{(N-1)L}{2}$ and this is the reason why $\ket{0}$ is called the ferromagnetic state.
From Eqs.(\ref{RMA},\ref{lr},\ref{mono}) it is not difficult to see that property (\ref{mono1}) is in fact satisfied and that
the expressions for functions $w_{a}(\lambda)$ are,
\EQ
w_{a}(\lambda)=\prod_{i=1}^{L} R(\lambda,\mu_{i})_{a, 1}^{a, 1} .
\EN

The next step would then be the construction of
other  eigenstates of $T(\lambda)$ besides $\ket{0}$. To this end we shall first
consider the action of the total spin operator
$\sum_{i=1}^{L} S_i^{z}$ on the monodromy matrix elements
${\cal{T}}_{a,b}(\lambda)$.  Considering the structure of the  Lax operators (\ref{RMA},\ref{lr})
it is possible to derive the
following relation,
\EQ
\left [
{\cal{T}}_{a,b}(\lambda), \sum_{i=1}^{L} S_i^{z} \right ]=
(b-a) {\cal{T}}_{a,b}(\lambda) .
\label{pro}
\EN

The above commutation relation is useful to illuminate the physical content of the
creation fields
${\cal{T}}_{a,b}(\lambda)$ for $a < b$. In fact,
by acting Eq.(\ref{pro}) on the reference state $\ket{0}$ one derives the property,
\EQ
\sum_{i=1}^{L} S_i^{z}
{\cal{T}}_{a,b}(\lambda) \ket{0}= \left [ s(L-1) +s-b+a \right ]
{\cal{T}}_{a,b}(\lambda) \ket{0}~~~\mathrm{for}~~ a<b .
\label{pro1}
\EN

From Eq.(\ref{pro1}) we conclude that the field
${\cal{T}}_{a,b}(\lambda)$ for $a < b$
behaves as raising operator of an excitation over $\ket{0}$ whose value of the
azimuthal spin component is $s_{a,b}=s-b+a$. This means that all monodromy
matrix elements of a given upper diagonal have the same azimuthal spin and therefore describe
the same type of excitations. As a consequence of that we end up with only $N-1$ linear
independent creations fields which is represented, in the simplest manner, by the first row of the
monodromy matrix
${\cal{T}}_{1,b}(\lambda)$ for $b=2,\cdots,N$.
The analogy with the Hilbert space of high spin
Heisenberg magnets strongly suggests that the eigenstates of
the transfer matrix (\ref{tran})  could be  build up in an algebraic fashion
by using the help of such $N-1$ independent creation fields. The whole construction, however,
will depend very
much on our ability to recast the Yang-Baxter algebra (\ref{fundre}) in the form of
convenient commutation rules between the diagonal and the off-diagonal monodromy matrix elements.
In next section we will
deal with this artistic part
of the quantum inverse scattering method.

\section{The fundamental commutation rules}
\label{sec3}

The purpose of this section is to discuss the structure of the
commutation relations that are relevant in the exact diagonalization of the transfer matrix (\ref{tran}).
In principle, these relations between the monodromy matrix elements
are derived considering the Yang-Baxter algebra (\ref{fundre}). Let us
use the symbol $[\alpha;\beta]$ to represent the $\alpha$-th row and the $\beta$-th
column of the $N^2 \times N^2$ matrix defined by Eq.(\ref{fundre}). The
projection of the Yang-Baxter algebra on the
rather general set of entries $[(\bar{a}-1)N+\bar{b};(\bar{c}-1)N+\bar{d}]$ can be written as,
\bear
\label{fundrelindex0}
\sum_{\bar{e}=M \lbrace 1,\bar{a}+\bar{b}-N \rbrace}^{m \lbrace \bar{a}+\bar{b}-1, N \rbrace}
R(\lambda,\mu)_{\bar{b}, \bar{a}}^{\bar{e}, \bar{a}+\bar{b}-\bar{e}}
{\cal T}_{\bar{e}, \bar{c}}(\lambda)
{\cal T}_{\bar{a}+\bar{b}-\bar{e}, \bar{d}}(\mu)&=&
\sum_{\bar{e}=M \lbrace 1,\bar{c}+\bar{d}-N \rbrace}^{m \lbrace \bar{c}+\bar{d}-1, N \rbrace}
{\cal T}_{\bar{a}, \bar{e}}(\mu)
{\cal T}_{\bar{b}, \bar{c}+\bar{d}-\bar{e}}(\lambda)
R(\lambda,\mu)_{\bar{c}+\bar{d}-\bar{e},\bar{e}}^{\bar{c},\bar{d}}
\nonumber \\
&& \bar{a},\bar{b},\bar{c},\bar{d}  = 1 \dots N,
\ear

In what follows we will discuss three different families of
commutation rules between the diagonal, creation and annihilation
operators which can be derived from Eq.(\ref{fundrelindex0}).

\subsection{The diagonal and creation fields}
\label{sub31}

The creation operators
$ {\cal T}_{1, b}(\lambda)$ for $b=2,\dots,N $ will provide us a basis to construct the
eigenvectors of the transfer matrix $T(\lambda)$.
Their commutation rules with the diagonal monodromy matrix elements $ {\cal T}_{a, a}(\lambda)$ for
$a=1,\dots,N $ are therefore essential in the transfer matrix eigenvalue
problem. These relations are obtained from the entries $[a;(a+c-1)N+b-c]$ upon
suitable choices of the extra variable $c$. By substituting
$\bar{a}=1$, $\bar{b}=a$, $\bar{c}=a+c$ and $\bar{d}=b-c$ in
Eq.(\ref{fundrelindex0}) we find that the respective equations are,
\bear
\label{fundrelindex1}
\sum_{\bar{e}=1}^{a} R(\lambda,\mu)_{a, 1}^{\bar{e}, a-\bar{e}+1}
{\cal T}_{\bar{e}, a+c}(\lambda)
{\cal T}_{a-\bar{e}+1, b-c}(\mu)=
\sum_{\bar{e}=M \lbrace 1,a+b-N \rbrace}^{m \lbrace a+b-1, N \rbrace}
{\cal T}_{1, \bar{e}}(\mu)
{\cal T}_{a, a+b-\bar{e}}(\lambda)
R(\lambda,\mu)_{a+b-\bar{e}, \bar{e}}^{a+c, b-c}.
\ear

In general, to obtain commutation rules that
are useful for the eigenvalue problem, we still have
to  elaborate on
Eq.(\ref{fundrelindex1}).  The additional manipulation consists in making
particular combinations of a number of equations derived from
Eq.(\ref{fundrelindex1}) with the help of the variable $c$. In Table 1
we have summarized in detail the linear combination required
for each diagonal field.
\begin{table}[ht]
\begin{center}
\begin{tabular}{|c|c|c|c|}
\hline
Operator  & Diagonal & Combination & Number of  \\
    & index & index & equations \\
\hline
${\cal T}_{1,1}(\mu) {\cal T}_{1,b}(\lambda)$ & $a=1$ & $c=b-1$ & $1$ \\
\hline
${\cal T}_{a,a}(\lambda) {\cal T}_{1,b}(\mu)$ & $ 2 \le a \le N+1-b $ & $ c=0, \dots, b-1 $  & $ b $ \\
\hline
${\cal T}_{a,a}(\lambda) {\cal T}_{1,b}(\mu)$  & $ N+2-b \le a < N $ & $ c=0, \dots, N-a $ & $N+1-a$ \\
\hline
${\cal T}_{N, N}(\lambda) {\cal T}_{1,b}(\mu)$  & $  a = N $ & $ c=0 $ & $1$ \\
\hline
\end{tabular}
\caption{\footnotesize{The linear combination made from Eq.(\ref{fundrelindex1})
to obtain commutation rules between the fields ${\cal T}_{a,a}(\lambda)$ and ${\cal T}_{1,b}(\lambda)$.
The number
of equations entering the linear combination is governed by the index $c$.}}
\end{center}
\end{table}

From Table 1 we see that only the commutation rules for the operators $
{\cal T}_{1,1}(\lambda)$ and ${\cal T}_{N,N}(\lambda)$ follows directly from
Eq.(\ref{fundrelindex1}). Indeed, by setting $c=b-1$ for
${\cal T}_{1,1}(\lambda)$ and $c=0$ for ${\cal T}_{N,N}(\lambda)$  one finds
that their commutation rules  with
${\cal T}_{1,b}(\mu)$  are,
\EQ
{\cal T}_{1, 1}(\lambda)
{\cal T}_{1, b}(\mu)
=
\frac{R(\mu,\lambda)_{1, 1}^{1, 1}}{R(\mu,\lambda)_{b, 1}^{b, 1}}
{\cal T}_{1, b}(\mu)
{\cal T}_{1, 1}(\lambda)
-\sum_{\bar{e}=2}^{b}
\frac{R(\mu,\lambda)_{1+b-\bar{e}, \bar{e}}^{b, 1}}{R(\mu,\lambda)_{b, 1}^{b, 1}}
{\cal T}_{1, \bar{e}}(\lambda)
{\cal T}_{1, 1+b-\bar{e}}(\mu)
\label{t11geral}
\EN
\bear
{\cal T}_{N, N}(\lambda)
{\cal T}_{1, b}(\mu) &=&
\frac{R(\lambda,\mu)_{N,b}^{N, b}}{R(\lambda,\mu)_{N, 1}^{N, 1}}
{\cal T}_{1, b}(\mu)
{\cal T}_{N, N}(\lambda)
+ \sum_{\bar{e}=b+1}^{N}
\frac{R(\lambda,\mu)_{N+b-\bar{e}, \bar{e}}^{N, b}}{R(\lambda,\mu)_{N, 1}^{N, 1}}
{\cal T}_{1, \bar{e}}(\mu)
{\cal T}_{N, N+b-\bar{e}}(\lambda)
\nonumber \\
&-&\sum_{\bar{e}=1}^{N-1}
\frac{R(\lambda,\mu)_{N, 1}^{\bar{e}, N-\bar{e}+1}}{R(\lambda,\mu)_{N, 1}^{N, 1}}
{\cal T}_{\bar{e}, N}(\lambda)
{\cal T}_{N-\bar{e}+1, b}(\mu)
\nonumber \\
\label{tNNgeral}
\ear

The commutation rules for the
remaining diagonal operators
${\cal T}_{a,a}(\lambda)$  for $ 2 \le a \le N-1$ demand a considerable amount of extra work.
In these cases we first have to implement the linear combination
according to the indices $c$ exhibited in Table 1. As an example let us consider the linear system
associated to the diagonal fields with $ 2 \le a \le N+1-b$. We find that the relations coming
from
Eq.(\ref{fundrelindex1})  for $c=0,\dots,b$ can be arranged in the following form,
\bear
\label{diagon1}
A^{(a,b)}_1(\lambda,\mu)
\left(\begin{array}{c}
{\cal T}_{1, 1}(\mu)
{\cal T}_{a, a+b-1}(\lambda) \\
{\cal T}_{1, 2}(\mu)
{\cal T}_{a, a+b-2}(\lambda) \\
\vdots \\
{\cal T}_{1, b}(\mu)
{\cal T}_{a, a}(\lambda)
\end{array}\right)
= \sum_{\bar{e}=1}^{a} R(\lambda,\mu)_{a, 1}^{\bar{e}, a-\bar{e}+1}
\left(\begin{array}{c}
{\cal T}_{\bar{e}, a}(\lambda)
{\cal T}_{a-\bar{e}+1, b}(\mu) \\
{\cal T}_{\bar{e}, a+1}(\lambda)
{\cal T}_{a-\bar{e}+1, b-1}(\mu) \\
\vdots \\
{\cal T}_{\bar{e}, a+b-1}(\lambda)
{\cal T}_{a-\bar{e}+1, 1}(\mu)
\end{array}\right)
\nonumber \\
%&-&
-
\sum_{\bar{e}=b+1}^{a+b-1}
v^{(a,b)}_1(\lambda,\mu)
{\cal T}_{1, \bar{e}}(\mu)
{\cal T}_{a, a+b-\bar{e}}(\lambda)
%\nonumber \\
~~\mathrm{for}~~~ 2 \le  a \le N+1-b,
\ear
where the $b \times b$ matrix $A^{(a,b)}_1(\lambda,\mu)$ is given in terms of the structure constants as
\EQ
\label{am1}
A^{(a,b)}_1(\lambda,\mu)=\left( \begin{array}{cccc}
R(\lambda,\mu)_{a+b-1, 1}^{a, b} & R(\lambda,\mu)_{a+b-2, 2}^{a, b} & \cdots & R(\lambda,\mu)_{a, b}^{a, b} \\
R(\lambda,\mu)_{a+b-1, 1}^{a+1, b-1} & R(\lambda,\mu)_{a+b-2, 2}^{a+1, b-1} & \cdots & R(\lambda,\mu)_{a, b}^{a+1, b-1} \\
\vdots & \vdots & \ddots & \vdots \\
R(\lambda,\mu)_{a+b-1, 1}^{a+b-1, 1} & R(\lambda,\mu)_{a+b-2, 2}^{a+b-1, 1} & \cdots & R(\lambda,\mu)_{a, b}^{a+b-1, 1}
\end{array}\right),
\EN
while the $b \times 1$ vector $v^{(a,b)}_1(\lambda,\mu)$ is
\EQ
\label{v1}
v^{(a,b)}_1(\lambda,\mu)=\left( \begin{array}{c}
R(\lambda,\mu)_{a+b-\bar{e}, \bar{e}}^{a, b}
\\
R(\lambda,\mu)_{a+b-\bar{e}, \bar{e}}^{a+1, b-1}
\\
\vdots
\\
R(\lambda,\mu)_{a+b-\bar{e}, \bar{e}}^{a+b-1, 1}
\end{array}\right).
\EN

We have now reached a point in which the properties of linear systems
of equations are applicable. In fact, the term
${\cal T}_{1, b}(\mu)
{\cal T}_{a, a}(\lambda)$  can be systematically single out from the left-hand side of Eq.(\ref{diagon1})
with the help of Cramer's rule. Besides that we note the first
non-homogeneous part of Eq.(\ref{diagon1}) contains
the reverse product ${\cal T}_{a, a}(\lambda) {\cal T}_{1, b}(\mu)$ when
the sum index is $\bar{e}=a$. Therefore, by computing determinants of $b \times b$ matrices
we are able to generate a unique linear equation relating the operator products
${\cal T}_{a, a}(\lambda) {\cal T}_{1, b}(\mu)$ and
${\cal T}_{1, b}(\mu)
{\cal T}_{a, a}(\lambda)$.  This equation is then easily solved
in order to establish the commutation rule between
${\cal T}_{a, a}(\lambda)$ and ${\cal T}_{1, b}(\mu)$
for $2 \le a \le N+1-b$.

Clearly, the same method described above can be used to obtain the commutation
rules for the other values $N+2-b \le a \le N-1$ of the index $a$.
The system of linear equations is however different and for sake of
completeness  we also present it here.
Following Table 1 one finds that the respective Eq.(\ref{fundrelindex1}) can be written as,
\bear
\label{diagon2}
A^{(a,b)}_2(\lambda,\mu)
\left(\begin{array}{c}
{\cal T}_{1, a+b-N}(\mu)
{\cal T}_{a, N}(\lambda) \\
{\cal T}_{1, a+b-N+1}(\mu)
{\cal T}_{a, N-1}(\lambda) \\
\vdots \\
{\cal T}_{1, b}(\mu)
{\cal T}_{a, a}(\lambda)
\end{array}\right)
=
\sum_{\bar{e}=1}^{a} R(\lambda,\mu)_{a, 1}^{\bar{e}, a-\bar{e}+1}
\left(\begin{array}{c}
{\cal T}_{\bar{e}, a}(\lambda)
{\cal T}_{a-\bar{e}+1, b}(\mu) \\
{\cal T}_{\bar{e}, a+1}(\lambda)
{\cal T}_{a-\bar{e}+1, b-1}(\mu) \\
\vdots \\
{\cal T}_{\bar{e}, N}(\lambda)
{\cal T}_{a-\bar{e}+1, a+b-N}(\mu)
\end{array}\right)
\nonumber \\
-
\sum_{\bar{e}=b+1}^{N}
v^{(a,b)}_2(\lambda,\mu)
{\cal T}_{1, \bar{e}}(\mu)
{\cal T}_{a, a+b-\bar{e}}(\lambda)
%\nonumber \\
~~\mathrm{for}~~~ N+2-b  \le  a \le N-1,
\ear
where the $(N+1-a) \times (N+1-a)$ matrix $A^{(a,b)}_2(\lambda,\mu)$ is given by
\EQ
\label{am2}
A^{(a,b)}_2(\lambda,\mu)=
\left( \begin{array}{cccc}
R(\lambda,\mu)_{N, a+b-N}^{a, b} & R(\lambda,\mu)_{N-1, a+b-N+1}^{a, b} & \cdots & R(\lambda,\mu)_{a, b}^{a, b} \\
R(\lambda,\mu)_{N, a+b-N}^{a+1, b-1} & R(\lambda,\mu)_{N-1, a+b-N+1}^{a+1, b-1} & \cdots & R(\lambda,\mu)_{a, b}^{a+1, b-1} \\
\vdots & \vdots & \ddots & \vdots \\
R(\lambda,\mu)_{N, a+b-N}^{N, a+b-N} & R(\lambda,\mu)_{N-1, a+b-N+1}^{N, a+b-N} & \cdots & R(\lambda,\mu)_{a, b}^{N, a+b-N}
\end{array}\right),
\EN
while the $(N+1-a) \times 1$ vector $v^{(a,b)}_2(\lambda,\mu)$ is
\EQ
\label{v2}
v^{(a,b)}_2(\lambda,\mu)=\left( \begin{array}{c}
R(\lambda,\mu)_{a+b-\bar{e}, \bar{e}}^{a, b}
\\
R(\lambda,\mu)_{a+b-\bar{e}, \bar{e}}^{a+1, b-1}
\\
\vdots
\\
R(\lambda,\mu)_{a+b-\bar{e}, \bar{e}}^{N, a+b-N}
\end{array}\right).
\EN

As before the desirable commutation relations between
${\cal T}_{a, a}(\lambda)$ and ${\cal T}_{1, b}(\mu)$
for
$N+2-b  \le  a \le N-1$  are obtained by using Cramer's rule
in the linear system (\ref{diagon2}-\ref{v2}). We emphasize that the commutation
rules discussed in this subsection have the property that the corresponding
${\cal T}_{1,\bar{e}}(\mu)$ fields that appear on the left-hand side of the products satisfy
the condition $ \bar{e} \ge b$. Here we recall that these type of fields are the ones with
clear potential to participate directly on the structure of the eigenstates. The
condition $\bar{e} \ge b$ guarantees that such
${\cal T}_{1,\bar{e}}(\mu)$ operators, generated by
taking ${\cal T}_{a,a}(\lambda)$ through
${\cal T}_{1,b}(\mu)$, will only contribute to eigenvectors made out creation fields
with spin lower or equal to $s_{1,b}$. In this manner we assured that the eigenvectors
constructed in terms of the $N-1$
${\cal T}_{1,b}(\mu)$ operators can indeed be thought as multiparticle states
ordered by their spin values.

Let us now illustrate how the general method
explained above works in practice.  To this end we shall present explicitly
the complete structure of the
commutation rules between the first two creation fields with all the diagonal
operators. The simplest example concerns to  the
${\cal T}_{1, 2}(\mu)$ operator.  Its commutation rules with the diagonal operators
play a fundamental role in the transfer matrix
eigenvalue problem since they dictate in which way
the transfer matrix eigenvalues depend
on the arbitrary $R$-matrix elements.  These relations
for $a=1$ and $a=N$ follows directly from
Eqs.(\ref{t11geral},\ref{tNNgeral}) while those
for $ 2 \le a \le N-1$ are obtained by solving the $2 \times 2$ linear system
of equations
(\ref{diagon1}-\ref{v1}). It turns out that the final results for the
commutation rules between
${\cal T}_{a,a}(\lambda)$  and
${\cal T}_{1,2}(\mu)$ are,

\bear
{\cal T}_{1,1}(\lambda)
{\cal T}_{1,2}(\mu) & = &\frac{R(\mu,\lambda)^{1,1}_{1,1}}
{R(\mu,\lambda)^{2,1}_{2,1}}
{\cal
T}_{1,2}(\mu) {\cal T}_{1,1}(\lambda)
-\frac{R(\mu,\lambda)_{1, 2}^{2, 1}}{R(\mu,\lambda)_{2, 1}^{2, 1}}
{\cal T}_{1,2}(\lambda) {\cal T}_{1,1}(\mu),
\label{1part1}
\ear
\bear
{\cal T}_{a,a}(\lambda)
{\cal T}_{1,2}(\mu) & =&
D_{2}^{(a,0)}(\lambda,\mu){\cal
T}_{1,2}(\mu) {\cal
T}_{a,a}(\lambda)
+\sum_{\bar{e}=3}^{a+1}
D_{2}^{(a,\bar{e}-2)}(\lambda,\mu){\cal
T}_{1,\bar{e}}(\mu) {\cal
T}_{a,a+2-\bar{e}}(\lambda)
\nonumber \\
&+& \sum_{\bar{e}=1}^{a}
\frac{R(\lambda,\mu)_{a+1,1}^{a,2}}{R(\lambda,\mu)_{a+1,1}^{a+1,1}}
 \frac{R(\lambda,\mu)_{a, 1}^{\bar{e}, a-\bar{e}+1}}{R(\lambda,\mu)_{a, 1}^{a, 1}}
{\cal T}_{\bar{e},a+1}(\lambda) {\cal T}_{a-\bar{e}+1,1}(\mu)
\nonumber \\
&-& \sum_{\bar{e}=1}^{a-1}
\frac{R(\lambda,\mu)_{a, 1}^{\bar{e}, a-\bar{e}+1}}{R(\lambda,\mu)_{a,1}^{a,1}}
{\cal T}_{\bar{e},a}(\lambda) {\cal T}_{a-\bar{e}+1,2}(\mu)
~~~~~\mathrm{for}~~~
2 \le a \le N-1,
\label{1parta}
\ear
\bear
{\cal T}_{N, N}(\lambda)
{\cal T}_{1, 2}(\mu) &=&
\frac{R(\lambda,\mu)_{N,2}^{N, 2}}{R(\lambda,\mu)_{N, 1}^{N, 1}}
{\cal T}_{1, 2}(\mu)
{\cal T}_{N, N}(\lambda)
+ \sum_{\bar{e}=3}^{N}
\frac{R(\lambda,\mu)_{N+2-\bar{e}, \bar{e}}^{N, 2}}{R(\lambda,\mu)_{N, 1}^{N, 1}}
{\cal T}_{1, \bar{e}}(\mu)
{\cal T}_{N, N+2-\bar{e}}(\lambda)
\nonumber \\
&-&\sum_{\bar{e}=1}^{N-1}
\frac{R(\lambda,\mu)_{N, 1}^{\bar{e}, N-\bar{e}+1}}{R(\lambda,\mu)_{N, 1}^{N, 1}}
{\cal T}_{\bar{e}, N}(\lambda)
{\cal T}_{N-\bar{e}+1, 2}(\mu)
\label{1partN}
\ear

Function
$D^{(a,\bar{e})}_{2}(\lambda,\mu)$ is defined by using the determinant of a matrix
whose elements combines the first column of $A_1^{(a,2)}(\lambda,\mu)$ with the
vector
$v_1^{(a,2)}(\lambda,\mu)$. Its expression  in terms of the $R$-matrix elements is,
\bear
D^{(a,\bar{e})}_{2}(\lambda,\mu) & = &
-\frac{\left| \begin{array}{cc}
R(\lambda,\mu)_{a+1,1}^{a,2} & R(\lambda,\mu)_{a-\bar{e},\bar{e}+2}^{a,2} \\
R(\lambda,\mu)_{a+1, 1}^{a+1, 1} & R(\lambda,\mu)_{a-\bar{e},\bar{e}+2}^{a+1,1}
\end{array} \right|}{R(\lambda,\mu)_{a, 1}^{a, 1}
R(\lambda,\mu)_{a+1,1}^{a+1,1}}, \nonumber \\
&\mathrm{for}&~~~\bar{e}=0,\cdots,a-1~~~\mathrm{and}~~~2 \le a \le N-1.
\label{pda}
\ear

The next simplest case is related to the
${\cal T}_{1, 3}(\mu)$ operator.  In order to get the complete set
of commutation relations we have now to solve two different linear systems
with sizes $3 \times 3$ and $2 \times 2$. According to Table 1 the former
is defined by
Eqs.(\ref{diagon1}-\ref{v1}) while the latter is associated to
Eqs.(\ref{diagon2}-\ref{v2}).  By solving such linear systems of
equations we find that the commutation relations between
${\cal T}_{a,a}(\lambda)$ and
${\cal T}_{1,3}(\mu)$ are given by,
\bear
{\cal T}_{1, 1}(\lambda)
{\cal T}_{1, 3}(\mu)
&=&
\frac{R(\mu,\lambda)_{1, 1}^{1, 1}}{R(\mu,\lambda)_{3, 1}^{3, 1}}
{\cal T}_{1, 3}(\mu)
{\cal T}_{1, 1}(\lambda)
-\frac{R(\mu,\lambda)_{2, 2}^{3, 1}}{R(\mu,\lambda)_{3, 1}^{3, 1}}
{\cal T}_{1, 2}(\lambda)
{\cal T}_{1, 2}(\mu)
\nonumber \\
&-&\frac{R(\mu,\lambda)_{1, 3}^{3, 1}}{R(\mu,\lambda)_{3, 1}^{3, 1}}
{\cal T}_{1, 3}(\lambda)
{\cal T}_{1, 1}(\mu)
\label{2part1}
\ear
\bear
{\cal T}_{a,a}(\lambda)
{\cal T}_{1,3}(\mu) & =&
D_{3}^{(a,0)}(\lambda,\mu){\cal
T}_{1,3}(\mu) {\cal
T}_{a,a}(\lambda)
+\sum_{\bar{e}=4}^{a+2}
D_{3}^{(a,\bar{e}-3)}
(\lambda,\mu){\cal T}_{1,\bar{e}}(\mu)
{\cal T}_{a,a+3-\bar{e}}(\lambda)
\nonumber \\
&-& \sum_{\bar{e}=1}^{a-1}
\frac{R(\lambda,\mu)_{a, 1}^{\bar{e}, a-\bar{e}+1}}{R(\lambda,\mu)_{a,1}^{a,1}}
{\cal T}_{\bar{e},a}(\lambda) {\cal T}_{a-\bar{e}+1,3}(\mu)
\nonumber \\
&+& \sum_{\bar{e}=1}^{a}
\frac{R(\lambda,\mu)_{a, 1}^{\bar{e}, a-\bar{e}+1}}{R(\lambda,\mu)_{a, 1}^{a, 1}}
\frac{\left| \begin{array}{cc}
R(\lambda,\mu)_{a+2,1}^{a,3} & R(\lambda,\mu)_{a+1,2}^{a,3} \\
R(\lambda,\mu)_{a+2,1}^{a+2,1} & R(\lambda,\mu)_{a+1,2}^{a+2,1}
\end{array} \right|}
{\left| \begin{array}{cc}
R(\lambda,\mu)_{a+2,1}^{a+1,2} & R(\lambda,\mu)_{a+1,2}^{a+1,2} \\
R(\lambda,\mu)_{a+2,1}^{a+2,1} & R(\lambda,\mu)_{a+1,2}^{a+2,1}
\end{array} \right|}
{\cal T}_{\bar{e},a+1}(\lambda) {\cal T}_{a-\bar{e}+1,2}(\mu)
\nonumber \\
&-& \sum_{\bar{e}=1}^{a}
\frac{R(\lambda,\mu)_{a, 1}^{\bar{e}, a-\bar{e}+1}}{R(\lambda,\mu)_{a, 1}^{a, 1}}
\frac{\left| \begin{array}{cc}
R(\lambda,\mu)_{a+2,1}^{a,3} & R(\lambda,\mu)_{a+1,2}^{a,3} \\
R(\lambda,\mu)_{a+2,1}^{a+1,2} & R(\lambda,\mu)_{a+1,2}^{a+1,2}
\end{array} \right|}
{\left| \begin{array}{cc}
R(\lambda,\mu)_{a+2,1}^{a+1,2} & R(\lambda,\mu)_{a+1,2}^{a+1,2} \\
R(\lambda,\mu)_{a+2,1}^{a+2,1} & R(\lambda,\mu)_{a+1,2}^{a+2,1}
\end{array} \right|}
{\cal T}_{\bar{e},a+2}(\lambda) {\cal T}_{a-\bar{e}+1,1}(\mu)
\nonumber \\
&&~~~~~\mathrm{for}~~~
2 \le a \le N-2,
\label{2parta}
\ear
\bear
{\cal T}_{N-1, N-1}(\lambda)
{\cal T}_{1, 3}(\mu) &=&
-\frac{\left| \begin{array}{cc}
R(\lambda,\mu)_{N,2}^{N-1,3} & R(\lambda,\mu)_{N-1,3}^{N-1,3} \\
R(\lambda,\mu)_{N,2}^{N,2} & R(\lambda,\mu)_{N-1,3}^{N,2}
\end{array} \right|}
{R(\lambda,\mu)_{N-1, 1}^{N-1, 1} R(\lambda,\mu)_{N, 2}^{N, 2}}
{\cal T}_{1, 3}(\mu) {\cal T}_{N-1, N-1}(\lambda)
\nonumber \\
&-&
\sum_{\bar{e}=4}^{N} \frac{\left| \begin{array}{cc}
R(\lambda,\mu)_{N,2}^{N-1,3} & R(\lambda,\mu)_{N+2-\bar{e},\bar{e}}^{N-1,3} \\
R(\lambda,\mu)_{N,2}^{N,2} & R(\lambda,\mu)_{N+2-\bar{e},\bar{e}}^{N,2}
\end{array} \right|}
{R(\lambda,\mu)_{N-1, 1}^{N-1, 1} R(\lambda,\mu)_{N, 2}^{N, 2}}
{\cal T}_{1, \bar{e}}(\mu) {\cal T}_{N-1, N+2-\bar{e}}(\lambda)
\nonumber \\
&+&
\sum_{\bar{e}=1}^{N-2}
\left[\frac{R(\lambda,\mu)_{N,2}^{N-1, 3}}{R(\lambda,\mu)_{N, 2}^{N, 2}}
{\cal T}_{\bar{e}, N}(\lambda) {\cal T}_{N-\bar{e}, 2}(\mu)
-{\cal T}_{\bar{e}, N-1}(\lambda) {\cal T}_{N-\bar{e}, 3}(\mu) \right]
\nonumber \\
&\times &
\frac{R(\lambda,\mu)_{N-1,1}^{\bar{e}, N-\bar{e}}}{R(\lambda,\mu)_{N-1, 1}^{N-1, 1}}
+
\frac{R(\lambda,\mu)_{N,2}^{N-1,3}} {R(\lambda,\mu)_{N, 2}^{N, 2}}
{\cal T}_{N-1, N}(\lambda) {\cal T}_{1, 2}(\mu)
 \nonumber \\
\label{2partN-1}
\ear
\bear
{\cal T}_{N, N}(\lambda)
{\cal T}_{1, 3}(\mu) &=&
\frac{R(\lambda,\mu)_{N,3}^{N, 3}}{R(\lambda,\mu)_{N, 1}^{N, 1}}
{\cal T}_{1, 3}(\mu)
{\cal T}_{N, N}(\lambda)
+ \sum_{\bar{e}=4}^{N}
\frac{R(\lambda,\mu)_{N+3-\bar{e}, \bar{e}}^{N, 3}}{R(\lambda,\mu)_{N, 1}^{N, 1}}
{\cal T}_{1, \bar{e}}(\mu)
{\cal T}_{N, N+3-\bar{e}}(\lambda)
\nonumber \\
&-&\sum_{\bar{e}=1}^{N-1}
\frac{R(\lambda,\mu)_{N, 1}^{\bar{e}, N-\bar{e}+1}}{R(\lambda,\mu)_{N, 1}^{N, 1}}
{\cal T}_{\bar{e}, N}(\lambda)
{\cal T}_{N-\bar{e}+1, 3}(\mu)
\nonumber \\
\label{2partN}
\ear
where function
$D^{(a,\bar{e})}_{3}(\lambda,\mu)$ is represented by  the ratio of $3 \times 3$  and $2 \times 2$
determinants, namely
\bear
D^{(a,\bar{e})}_{3}(\lambda,\mu) & = &
\frac{\left| \begin{array}{ccc}
R(\lambda,\mu)_{a+2,1}^{a, 3} & R(\lambda,\mu)_{a+1,2}^{a, 3} & R(\lambda,\mu)_{a-\bar{e},3+\bar{e}}^{a, 3} \\
R(\lambda,\mu)_{a+2,1}^{a+1,2} & R(\lambda,\mu)_{a+1,2}^{a+1,2} & R(\lambda,\mu)_{a-\bar{e},3+\bar{e}}^{a+1,2} \\
R(\lambda,\mu)_{a+2,1}^{a+2,1} & R(\lambda,\mu)_{a+1,2}^{a+2,1} & R(\lambda,\mu)_{a-\bar{e},3+\bar{e}}^{a+2,1}
\end{array} \right|}
{R(\lambda,\mu)_{a,1}^{a,1}
\left| \begin{array}{cc}
R(\lambda,\mu)_{a+2,1}^{a+1,2} & R(\lambda,\mu)_{a+1,2}^{a+1,2} \\
R(\lambda,\mu)_{a+2,1}^{a+2,1} & R(\lambda,\mu)_{a+1,2}^{a+2,1}
\end{array} \right|}, \nonumber \\
&\mathrm{for}&~~~\bar{e}=0,\cdots,a-1~~~\mathrm{and}~~~2 \le a \le N-2.
\label{pda1}
\ear

\subsection{The creation fields}
\label{sub32}

The results from previous subsection reveal us that the
commutation rules between the operators
${\cal T}_{a,a}(\lambda)$ and
${\cal T}_{1,b}(\lambda)$ are able to produce additional creation fields
other than the basis vectors
${\cal T}_{1,b}(\lambda)$ or
${\cal T}_{1,b}(\mu)$.
Though these extra creation operators do not take a direct part on the multiparticle state basis
they are essential for the solution of the transfer matrix eigenvalue problem.
It is therefore necessary to disentangle not only the commutation rules among the
basis vectors themselves
but also those involving the operators
${\cal T}_{1,b}(\lambda)$
with the remaining creation fields
${\cal T}_{{a},{b}}(\mu)$  for  $ {b} > {a}=2,\cdots,N-1$.
In the course of our analysis we find convenient
to describe such commutation relations by using the fields
${\cal T}_{1,b_1-d_1}(\lambda)$ and
${\cal T}_{a_1-1,a_1+d_1}(\mu)$ where the underlying indices belong to the following
intervals,
\EQ
2 \le a_1 \le N,~~~ 0 \le d_1 \le N-a_1,~~~ 2\le b=b_1-d_1 \le N
\label{IND}
\EN

The above arrangement of indices bring us at least two technical advantages. First, we
assure that all the mentioned commutations rules between the creation fields will be
considered without
unnecessary repetition.  Next, Eq.(\ref{pro1}) implies that the effective azimuthal
spin component associated to the term
${\cal T}_{1,b}(\lambda) {\cal T}_{a_1-1,a_1+d_1}(\mu)$  should be indexed
by the composed variable $b+d_1$. The parameterization $b=b_1-d_1$ has therefore the merit of
allowing us to keep track of the azimuthal spin of such product of creation fields in terms
of a unique index $b_1$.
As a consequence of that the number of
the distinct commutation relations that are
needed in a given multiparticle state sector can now be controlled with the
help of the remaining indices $a_1$ and $d_1$. A systematic analysis of the Yang-Baxter
relation (\ref{fundrelindex0}) reveals us that these commutation rules are derived from
the entries $[a_1-1;(a_1+c-1)N+b_1-c]$ for selected values of the variable $c$. The
corresponding relations are obtained from Eq.(\ref{fundrelindex0}) by choosing
$\bar{a}=1$, $\bar{b}=a_1-1$, $\bar{c}=a_1+c$ e $\bar{d}=b_1-c$, namely
\EQ
\label{fundrelindex2}
\sum_{\bar{e}=1}^{a_1-1} R(\lambda,\mu)_{a_1-1, 1}^{\bar{e}, a_1-\bar{e}}
{\cal T}_{\bar{e}, a_1+c}(\lambda)
{\cal T}_{a_1-\bar{e}, b_1-c}(\mu)=
\sum_{\bar{e}=M \lbrace 1,a_1+b_1-N \rbrace}^{m \lbrace a_1+b_1-1, N \rbrace}
{\cal T}_{1, \bar{e}}(\mu)
{\cal T}_{a_1-1, a_1+b_1-\bar{e}}(\lambda)
R(\lambda,\mu)_{a_1+b_1-\bar{e}, \bar{e}}^{a_1+c, b_1-c}.
\EN

The majority of the suitable commutation rules between the fields
${\cal T}_{1,b_1-d_1}(\lambda)$ and \newline
 ${\cal T}_{a_1-1,a_1+d_1}(\mu)$  require a considerable amount
of manipulations among the
relations (\ref{fundrelindex2}).  As before we still have to implement certain linear combinations by using
the freedom of the index $c$. This procedure is highly dependent on the variables $a_1,b_1$ and it has
been detailed in Table 2.
\begin{table}[ht]
\begin{center}
\begin{tabular}{|c|c|c|c|}
\hline
Operator  & Creation & Combination & Number of  \\
    & index & index & equations \\
\hline
${\cal T}_{1, b_1-d_1}(\lambda) {\cal T}_{1, 2+d_1}(\mu)$
   & $ b_1 \ge N $ & $ c=b_1-d_1-2$ & $1$ \\
%\cline{2-4}
  $ a_1 = 2 $ & & & \\
\hline
 ${\cal T}_{1, b_1-d_1}(\lambda) {\cal T}_{1, 2+d_1}(\mu)$ & $ b_1 < N $ & $ c=b_1-d_1-2,b_1-1 $ & $ 2 $ \\
 $ a_1 = 2 $ & & & \\
\hline
 ${\cal T}_{1, b_1-d_1}(\mu) {\cal T}_{a_1-1, a_1+d_1}(\lambda)$& $b_1 <N $ & $ c=0, \dots, b_1-1 $ & $ b_1 $ \\
$ 3 \le a_1 \le N+1-b_1 $ & & & \\
\hline
%\cline{3-4}
  ${\cal T}_{1, b_1-d_1}(\mu) {\cal T}_{a_1-1, a_1+d_1}(\lambda)$ & $b_1 <N$  & $ c=0, \dots, N-a_1 $ & $N+1-a_1 $ \\
  $ N+2-b_1 \le a_1 \le N $ & & & \\
  \hline
${\cal T}_{1, b_1-d_1}(\mu) {\cal T}_{a_1-1, a_1+d_1}(\lambda)$
  & $ b_1 \ge N $ & $ c=b_1-N, \dots, N-a_1 $ & $2 N+1-a_1-b_1 $ \\
 $ 3 \le a_1 \le N $ & & &
 \\
\hline
\end{tabular}
\end{center}
\caption{\footnotesize{The linear combination derived from Eq.(\ref{fundrelindex2}) to achieve appropriate
suitable commutation rules between the fields ${\cal T}_{1,b_1-d_1}(\lambda)$} and ${\cal T}_{a_1-1,a_1+d_1}(\lambda)$.
The indices $a_1$, $b_1$ and $d_1$ belong to the intervals defined by Eq.(\ref{IND}).}
\end{table}

The simplest type of commutation rules are those involving the creation fields that has direct
participation on the eigenvector basis.
These relations are sorted out by choosing the index $a_1=2$. From Table 2 we see that the commutation rule
between
${\cal T}_{1, b_1-d_1}(\lambda)$ and ${\cal T}_{1, 2+d_1}(\mu)$ for $b_1 \ge N$ follows directly from
the entry $[1;(b_1-d_1-1)N+2+d_1]$ of Eq.(\ref{fundrelindex2}), namely
\bear
{\cal T}_{1, b_1-d_1}(\lambda) {\cal T}_{1, 2+d_1}(\mu)
&=&
\frac{R(\lambda,\mu)_{b_1-d_1, 2+d_1}^{b_1-d_1, 2+d_1}}{R(\lambda,\mu)_{1, 1}^{1, 1}}
{\cal T}_{1, 2+d_1}(\mu) {\cal T}_{1, b_1-d_1}(\lambda)
\nonumber\\
&+&
\sum_{\stackrel{\bar{e}=2+b_1-N}{\bar{e} \ne 2+d_1}}^{N}
\frac{R(\lambda,\mu)_{2+b_1-\bar{e}, \bar{e}}^{b_1-d_1, 2+d_1}}{R(\lambda,\mu)_{1, 1}^{1, 1}}
{\cal T}_{1, \bar{e}}(\mu) {\cal T}_{1, 2+b_1-\bar{e}}(\lambda),~~~\mathrm{for}~~ b_1 \ge N. \nonumber \\
\label{eq34}
\ear

On the other hand for $b_1 < N$ we have to make a combination of two entries
$[1;(b_1-d_1-1)N+2+d_1]$ and $[1;b_1 N+1]$ of Eq.(\ref{fundrelindex2}). These equations can be combined
to produce a
$2 \times 2$ system of linear equations given by,
\bear
A^{(b_1,d_1)}_3(\lambda,\mu)
\left(\begin{array}{c}
{\cal T}_{1, 1}(\mu) {\cal T}_{1, b_1+1}(\lambda) \\
{\cal T}_{1, 2}(\mu) {\cal T}_{1, b_1}(\lambda)
\end{array}\right)
= R(\lambda,\mu)_{1, 1}^{1, 1} \left(\begin{array}{c}
{\cal T}_{1, b_1-d_1}(\lambda)
{\cal T}_{1, 2+d_1}(\mu) \\
{\cal T}_{1, 1+b_1}(\lambda)
{\cal T}_{1, 1}(\mu)
\end{array}\right)
\nonumber \\
- \sum_{\bar{e}=3}^{b_1+1} \left(\begin{array}{c}
R(\lambda,\mu)_{2+b_1-\bar{e}, \bar{e}}^{b_1-d_1, 2+d_1} \\
R(\lambda,\mu)_{2+b_1-\bar{e}, \bar{e}}^{b_1+1, 1}
\end{array}\right)
{\cal T}_{1, \bar{e}}(\mu) {\cal T}_{1, 2+b_1-\bar{e}}(\lambda)
%\nonumber \\
~~~\mathrm{for} ~~ b_1 < N,
\label{2two}
\ear
where the $2 \times 2$ matrix $A^{(a_1,b_1)}_3(\lambda,\mu)$ is,
\bear
A^{(b_1,d_1)}_3(\lambda,\mu)=\left(\begin{array}{c c}
R(\lambda,\mu)_{b_1+1, 1}^{b_1-d_1, 2+d_1} & R(\lambda,\mu)_{b_1, 2}^{b_1-d_1, 2+d_1} \\
R(\lambda,\mu)_{b_1+1, 1}^{b_1+1, 1} & R(\lambda,\mu)_{b_1, 2}^{b_1+1, 1}
\end{array}\right).
\ear

The desirable commutation relations are derived by solving Eq.(\ref{2two}) for the
product
${\cal T}_{1, 2}(\mu)$ ${\cal T}_{1, b_1}(\lambda)$
with the assistance of Cramer's rule.
As a result we obtain a single linear equation relating the terms
${\cal T}_{1,b_1-d_1}(\lambda)  {\cal T}_{1,2+d_1}(\mu)$ and
${\cal T}_{1,2+d_1}(\mu)  {\cal T}_{1,b_1-d_1}(\lambda)$
\footnote{ We recall that the product
${\cal T}_{1,2+d_1}(\mu)  {\cal T}_{1,b_1-d_1}(\lambda)$ appears in the sum of Eq.(\ref{2two}) for $\bar{e}=2+d_1 \leq b_1$.} whose
solution gives us the following expression,
\bear
{\cal T}_{1, b_1-d_1}(\lambda) {\cal T}_{1, 2+d_1}(\mu)
&=&
-\frac{\left|\begin{array}{c c}
R(\lambda,\mu)_{b_1+1, 1}^{b_1-d_1, 2+d_1} & R(\lambda,\mu)_{b_1-d_1, 2+d_1}^{b_1-d_1, 2+d_1} \\
R(\lambda,\mu)_{b_1+1, 1}^{b_1+1, 1} & R(\lambda,\mu)_{b_1-d_1, 2+d_1}^{b_1+1, 1}
\end{array}\right|}
{R(\lambda,\mu)_{1,1}^{1,1} R(\lambda,\mu)_{b_1+1, 1}^{b_1+1, 1}}
{\cal T}_{1, 2+d_1}(\mu) {\cal T}_{1, b_1-d_1}(\lambda)
\nonumber \\
&+&
\frac{R(\lambda,\mu)_{b_1+1, 1}^{b_1-d_1, 2+d_1}}
{R(\lambda,\mu)_{b_1+1, 1}^{b_1+1, 1}}
{\cal T}_{1, b_1+1}(\lambda) {\cal T}_{1, 1}(\mu)
\nonumber \\
&-& \sum_{\stackrel{\bar{e}=2}{\bar{e} \ne 2+d_1}}^{b_1+1}
\frac{\left|\begin{array}{c c}
R(\lambda,\mu)_{b_1+1, 1}^{b_1-d_1, 2+d_1} & R(\lambda,\mu)_{2+b_1-\bar{e}, \bar{e}}^{b_1-d_1, 2+d_1} \\
R(\lambda,\mu)_{b_1+1, 1}^{b_1+1, 1} & R(\lambda,\mu)_{2+b_1-\bar{e}, \bar{e}}^{b_1+1, 1}
\end{array}\right|}
{R(\lambda,\mu)_{1,1}^{1,1} R(\lambda,\mu)_{b_1+1, 1}^{b_1+1, 1}}
{\cal T}_{1, \bar{e}}(\mu) {\cal T}_{1, 2+b_1-\bar{e}}(\lambda),
\nonumber \\
&& ~~~~~~~~~~~~~~~~~~\mathrm{for}~~ b_1 < N.
\label{eq37}
\ear

Let us now discuss the strategy we have used so far in order to obtain
the commutation relations for the creation fields.
The basic idea is that any possible product
of creation operators on the right-hand side of the commutation relations
should be equally ordered as far as the rapidities $\lambda$ and $\mu$ are concerned.
This ordering is certainly the opposite of that we have started with for
the left-hand side product of creation operators. In addition to that, the
specific ordering choice is actually dictated by the results established in  section \ref{sub31} for the
commutation rules among the diagonal fields and the basis vectors. The commutation
relations for the creation fields are then constructed to bring a given creation operator
with the transfer
matrix spectral parameter to the further left position
in products of monodromy matrix elements that are not proportional to the eigenvectors.

The above procedure is of special importance when we deal with the commutation rules between the operators
${\cal T}_{1, b_1-d_1}(\mu)$ and ${\cal T}_{a_1-1,a_1+d_1}(\lambda)$ for $3 \le a_1 \le N$. In this
situation the right-hand side of the respective commutation relations is not allowed to
possess terms of the form
${\cal T}_{1, b_1-\bar{e}}(\mu) {\cal T}_{a_1-1,a_1+\bar{e}}(\lambda)$ with $\bar{e} \geq 0$.
This can be clearly seen by considering in an explicit way the linear systems associated
to such commutation rules. From Table 2 we observe they depend very much whether $b_1 < N$ or
$b_1 \geq N$. The former case is rather similar to the linear combination discussed in
previous subsection and we find that,
\bear
\label{creation1}
A^{(a_1,b_1)}_1(\lambda,\mu)
\left(\begin{array}{c}
{\cal T}_{1, 1}(\mu)
{\cal T}_{a_1-1, a_1+b_1-1}(\lambda) \\
{\cal T}_{1, 2}(\mu)
{\cal T}_{a_1-1, a_1+b_1-2}(\lambda) \\
\vdots \\
{\cal T}_{1, b_1}(\mu)
{\cal T}_{a_1-1, a_1}(\lambda)
\end{array}\right)
= \sum_{\bar{e}=1}^{a_1-1} R(\lambda,\mu)_{a_1-1, 1}^{\bar{e}, a_1-\bar{e}}
\left(\begin{array}{c}
{\cal T}_{\bar{e}, a_1}(\lambda)
{\cal T}_{a_1-\bar{e}, b_1}(\mu) \\
{\cal T}_{\bar{e}, a_1+1}(\lambda)
{\cal T}_{a_1-\bar{e}, b_1-1}(\mu) \\
\vdots \\
{\cal T}_{\bar{e}, a_1+b_1-1}(\lambda)
{\cal T}_{a_1-\bar{e}, 1}(\mu)
\end{array}\right)
\nonumber \\
-\sum_{\bar{e}=b_1+1}^{a_1+b_1-1}
v^{(a_1,b_1)}_1(\lambda,\mu)
{\cal T}_{1, \bar{e}}(\mu)
{\cal T}_{a_1-1, a_1+b_1-\bar{e}}(\lambda)
% \nonumber \\
~~~\mathrm{for}~~b_1<N~~~\mathrm{and}~~~3 \le a_1 \le N+1-b_1
\ear
\bear
\label{creation2}
\lefteqn{A^{(a_1,b_1)}_2(\lambda,\mu)
\left(\begin{array}{c}
{\cal T}_{1, a_1+b_1-N}(\mu)
{\cal T}_{a_1-1, N}(\lambda) \\
{\cal T}_{1, a_1+b_1-N+1}(\mu)
{\cal T}_{a_1-1, N-1}(\lambda) \\
\vdots \\
{\cal T}_{1, b_1}(\mu)
{\cal T}_{a_1-1, a_1}(\lambda)
\end{array}\right)
=
\sum_{\bar{e}=1}^{a_1-1} R(\lambda,\mu)_{a_1-1, 1}^{\bar{e}, a_1-\bar{e}}}
\nonumber \\
&\times&
\left(\begin{array}{c}
{\cal T}_{\bar{e}, a_1}(\lambda)
{\cal T}_{a_1-\bar{e}, b_1}(\mu) \\
{\cal T}_{\bar{e}, a_1+1}(\lambda)
{\cal T}_{a_1-\bar{e}, b_1-1}(\mu) \\
\vdots \\
{\cal T}_{\bar{e}, N}(\lambda)
{\cal T}_{a_1-\bar{e}, a_1+b_1-N}(\mu)
\end{array}\right)
%\nonumber \\
-
\sum_{\bar{e}=b_1+1}^{N}
v^{(a_1,b_1)}_2(\lambda,\mu)
{\cal T}_{1, \bar{e}}(\mu)
{\cal T}_{a_1-1, a_1+b_1-\bar{e}}(\lambda)
\nonumber \\
&\mathrm{for}&~~~b_1 < N~~~\mathrm{and}~~~N+2-b_1 \le a_1 \le N
\nonumber \\
\ear
where the $b_1 \times b_1$ matrix $A^{(a_1,b_1)}_1(\lambda,\mu)$, the
$b_1 \times 1$ vector $v^{(a_1,b_1)}_1(\lambda,\mu)$,
the $(N+1-a_1) \times (N+1-a_1)$ matrix $A^{(a_1,b_1)}_2(\lambda,\mu)$ and
the $(N+1-a_1) \times 1$ vector $v^{(a_1,b_1)}_2(\lambda,\mu)$  are given by Eqs.(\ref{am1},\ref{v1},\ref{am2},\ref{v2}), respectively.

For $b_1 \geq N$ we have instead a different linear system of equations. In this case the relations coming
from Eq.(\ref{fundrelindex2}) for $c=b_1-N, \cdots, N-a_1$ are organized as follows,
\bear
\label{creation3}
\lefteqn{A^{(a_1,b_1)}_4(\lambda,\mu)
\left(\begin{array}{c}
{\cal T}_{1, a_1+b_1-N}(\mu)
{\cal T}_{a_1-1, N}(\lambda) \\
{\cal T}_{1, a_1+b_1-N+1}(\mu)
{\cal T}_{a_1-1, N-1}(\lambda) \\
\vdots \\
{\cal T}_{1, N}(\mu)
{\cal T}_{a_1-1, a_1+b_1-N}(\lambda)
\end{array}\right)
=
\sum_{\bar{e}=1}^{a_1-1} R(\lambda,\mu)_{a_1-1, 1}^{\bar{e}, a_1-\bar{e}}}
\nonumber \\
& \times &
\left(\begin{array}{c}
{\cal T}_{\bar{e}, a_1+b_1-N}(\lambda)
{\cal T}_{a_1-\bar{e}, N}(\mu) \\
{\cal T}_{\bar{e}, a_1+b_1-N+1}(\lambda)
{\cal T}_{a_1-\bar{e}, N-1}(\mu) \\
\vdots \\
{\cal T}_{\bar{e}, N}(\lambda)
{\cal T}_{a_1-\bar{e}, a_1+b_1-N}(\mu)
\end{array}\right)
%\nonumber \\
~~~\mathrm{for}~~b_1 \ge N~~~\mathrm{and}~~~3 \le a_1 \le N,
\ear
where
the $(2N+1-a_1-b_1) \times (2N+1-a_1-b_1)$ matrix $A^{(a_1,b_1)}_4(\lambda,\mu)$  is given by
\EQ
\label{am3}
A^{(a_1,b_1)}_4(\lambda,\mu)=
\left( \begin{array}{cccc}
R(\lambda,\mu)_{N, a_1+b_1-N}^{a_1+b_1-N, N} & R(\lambda,\mu)_{N-1, a_1+b_1-N+1}^{a_1+b_1-N, N} & \cdots & R(\lambda,\mu)_{a_1+b_1-N, N}^{a_1+b_1-N, N} \\
R(\lambda,\mu)_{N, a_1+b_1-N}^{a_1+b_1-N+1, N-1} & R(\lambda,\mu)_{N-1, a_1+b_1-N+1}^{a_1+b_1-N+1, N-1} & \cdots & R(\lambda,\mu)_{a_1+b_1-N, N}^{a_1+b_1-N+1, N-1} \\
\vdots & \vdots & \ddots & \vdots \\
R(\lambda,\mu)_{N, a_1+b_1-N}^{N, a_1+b_1-N} & R(\lambda,\mu)_{N-1, a_1+b_1-N+1}^{N, a_1+b_1-N} & \cdots & R(\lambda,\mu)_{a_1+b_1-N, N}^{N, a_1+b_1-N} \\
\end{array}\right).
\EN

Direct inspection of
Eqs.(\ref{creation1},\ref{creation2},\ref{creation3})  reveal us that the  variables of the corresponding
system of linear equations are indeed the products
${\cal T}_{1, b_1-d_1}(\mu) {\cal T}_{a_1-1, a_1+d_1}(\lambda)$. The origin of  these
independent linear systems
is directly related to the existence of three distinct intervals for the index $d_1$ once the variables
$a_1$ and $b_1$ are fixed. Their structure
have been constructed to collect together
products of creation fields whose respective values for the index $d_1$
belong to one of the three possible such intervals.
The commutation relations between the operators
${\cal T}_{1, b_1-d_1}(\mu)$ and  ${\cal T}_{a_1-1, a_1+d_1}(\lambda)$ are then determined by means of
the systematic application of Cramer's rule  to solve the systems of linear equations
(\ref{creation1},\ref{creation2},\ref{creation3}). The procedure is similar to that already
described in section (\ref{sub31}). For instance, we note that the reversed product terms
appear on the first non-homogeneous part of
Eqs.(\ref{creation1},\ref{creation2},\ref{creation3}) when the sum index is $\bar{e}=a_1-1$.
The task is however much more cumbersome since we have
to apply Cramer's rule for each
left-hand side product of creation fields entering
Eqs.(\ref{creation1},\ref{creation2},\ref{creation3}).
The number of different commutation rules associated to each linear system for a given multiparticle
state depends strongly on the index $b_1$. This feature and the dependence
of the total number of relations for a given $N$ is illustrated in Table 3.
\begin{table}[h]
\begin{center}
\begin{tabular}{|c|c|c|}
\hline
Linear   & \multicolumn{2}{c}{Number of Commutation Rules } \vline \\
\cline{2-3}
System & fixed $b_1$ & fixed $N$ \\
\hline
 $A^{(a_1,b_1)}_1(\lambda,\mu)$ & $\displaystyle \sum_{a_1=3}^{N+1-b_1}(b_1-1)$ & $\displaystyle \sum_{b_1=2}^{N-2}(N-b_1-1)(b_1-1)$ \\
 & $=(N-b_1-1)(b_1-1)$ & $\displaystyle =\frac{(N-1)(N-2)(N-3)}{6}$ \\
 \hline
$A^{(a_1,b_1)}_2(\lambda,\mu)$  & $\displaystyle \sum_{a_1=N+2-b_1}^{N}(N+1-a_1)$ &
$\displaystyle \sum_{b_1=2}^{N-1}\frac{(b_1-1)b_1}{2}$ \\
 & $\displaystyle =\frac{(b_1-1)b_1}{2}$ & $\displaystyle =\frac{N(N-1)(N-2)}{6}$ \\
 \hline
$A^{(a_1,b_1)}_4(\lambda,\mu)$  & $\displaystyle \sum_{a_1=3}^{2N-b_1}(2N+1-a_1-b_1)$ &
$\displaystyle \sum_{b_1=N}^{2N-2}\frac{(2N-b_1-1)(2N-b_1-2)}{2}$ \\
 & $\displaystyle=\frac{(2N-b_1-1)(2N-b_1-2)}{2}$ & $\displaystyle=\frac{N(N-1)(N-2)}{6}$ \\
 \hline
\end{tabular}
\end{center}
\caption{The number of commutations relations between the creation fields for fixed $b_1$ as well as
for fixed $N$.}
\end{table}

We would like to conclude by presenting an example of a complete set of commutation rules
between the creation operators. The simplest
non-trivial situation occurs when the respective azimuthal spin component is indexed
by $b_1=2$.  The intervals (\ref{IND}) tell us that such relations are those
between the operators
${\cal T}_{1, 2}(\mu)$ and  ${\cal T}_{a_1-1, a_1}(\lambda)$  for $2 \leq a_1 \leq N$. The case
$a_1=2$ follows from the general expression (\ref{eq37}) while for the remaining values we need to
solve the linear systems
(\ref{creation1},\ref{creation2}). Putting the results together we find that,
\bear
\lefteqn{D_{2}^{(a_1,0)}(\lambda,\mu) {\cal T}_{1,2}(\mu) {\cal T}_{a_1-1,a_1}(\lambda) =
\frac{R(\lambda,\mu)_{a_1-1,1}^{a_1-1,1}}{R(\lambda,\mu)_{a_1,1}^{a_1,1}} {\cal T}_{a_1-1,a_1}(\lambda) {\cal T}_{1,2}(\mu)
-\frac{R(\lambda,\mu)_{a_1-1,1}^{a_1-1,1}}{R(\lambda,\mu)_{a_1,1}^{a_1,1}}
\frac{R(\lambda,\mu)_{a_1+1,1}^{a_1,2}}{R(\lambda,\mu)_{a_1+1,1}^{a_1+1,1}} }
\nonumber \\
&\times &
{\cal T}_{a_1-1,a_1+1}(\lambda) {\cal T}_{1,1}(\mu)
%\nonumber \\
-\sum_{\bar{e}=1}^{a_1-2}
\frac{R(\lambda,\mu)_{a_1-1,1}^{\bar{e},a_1-\bar{e}}}{R(\lambda,\mu)_{a_1,1}^{a_1,1}}
\frac{R(\lambda,\mu)_{a_1+1,1}^{a_1,2}}{R(\lambda,\mu)_{a_1+1,1}^{a_1+1,1}}
{\cal T}_{\bar{e},a_1+1}(\lambda) {\cal T}_{a_1-\bar{e},1}(\mu)
\nonumber \\
&+& \sum_{\bar{e}=1}^{a_1-2}
\frac{R(\lambda,\mu)_{a_1-1,1}^{\bar{e},a_1-\bar{e}}}{R(\lambda,\mu)_{a_1,1}^{a_1,1}}
{\cal T}_{\bar{e},a_1}(\lambda) {\cal T}_{a_1-\bar{e},2}(\mu)
%\nonumber \\
- \sum_{\bar{e}=3}^{a_1+1}
D_{2}^{(a_1,\bar{e}-2)}(\lambda,\mu)
{\cal T}_{1,\bar{e}}(\mu) {\cal T}_{a_1-1,a_1+2-\bar{e}}(\lambda),
\nonumber \\
&& ~~~~~~~~~~~~~~~~~~~~~~~~~~~~~~~~~~~~~~~~~~~~~~~~~~~~ \mathrm{for} ~~2 \le a_1 \le N-1
\label{b12ba-1a}
\ear
\bear
{\cal T}_{1,2}(\mu) {\cal T}_{N-1,N}(\lambda) &=&
\frac{R(\lambda,\mu)_{N-1,1}^{N-1,1}}{R(\lambda,\mu)_{N,2}^{N,2}}
{\cal T}_{N-1,N}(\lambda) {\cal T}_{1,2}(\mu)
+\sum_{\bar{e}=1}^{N-2}
\frac{R(\lambda,\mu)_{N-1,1}^{\bar{e},N-\bar{e}}}{R(\lambda,\mu)_{N,2}^{N,2}}
{\cal T}_{\bar{e},N}(\lambda) {\cal T}_{N-\bar{e},2}(\mu)
\nonumber \\
&-&\sum_{\bar{e}=3}^{a_1+1}
\frac{R(\lambda,\mu)_{N+2-\bar{e},\bar{e}}^{N,2}}{R(\lambda,\mu)_{N,2}^{N,2}}
{\cal T}_{1,\bar{e}}(\mu) {\cal T}_{N-1,N+2-\bar{e}}(\lambda),
\label{b12bN-1N}
\ear
where we recall that function $D_{2}^{(a_1,\bar{e})}(\lambda,\mu)$ is given by Eq.(\ref{pda}).

\subsection{The creation and annihilation fields}
\label{sub33}

The third class of commutation rules that still need to be considered are those between the basis vectors ${\cal T}_{1,b}(\mu)$ and all possible annihilation operator ${\cal T}_{a_1+d_1,a_1-1}(\lambda)$.
We stress here that the indices $a_1$, $d_1$ and $b$ are the same used in subsection (\ref{sub32}) and therefore they belong to the intervals defined by Eq.(\ref{IND}).
In general, the result of taking the operator ${\cal T}_{a_1+d_1,a_1-1}(\lambda)$ through the creation fields ${\cal T}_{1,b}(\mu)$ is the generation of several distinct annihilation fields besides ${\cal T}_{a_1+d_1,a_1-1}(\lambda)$.
The basic strategy we shall use to construct these relations is to enforce that all such resulting annihilation
fields must be kept in the right-hand side position on the respective commutation rule.
A detailed study of the Yang-Baxter algebra reveals us that these commutation rules can be built up from the entries
$[a_1+d_1+c_1(N-1);(a_1-2)N+b+c_2(N-1)]$ of Eq.(\ref{fundrelindex0}).
The structure of the corresponding equations are now more complicated since they carry a dependence on two independent indices $c_1$ and $c_2$ as well as on the particular combination $a_1+d_1$.
Taking into account the latter feature we find convenient here to express the expressions by means of the following auxiliary indices,
\EQ
f_1=a_1+d_1~~~\mathrm{and}~~~f_2=b-d_1-2
\EN

The starting point relations are then obtained from Eq.(\ref{fundrelindex0}) by choosing
$\bar{a}=c_1+1$, $\bar{b}=f_1-c_1$, $\bar{c}=a_1+c_2-1$ and $\bar{d}=b-c_2$, namely
\bear
&&\sum_{\bar{e}=1}^{f_1}
R(\lambda,\mu)_{f_1-c_1 , c_1 + 1}^{\bar{e}, f_1+1-\bar{e}}
{\cal T}_{\bar{e}, a_1+c_2-1}(\lambda)
{\cal T}_{f_1+1-\bar{e}, b-c_2}(\mu)
=
\nonumber \\
&&\sum_{\bar{e}=M \lbrace 1,a_1+b-N-1 \rbrace}^{m \lbrace a_1+b-2, N \rbrace}
{\cal T}_{c_1+1, \bar{e}}(\mu)
{\cal T}_{f_1-c_1, a_1+b-1-\bar{e}}(\lambda)
%\nonumber \\
%& \times &
R(\lambda,\mu)_{a_1+b-1-\bar{e}, \bar{e}}^{a_1+c_2-1, b-c_2}.
\label{fundrelindex3}
\ear

The analysis of Eq.(\ref{fundrelindex3}) turns out to be far more involved than that discussed in the two last subsections.
The respective linear combinations have to be performed considering two distinct steps and they will culminated
in two coupled linear systems.
It is rather illuminating to start this discussion by examining Eq.(\ref{fundrelindex3}) in the particular situation $c_1=c_2=0$,
\bear
&& \sum_{\bar{e}=1}^{f_1}
R(\lambda,\mu)_{f_1 , 1}^{\bar{e}, f_1+1-\bar{e}}
{\cal T}_{\bar{e}, a_1-1}(\lambda)
{\cal T}_{f_1+1-\bar{e}, b}(\mu)
=
\nonumber \\
&& \sum_{\bar{e}=M \lbrace 1,a_1+b-N-1 \rbrace}^{m \lbrace a_1+b-2, N \rbrace}
{\cal T}_{1, \bar{e}}(\mu)
{\cal T}_{f_1, a_1+b-1-\bar{e}}(\lambda)
R(\lambda,\mu)_{a_1+b-1-\bar{e}, \bar{e}}^{a_1-1, b},
\label{aniq01}
\ear

Let us now concentrate our attention on the left-hand side of Eq.(\ref{aniq01}).
For $\bar{e}=f_1$ it gives us the product ${\cal T}_{a_1+d_1,a_1-1}(\lambda) {\cal T}_{1,b}(\mu)$ which is indispensable to
build up the desirable commutation rules.
Unfortunately, for arbitrary values of $b$, this term is not the only one involving annihilation and creation operators generated by Eq.(\ref{aniq01}).
From the left-hand side of Eq.(\ref{aniq01}) we see indeed that ${\cal T}_{\bar{e},a_1-1}(\lambda)$
behaves as annihilation or diagonal operator
when $\bar{e} \ge a_1-1$ while ${\cal T}_{a_1+d_1+1-\bar{e}, b}(\mu)$ play
the role of creation fields for $\bar{e} \ge a_1-f_2=a_1-(b-d_1-2)$.
This product combination clearly spoil the main characteristic one would expect from the commutation rule among ${\cal T}_{a_1+d_1,a_1-1}(\lambda)$ and ${\cal T}_{1,b}(\mu)$.
The first task therefore is to eliminate the left-hand side products ${\cal T}_{\bar{e}, a_1-1}(\lambda) {\cal T}_{a_1+d_1+1-\bar{e}, b}(\mu)$ when the index $\bar{e}$ takes values on the intersection of the above intervals.
Considering the upper value of $\bar{e}$ in Eq.(\ref{aniq01}) it is not difficult to see that the mentioned interval is
\EQ
a_1 -m \lbrace 1, f_2=b-d_1-2 \rbrace \le  \bar{e} < f_1=a_1+d_1.
\label{indaniq1}
\EN

The constraint (\ref{indaniq1}) tells us that the case $b=2$ is a fortunate exception and therefore there is no need of considering the above mentioned cancellations.
This means that the appropriate commutation rules between ${\cal T}_{a_1+d_1,a_1-1}(\lambda)$ and ${\cal T}_{1,2}(\mu)$ can be derived directly from Eq.(\ref{aniq01}).
After few manipulations we find,
\bear
\label{aniq0}
{\cal T}_{f_1, a_1-1}(\lambda)
{\cal T}_{1, 2}(\mu)
=
\frac{R(\lambda,\mu)_{a_1-1, 2}^{a_1-1, 2}}{R(\lambda,\mu)_{f_1 , 1}^{f_1, 1}}
{\cal T}_{1, 2}(\mu)
{\cal T}_{f_1, a_1-1}(\lambda)
+
\sum_{\stackrel{\bar{e}=1}{\bar{e} \ne 2}}^{a_1}
\frac{R(\lambda,\mu)_{a_1+1-\bar{e}, \bar{e}}^{a_1-1, 2}}{R(\lambda,\mu)_{f_1 , 1}^{f_1, 1}}
{\cal T}_{1, \bar{e}}(\mu)
{\cal T}_{f_1, a_1+1-\bar{e}}(\lambda)
\nonumber \\
-
\sum_{\bar{e}=1}^{f_1-1}
\frac{R(\lambda,\mu)_{f_1 , 1}^{\bar{e}, f_1+1-\bar{e}}}{R(\lambda,\mu)_{f_1 , 1}^{f_1, 1}}
{\cal T}_{\bar{e}, a_1-1}(\lambda)
{\cal T}_{f_1+1-\bar{e}, 2}(\mu).
\ear

On the other hand, for $b \ge 3$ we are forced to get rid of the products ${\cal T}_{\bar{e}, a_1-1}(\lambda) {\cal T}_{a_1+d_1+1-\bar{e}, b}(\mu)$ when the index $\bar{e}$ belongs to the interval (\ref{indaniq1}).
This can be done by exploring the linear combination of equations derived from Eq.(\ref{fundrelindex3}) for suitable values of the index $c_1$ while $c_2$ is fixed at $c_2=0$.
The structure of the linear combination depends on the sign of the azimuthal spin of the product
${\cal T}_{f_1, a_1-1}(\lambda) {\cal T}_{1, b}(\mu)$\footnote{Recall that from Eq.(\ref{pro})
the corresponding spin value is $f_2=b-2-d_1$} as it is illustrated in Table 4.

\begin{table}[h]
\begin{center}
\begin{tabular}{|c|c|c|c|}
\hline
Operator & Creation & Combination & Number of  \\
eliminated & index & index $c_1$ & Equations \\
\hline
${\cal T}_{\bar{e}, a_1-1}(\lambda) {\cal T}_{f_1+1-\bar{e}, b}(\mu)$ & $b-2 \le d_1$ & $c_1= 0, \dots, b-2$ & $b-1$ \\
\hline
${\cal T}_{\bar{e}, a_1-1}(\lambda) {\cal T}_{f_1+1-\bar{e}, b}(\mu)$ & $b-2 > d_1$ & $c_1= 0, \dots, d_1+1$ & $d_1+2$ \\
\hline
\end{tabular}
\end{center}
\caption{\footnotesize{The linear combination from Eq.(\ref{fundrelindex3}) for $c_2=0$ to cancel the operators ${\cal T}_{\bar{e}, a_1-1}(\lambda) {\cal T}_{f_1+1-\bar{e}, b}(\mu)$ when $\bar{e}$ belongs to the interval (\ref{indaniq1})}}
\end{table}

Taking into account Table 4 it is not difficult to start the construction of the first needed linear combination for $b \ge 3$.
Its main structure is as follows,
\bear
\label{aniq1}
\lefteqn{ A^{(f_1,b)}_5(\lambda,\mu)
\left(\begin{array}{c}
{\cal T}_{f_1-b+2, a_1-1}(\lambda)
{\cal T}_{b-1, b}(\mu) \\
{\cal T}_{f_1-b+3, a_1-1}(\lambda)
{\cal T}_{b-2, b}(\mu) \\
\vdots \\
{\cal T}_{f_1, a_1-1}(\lambda)
{\cal T}_{1, b}(\mu)
\end{array}\right)
=
 \sum_{\bar{e}=1}^{a_1+b-2} R(\lambda,\mu)_{a_1+b-1-\bar{e}, \bar{e}}^{a_1-1,b} }
\nonumber \\
& \times &
\left(\begin{array}{c}
{\cal T}_{1, \bar{e}}(\mu)
{\cal T}_{f_1, a_1+b-1-\bar{e}}(\lambda) \\
{\cal T}_{2, \bar{e}}(\mu)
{\cal T}_{f_1-1, a_1+b-1-\bar{e}}(\lambda) \\
\vdots \\
{\cal T}_{b-1, \bar{e}}(\mu)
{\cal T}_{f_1-b+2, a_1+b-1-\bar{e}}(\lambda)
\end{array}\right)
-
\sum_{\bar{e}=1}^{f_1-b+1}
v^{(f_1,b)}_5(\lambda,\mu)
{\cal T}_{\bar{e}, a_1-1}(\lambda)
{\cal T}_{f_1+1-\bar{e}, b}(\mu)
\nonumber \\
&&~~~~~~~~~~~~~~~~~~~~~~~~~~~~~~~~~~~~~~~~~~~~~~~~ \mathrm{for}~~~ b-2 \le d_1,
\ear
and
\bear
\label{aniq2}
\lefteqn{ A^{(a_1,d_1)}_6(\lambda,\mu)
\left(\begin{array}{c}
{\cal T}_{a_1-1, a_1-1}(\lambda)
{\cal T}_{d_1+2, b}(\mu) \\
{\cal T}_{a_1, a_1-1}(\lambda)
{\cal T}_{d_1+1, b}(\mu) \\
\vdots \\
{\cal T}_{f_1, a_1-1}(\lambda)
{\cal T}_{1, b}(\mu)
\end{array}\right)
=
\sum_{\bar{e}=M \lbrace 1,a_1+b-N-1 \rbrace }^{m \lbrace a_1+b-2,N \rbrace} R(\lambda,\mu)_{a_1+b-1-\bar{e}, \bar{e}}^{a_1-1,b} }
\nonumber \\
& \times &
\left(\begin{array}{c}
{\cal T}_{1, \bar{e}}(\mu)
{\cal T}_{f_1, a_1+b-1-\bar{e}}(\lambda) \\
{\cal T}_{2, \bar{e}}(\mu)
{\cal T}_{f_1-1, a_1+b-1-\bar{e}}(\lambda) \\
\vdots \\
{\cal T}_{d_1+2, \bar{e}}(\mu)
{\cal T}_{a_1-1, a_1+b-1-\bar{e}}(\lambda)
\end{array}\right)
-
\sum_{\bar{e}=1}^{a_1-2}
v^{(a_1,d_1)}_6(\lambda,\mu)
{\cal T}_{\bar{e}, a_1-1}(\lambda)
{\cal T}_{f_1+1-\bar{e}, b}(\mu)
\nonumber \\
&&~~~~~~~~~~~~~~~~~~~~~~~~~~~~~~~~~~~~~~~~~~~~~~~~~ \mathrm{for}~~~ b-2 > d_1.
\ear

The $(b-1) \times (b-1)$ matrix $A^{(f_1,b)}_5(\lambda,\mu)$ as well as the $(d_1+2) \times (d_1+2)$ matrix $A^{(a_1,d_1)}_6(\lambda,\mu)$ are given by
\EQ
\label{am5}
A^{(f_1,b)}_5(\lambda,\mu)=\left( \begin{array}{cccc}
R(\lambda,\mu)_{f_1, 1}^{f_1-b+2, b-1} & R(\lambda,\mu)_{f_1, 1}^{f_1-b+3, b-2} & \cdots & R(\lambda,\mu)_{f_1, 1}^{f_1, 1} \\
R(\lambda,\mu)_{f_1-1, 2}^{f_1-b+2, b-1} & R(\lambda,\mu)_{f_1-1, 2}^{f_1-b+3, b-2} & \cdots & R(\lambda,\mu)_{f_1-1, 2}^{f_1, 1} \\
\vdots & \vdots & \ddots & \vdots \\
R(\lambda,\mu)_{f_1-b+2, b-1}^{f_1-b+2, b-1} & R(\lambda,\mu)_{f_1-b+2, b-1}^{f_1-b+3, b-2} & \cdots & R(\lambda,\mu)_{f_1-b+2, b-1}^{f_1, 1} \\
\end{array}\right)
\EN
and
\EQ
\label{am6}
A^{(a_1,d_1)}_6(\lambda,\mu)=\left( \begin{array}{cccc}
R(\lambda,\mu)_{f_1, 1}^{a_1-1, d_1+2} & R(\lambda,\mu)_{f_1, 1}^{a_1, d_1+1} & \cdots & R(\lambda,\mu)_{f_1, 1}^{f_1, 1} \\
R(\lambda,\mu)_{f_1-1, 2}^{a_1-1, d_1+2} & R(\lambda,\mu)_{f_1-1, 2}^{a_1, d_1+1} & \cdots & R(\lambda,\mu)_{f_1-1, 2}^{f_1, 1} \\
\vdots & \vdots & \ddots & \vdots \\
R(\lambda,\mu)_{a_1-1, d_1+2}^{a_1-1, d_1+2} & R(\lambda,\mu)_{a_1-1, d_1+2}^{a_1, d_1+1} & \cdots & R(\lambda,\mu)_{a_1-1, d_1+2}^{f_1, 1} \\
\end{array}\right),
\EN
while the $(b-1) \times 1$ vector $v^{(f_1,b)}_5(\lambda,\mu)$ and the $(d_1+2) \times 1$ vector $v^{(a_1,d_1)}_6(\lambda,\mu)$ are
\EQ
\label{v5}
v^{(f_1,b)}_5(\lambda,\mu)=\left( \begin{array}{c}
R(\lambda,\mu)_{f_1, 1}^{\bar{e}, f_1+1-\bar{e}}
\\
R(\lambda,\mu)_{f_1-1, 2}^{\bar{e}, f_1+1-\bar{e}}
\\
\vdots
\\
R(\lambda,\mu)_{f_1-b+2, b-1}^{\bar{e}, f_1+1-\bar{e}}
\end{array}\right),
~~~~ v^{(a_1,d_1)}_6(\lambda,\mu)=\left( \begin{array}{c}
R(\lambda,\mu)_{f_1, 1}^{\bar{e}, f_1+1-\bar{e}}
\\
R(\lambda,\mu)_{f_1-1, 2}^{\bar{e}, f_1+1-\bar{e}}
\\
\vdots
\\
R(\lambda,\mu)_{a_1-1, d_1+2}^{\bar{e}, f_1+1-\bar{e}}
\end{array}\right).
\EN

By construction the linear systems defined by Eqs.(\ref{aniq1},\ref{aniq2})
contain explicitly the product ${\cal T}_{f_1, a_1-1}(\lambda) {\cal T}_{1, b}(\mu)$
which can be calculated by means of Cramer's rule.
From the right-hand side of Eqs.(\ref{aniq1},\ref{aniq2}) we observe that this solution is able to produce products having the general form ${\cal T}_{c_1+1, \bar{e}}(\mu) {\cal T}_{f_1-c_1, a_1+b-1-\bar{e}}(\lambda)$.
This type of terms include the desirable reversed product ${\cal T}_{1, b}(\mu) {\cal T}_{f_1, a_1-1}(\lambda)$\footnote{This is the first component of the right-hand side vector ($c_1=0$) in Eqs.(\ref{aniq1},\ref{aniq2}) when $\bar{e}=b$.} but
also several other products having rather undesirable properties.
The latter feature occurs when $\bar{e} \le c_1+f_2=b+c_1-d_1-2$
because the operator  ${\cal T}_{f_1-c_1, a_1+b-1-\bar{e}}(\lambda)$ turns out
to be a creation field and its presence on the further
right position of the right-hand side commutation rule should definitely not be permitted.
We stress here that this problem is independent of the character of the companion field ${\cal T}_{c_1+1, \bar{e}}(\mu)$.
This is clear when ${\cal T}_{c_1+1, \bar{e}}(\mu)$ plays the role of either annihilation
or diagonal operator since we have exactly the same situation
we managed to handle for the left-hand side of the commutation rules.
The case when ${\cal T}_{c_1+1, \bar{e}}(\mu)$ play the role of creation operators is more subtle because the product
${\cal T}_{c_1+1, \bar{e}}(\mu) {\cal T}_{f_1-c_1, a_1+b-1-\bar{e}}(\lambda)$ could, in principle, contribute to the eigenvector basis.
The order of the rapidities in such product is however opposite to that already chosen in section (\ref{sub32}).
For this reason such remaining type of product combination has also to be avoided.
Therefore, no matter the role of ${\cal T}_{c_1+1, \bar{e}}(\mu)$, a second step is still necessary
to provide us the means to compute the product ${\cal T}_{c_1+1, \bar{e}}(\mu)
{\cal T}_{f_1-c_1, a_1+b-1-\bar{e}}(\lambda)$ as long as the index $\bar{e}$ satisfies the relation,
\EQ
\bar{e} \le c_1+f_2=b+c_1-d_1-2.
\EN

This additional task has to be implemented without spoiling
the main construction underlying the first linear system.
This can be done by using the freedom of the extra index $c_2$ since so far
we have kept it fixed at $c_2=0$.
In Table 5 we have summarized the second linear combination for a given $c_1$ with the help of index $c_2$.
The equations derived from such linear combination are suitable to
calculate the products ${\cal T}_{c_1+1, \bar{e}}(\mu) {\cal T}_{f_1-c_1, a_1+b-1-\bar{e}}(\lambda)$ on the interval
$ M \lbrace 1, a_1+b-N-1 \rbrace \le \bar{e} \le m \lbrace a_1+b-2, N \rbrace$. Recall here that this interval is the total
range of the index $\bar{e}$ in Eqs.(\ref{aniq1},\ref{aniq2}).
\begin{table}[h]
\begin{center}
\begin{tabular}{|c|c|c|c|}
\hline
Operators  & Creation & Combination  & Number of \\
calculated & index & indexes $c_2$ & equations \\
\hline
${\cal T}_{c_1+1, \bar{e}}(\mu) {\cal T}_{f_1-c_1, a_1+b-1-\bar{e}}(\lambda)$ &
$ b-2 \le N-a_1 $ & $c_2= d_1-c_1+2, \dots, b-1$ & $c_1+b-d_1-2$ \\
\hline
${\cal T}_{c_1+1, \bar{e}}(\mu) {\cal T}_{f_1-c_1, a_1+b-1-\bar{e}}(\lambda)$ &
$ b-2 > N-a_1 $ & $c_2= d_1-c_1+2, \dots, $ & $c_1+N-a_1-d_1$ \\
\hline
 & & $N-a_1+1 $ & \\
\hline
\end{tabular}
\end{center}
\caption{\footnotesize{The linear combination from Eq.(\ref{fundrelindex3}) to compute
products of the form ${\cal T}_{c_1+1, \bar{e}}(\mu) {\cal T}_{f_1-c_1, a_1+b-1-\bar{e}}(\lambda)$
}}
\end{table}

By substituting the data of Table 5 in Eq.(\ref{fundrelindex3}) we find that the second system of
linear equations are given by,

\bear
\label{aniq3}
\lefteqn{ A^{(f_1-c_1,f_2+c_1)}_7(\lambda,\mu)
\left(\begin{array}{c}
{\cal T}_{c_1+1, 1}(\mu)
{\cal T}_{f_1-c_1, a_1+b-2}(\lambda) \\
{\cal T}_{c_1+1, 2}(\mu)
{\cal T}_{f_1-c_1, a_1+b-3}(\lambda) \\
\vdots \\
{\cal T}_{c_1+1, f_2+c_1}(\mu)
{\cal T}_{f_1-c_1, f_1-c_1+1}(\lambda)
\end{array}\right)
=
\sum_{\bar{e}=1}^{f_1} R(\lambda,\mu)_{f_1-c_1, c_1+1}^{\bar{e},f_1+1-\bar{e}} }
\nonumber \\
& \times &
\left(\begin{array}{c}
{\cal T}_{\bar{e}, f_1-c_1+1}(\lambda)
{\cal T}_{f_1+1-\bar{e}, f_2+c_1}(\mu) \\
{\cal T}_{\bar{e}, f_1-c_1+2}(\lambda)
{\cal T}_{f_1+1-\bar{e}, f_2+c_1-1}(\mu) \\
\vdots \\
{\cal T}_{\bar{e}, a_1+b-2}(\lambda)
{\cal T}_{f_1+1-\bar{e}, 1}(\mu)
\end{array}\right)
- \sum_{\bar{e}=b+c_1-d_1-1}^{a_1+b-2}
v^{(f_1-c_1,f_2+c_1)}_7(\lambda,\mu)
\nonumber \\
&\times&
{\cal T}_{c_1+1, \bar{e}}(\mu)
{\cal T}_{f_1-c_1,a_1+b-1-\bar{e}}(\lambda)
%\nonumber \\
~~~~~~~~~~~~~~~~~~~~~~~~~~~~~ \mathrm{for}~~~ b \le N-a_1+2,
\ear
and
\bear
\label{aniq4}
\lefteqn{A^{(f_1-c_1,f_2+c_1)}_8(\lambda,\mu)
\left(\begin{array}{c}
{\cal T}_{c_1+1, a_1+b-N-1}(\mu)
{\cal T}_{f_1-c_1, N}(\lambda) \\
{\cal T}_{c_1+1, a_1+b-N}(\mu)
{\cal T}_{f_1-c_1, N-1}(\lambda) \\
\vdots \\
{\cal T}_{c_1+1, f_2+c_1}(\mu)
{\cal T}_{f_1-c_1, f_1-c_1+1}(\lambda)
\end{array}\right)
=
\sum_{\bar{e}=1}^{f_1} R(\lambda,\mu)_{f_1-c_1, c_1+1}^{\bar{e},f_1+1-\bar{e}}}
\nonumber \\
&\times&
\left(\begin{array}{c}
{\cal T}_{\bar{e}, f_1-c_1+1}(\lambda)
{\cal T}_{f_1+1-\bar{e}, f_2+c_1}(\mu) \\
{\cal T}_{\bar{e}, f_1-c_1+2}(\lambda)
{\cal T}_{f_1+1-\bar{e}, f_2+c_1-1}(\mu) \\
\vdots \\
{\cal T}_{\bar{e}, N}(\lambda)
{\cal T}_{f_1+1-\bar{e}, a_1+b-N-1}(\mu)
\end{array}\right)
%\nonumber \\
-
\sum_{\bar{e}=b+c_1-d_1-1}^{N}
v^{(f_1-c_1,f_2+c_1)}_8(\lambda,\mu)
\nonumber \\
&\times&
{\cal T}_{c_1+1, \bar{e}}(\mu)
{\cal T}_{f_1-c_1,a_1+b-1-\bar{e}}(\lambda)
% \nonumber \\
~~~~~~~~~~~~~~~~~~~~~~~~~~~~~ \mathrm{for}~~~ b > N-a_1+2,
\ear
where the $b \times b$ matrix $A^{(a,b)}_7(\lambda,\mu)$ and the $N+1-a \times N+1-a$ matrix $A^{(a,b)}_8(\lambda,\mu)$ are given in terms of the structure constants as
\bear
\label{am7}
A^{(a,b)}_7(\lambda,\mu)=
\left( \begin{array}{cccc}
R(\lambda,\mu)_{a+b, 1}^{a+1, b} & R(\lambda,\mu)_{a+b-1, 2}^{a+1, b} & \cdots & R(\lambda,\mu)_{a+1,b}^{a+1, b} \\
R(\lambda,\mu)_{a+b, 1}^{a+2, b-1} & R(\lambda,\mu)_{a+b-1, 2}^{a+2, b-1} & \cdots & R(\lambda,\mu)_{a+1,b}^{a+2, b-1} \\
\vdots & \vdots & \ddots & \vdots \\
R(\lambda,\mu)_{a+b, 1}^{a+b, 1} & R(\lambda,\mu)_{a+b-1, 2}^{a+b, 1} & \cdots & R(\lambda,\mu)_{a+1,b}^{a+b, 1} \\
\end{array}\right)
\ear
and
\bear
\label{am8}
A^{(a,b)}_8(\lambda,\mu)=
\left( \begin{array}{cccc}
R(\lambda,\mu)_{N,a+b-N+1}^{a+1, b} & R(\lambda,\mu)_{N-1,a+b-N+2}^{a+1, b} & \cdots & R(\lambda,\mu)_{a+1,b}^{a+1, b} \\
R(\lambda,\mu)_{N,a+b-N+1}^{a+2, b-1} & R(\lambda,\mu)_{N-1,a+b-N+2}^{a+2, b-1} & \cdots & R(\lambda,\mu)_{a+1,b}^{a+2, b-1} \\
\vdots & \vdots & \ddots & \vdots \\
R(\lambda,\mu)_{N,a+b-N+1}^{N,a+b-N+1} & R(\lambda,\mu)_{N-1,a+b-N+2}^{N,a+b-N+1} & \cdots & R(\lambda,\mu)_{a+1,b}^{N,a+b-N+1} \\
\end{array}\right),
\ear
while the $b \times 1$ vector $v^{(a,b)}_7(\lambda,\mu)$ and the $N+1-a \times 1$ vector $v^{(a,b)}_8(\lambda,\mu)$ are
\EQ
\label{v7}
v^{(a,b)}_7(\lambda,\mu)=\left( \begin{array}{c}
R(\lambda,\mu)_{a+b+1-\bar{e}, \bar{e}}^{a+1, b}
\\
R(\lambda,\mu)_{a+b+1-\bar{e}, \bar{e}}^{a+2, b-1}
\\
\vdots
\\
R(\lambda,\mu)_{a+b+1-\bar{e}, \bar{e}}^{a+b, 1}
\end{array}\right)~,~~~
v^{(a,b)}_8(\lambda,\mu)=\left( \begin{array}{c}
R(\lambda,\mu)_{a+b+1-\bar{e}, \bar{e}}^{a+1, b}
\\
R(\lambda,\mu)_{a+b+1-\bar{e}, \bar{e}}^{a+2, b-1}
\\
\vdots
\\
R(\lambda,\mu)_{a+b+1-\bar{e}, \bar{e}}^{N, a+b-N+1}
\end{array}\right).
\EN

\begin{table}[ht]
\begin{center}
\begin{tabular}{|c c|c|c|}
\hline
\multicolumn{2}{|c|}{Annihilation} & Combination & Combination   \\
\multicolumn{2}{|c|}{index} & index 1 & index 2  \\
\hline
\multicolumn{2}{|c|}{$ b-2 \le d_1 $} & $ c_1=0, \dots, b-2 $ & $c_2= d_1-c_1+2, \dots, b-1$ \\
\hline
$ b-2 > d_1 $ & $ b-1 \le N-a_1 $ & $ c_1=0, \dots, d_1+1 $ & $c_2= d_1-c_1+2, \dots, b-1$ \\
\hline
$ b-2 > d_1 $ & $ b-1 > N-a_1 $ & $ c_1=0, \dots, d_1+1 $ & $c_2= d_1-c_1+2, \dots, N-a_1+1$ \\
\hline
\end{tabular}
\end{center}
\caption{\footnotesize{The linear combination from equation (\ref{fundrelindex3}) necessary to obtain suitable commutation rules between the fields ${\cal T}_{a_1+d_1,a_1-1}(\lambda)$ and ${\cal T}_{1,b}(\mu)$}}
\end{table}

Let us now show how the first and the second systems of linear equations work in practice together. In order
to see that we shall discuss in detail the commutation rule among the fields
${\cal T}_{f_1,a_1-1}(\lambda)$ and ${\cal T}_{1,3}(\mu)$. We start by considering
the first linear system of equations (\ref{aniq1},\ref{aniq2}) which for $b=3$ gives us the following
expressions,
\bear
\label{aniq131}
A^{(f_1,3)}_5(\lambda,\mu)
\left(\begin{array}{c}
{\cal T}_{f_1-1, a_1-1}(\lambda)
{\cal T}_{2, 3}(\mu) \\
{\cal T}_{f_1, a_1-1}(\lambda)
{\cal T}_{1, 3}(\mu)
\end{array}\right)
=
 \sum_{\bar{e}=1}^{a_1+1} R(\lambda,\mu)_{a_1+2-\bar{e}, \bar{e}}^{a_1-1,3}
\left(\begin{array}{c}
{\cal T}_{1, \bar{e}}(\mu)
{\cal T}_{f_1, a_1+2-\bar{e}}(\lambda) \\
{\cal T}_{2, \bar{e}}(\mu)
{\cal T}_{f_1-1, a_1+2-\bar{e}}(\lambda)
\end{array}\right)
\nonumber \\
- \sum_{\bar{e}=1}^{f_1-2}
v^{(f_1,3)}_5(\lambda,\mu)
{\cal T}_{\bar{e}, a_1-1}(\lambda)
{\cal T}_{f_1+1-\bar{e}, 3}(\mu)
~~~~~~~~~~~~~~ \mathrm{for}~~~  d_1 \ge 1,
\ear
and
\bear
\label{aniq132}
A^{(a_1,0)}_6(\lambda,\mu)
\left(\begin{array}{c}
{\cal T}_{a_1-1, a_1-1}(\lambda)
{\cal T}_{2, 3}(\mu) \\
{\cal T}_{a_1, a_1-1}(\lambda)
{\cal T}_{1, 3}(\mu)
\end{array}\right)
= \sum_{\bar{e}=M \lbrace 1,a_1+2-N \rbrace }^{m \lbrace a_1+1,N \rbrace}
\left(\begin{array}{c}
{\cal T}_{1, \bar{e}}(\mu)
{\cal T}_{a_1, a_1+2-\bar{e}}(\lambda) \\
{\cal T}_{2, \bar{e}}(\mu)
{\cal T}_{a_1-1, a_1+2-\bar{e}}(\lambda)
\end{array}\right)
\nonumber \\
\times
R(\lambda,\mu)_{a_1+2-\bar{e}, \bar{e}}^{a_1-1,3}
-
\sum_{\bar{e}=1}^{a_1-2}
v^{(a_1,0)}_6(\lambda,\mu)
{\cal T}_{\bar{e}, a_1-1}(\lambda)
{\cal T}_{a_1+1-\bar{e}, 3}(\mu)
%\nonumber \\
 ~~~~~ \mathrm{for}~~~ d_1=0.
\ear

The matrices $A^{(f_1,3)}_5(\lambda,\mu)$ and $A^{(a_1,0)}_6(\lambda,\mu)$ are given by
\EQ
A^{(f_1,3)}_5(\lambda,\mu)=
\bigg( \begin{array}{cc}
R(\lambda,\mu)_{f_1, 1}^{f_1-1, 2} & R(\lambda,\mu)_{f_1, 1}^{f_1, 1} \\
R(\lambda,\mu)_{f_1-1, 2}^{f_1-1, 2} & R(\lambda,\mu)_{f_1-1, 2}^{f_1, 1}
\end{array}\bigg),~
A^{(a_1,0)}_6(\lambda,\mu)=
\bigg( \begin{array}{cc}
R(\lambda,\mu)_{a_1, 1}^{a_1-1, 2} & R(\lambda,\mu)_{a_1, 1}^{a_1, 1} \\
R(\lambda,\mu)_{a_1-1, 2}^{a_1-1, 2} & R(\lambda,\mu)_{a_1-1, 2}^{a_1, 1}
\end{array}\bigg),
\EN
while the vectors $v^{(f_1,3)}_5(\lambda,\mu)$ and $v^{(a_1,0)}_6(\lambda,\mu)$ are
\EQ
v^{(f_1,3)}_5(\lambda,\mu)=\left( \begin{array}{c}
R(\lambda,\mu)_{f_1, 1}^{\bar{e}, f_1+1-\bar{e}}
\\
R(\lambda,\mu)_{f_1-1, 2}^{\bar{e}, f_1+1-\bar{e}}
\end{array}\right),
~~~~ v^{(a_1,0)}_6(\lambda,\mu)=\left( \begin{array}{c}
R(\lambda,\mu)_{a_1, 1}^{\bar{e}, a_1+1-\bar{e}}
\\
R(\lambda,\mu)_{a_1-1, 2}^{\bar{e}, a_1+1-\bar{e}}
\end{array}\right).
\EN

By using  Cramer's rule in
Eqs.(\ref{aniq131},\ref{aniq132}) we can compute the products ${\cal T}_{f_1, a_1-1}(\lambda)
{\cal T}_{1, 3}(\mu)$ and the results are,
\bear
\label{saniq131}
{\cal T}_{f_1, a_1-1}(\lambda)
{\cal T}_{1, 3}(\mu)
&=&
\sum_{\bar{e}=1}^{a_1+1}
\frac{R(\lambda,\mu)_{a_1+2-\bar{e}, \bar{e}}^{a_1-1,3} R(\lambda,\mu)_{f_1, 1}^{f_1-1, 2}}
{\left| \begin{array}{cc}
R(\lambda,\mu)_{f_1, 1}^{f_1-1, 2} & R(\lambda,\mu)_{f_1, 1}^{f_1, 1} \\
R(\lambda,\mu)_{f_1-1, 2}^{f_1-1, 2} & R(\lambda,\mu)_{f_1-1, 2}^{f_1, 1}
\end{array}\right|} {\cal T}_{2, \bar{e}}(\mu)
{\cal T}_{f_1-1, a_1+2-\bar{e}}(\lambda)
\nonumber \\
&-&
\sum_{\bar{e}=1}^{a_1+1}
\frac{R(\lambda,\mu)_{a_1+2-\bar{e}, \bar{e}}^{a_1-1,3} R(\lambda,\mu)_{f_1-1, 2}^{f_1-1, 2}}
{\left| \begin{array}{cc}
R(\lambda,\mu)_{f_1, 1}^{f_1-1, 2} & R(\lambda,\mu)_{f_1, 1}^{f_1, 1} \\
R(\lambda,\mu)_{f_1-1, 2}^{f_1-1, 2} & R(\lambda,\mu)_{f_1-1, 2}^{f_1, 1}
\end{array}\right|}
{\cal T}_{1, \bar{e}}(\mu)
{\cal T}_{f_1, a_1+2-\bar{e}}(\lambda)
\nonumber \\
&-&\sum_{\bar{e}=1}^{f_1-2}
\frac{\left| \begin{array}{cc}
R(\lambda,\mu)_{f_1, 1}^{f_1-1, 2} & R(\lambda,\mu)_{f_1, 1}^{\bar{e}, f_1+1-\bar{e}} \\
R(\lambda,\mu)_{f_1-1, 2}^{f_1-1, 2} & R(\lambda,\mu)_{f_1-1, 2}^{\bar{e}, f_1+1-\bar{e}}
\end{array}\right|}
{\left| \begin{array}{cc}
R(\lambda,\mu)_{f_1, 1}^{f_1-1, 2} & R(\lambda,\mu)_{f_1, 1}^{f_1, 1} \\
R(\lambda,\mu)_{f_1-1, 2}^{f_1-1, 2} & R(\lambda,\mu)_{f_1-1, 2}^{f_1, 1}
\end{array}\right|}
{\cal T}_{\bar{e}, a_1-1}(\lambda)
{\cal T}_{f_1+1-\bar{e}, 3}(\mu)
\nonumber \\
&&~~~~~~~~~~~~~~~~~~~~~~~~ \mathrm{for}~~~  d_1 \ge 1,
\ear
and
\bear
\label{saniq132}
{\cal T}_{a_1, a_1-1}(\lambda)
{\cal T}_{1, 3}(\mu)
&=&
\sum_{\bar{e}=M \lbrace 1,a_1+2-N \rbrace }^{m \lbrace a_1+1,N \rbrace}
\frac{R(\lambda,\mu)_{a_1+2-\bar{e}, \bar{e}}^{a_1-1,3} R(\lambda,\mu)_{a_1, 1}^{a_1-1, 2}}
{\left| \begin{array}{cc}
R(\lambda,\mu)_{a_1, 1}^{a_1-1, 2} & R(\lambda,\mu)_{a_1, 1}^{a_1, 1} \\
R(\lambda,\mu)_{a_1-1, 2}^{a_1-1, 2} & R(\lambda,\mu)_{a_1-1, 2}^{a_1, 1}
\end{array}\right|}
{\cal T}_{2, \bar{e}}(\mu) {\cal T}_{a_1-1, a_1+2-\bar{e}}(\lambda)
\nonumber \\
&-&
\sum_{\bar{e}=M \lbrace 1,a_1+2-N \rbrace }^{m \lbrace a_1+1,N \rbrace}
\frac{R(\lambda,\mu)_{a_1+2-\bar{e}, \bar{e}}^{a_1-1,3} R(\lambda,\mu)_{a_1-1, 2}^{a_1-1, 2}}
{\left| \begin{array}{cc}
R(\lambda,\mu)_{a_1, 1}^{a_1-1, 2} & R(\lambda,\mu)_{a_1, 1}^{a_1, 1} \\
R(\lambda,\mu)_{a_1-1, 2}^{a_1-1, 2} & R(\lambda,\mu)_{a_1-1, 2}^{a_1, 1}
\end{array}\right|}
{\cal T}_{1, \bar{e}}(\mu) {\cal T}_{a_1, a_1+2-\bar{e}}(\lambda)
\nonumber \\
&-&\sum_{\bar{e}=1}^{a_1-2}
\frac{\left| \begin{array}{cc}
R(\lambda,\mu)_{a_1, 1}^{a_1-1, 2} & R(\lambda,\mu)_{a_1, 1}^{\bar{e}, a_1+1-\bar{e}} \\
R(\lambda,\mu)_{a_1-1, 2}^{a_1-1, 2} & R(\lambda,\mu)_{a_1-1, 2}^{\bar{e}, a_1+1-\bar{e}}
\end{array}\right|}
{\left| \begin{array}{cc}
R(\lambda,\mu)_{a_1, 1}^{a_1-1, 2} & R(\lambda,\mu)_{a_1, 1}^{a_1, 1} \\
R(\lambda,\mu)_{a_1-1, 2}^{a_1-1, 2} & R(\lambda,\mu)_{a_1-1, 2}^{a_1, 1}
\end{array}\right|}
{\cal T}_{\bar{e}, a_1-1}(\lambda)
{\cal T}_{a_1+1-\bar{e}, 3}(\mu)
\nonumber \\
&& ~~~~~~~~~~~~~~~~~~~~~~~~ \mathrm{for}~~~ d_1=0.
\ear

At this point we note that the commutation rules (\ref{saniq131},\ref{saniq132}) between
${\cal T}_{f_1=a_1+d_1,a_1-1}(\lambda)$  and
${\cal T}_{1,3}(\mu)$  possess the appropriate form only when
$d_1>1$.  In these cases the right-hand side of
Eqs.(\ref{saniq131},\ref{saniq132})  does not generate
undesirable products of fields and the second system of linear equations is not needed.
This is not the situation of the remaining  values $d_1=1$ or $d_1=0$.  For example, when $d_1=1$
we see that the first term of Eq.(\ref{saniq131}) for $\bar{e}=1$  produces the undesirable product
${\cal T}_{2, 1}(\mu) {\cal T}_{a_1, a_1+1}(\lambda)$. This product can, however, be eliminated thanks to
the second linear system of equations. Indeed, by considering Eq.(\ref{aniq3}) for
$c_1=1$ we can compute the product
${\cal T}_{2, 1}(\mu) {\cal T}_{a_1, a_1+1}(\lambda)$ as,
\bear
\label{extraeq}
{\cal T}_{2, 1}(\mu)
{\cal T}_{a_1, a_1+1}(\lambda)
&=&
\sum_{\bar{e}=1}^{a_1+1}
\frac{R(\lambda,\mu)_{a_1, 2}^{\bar{e},a_1+2-\bar{e}}}{R(\lambda,\mu)_{a_1+1, 1}^{a_1+1, 1}}
{\cal T}_{\bar{e}, a_1+1}(\lambda)
{\cal T}_{a_1+2-\bar{e}, 1}(\mu)
\nonumber \\
&-&
\sum_{\bar{e}=2}^{a_1+1}
\frac{R(\lambda,\mu)_{a_1+2-\bar{e}, \bar{e}}^{a_1+1, 1}}{R(\lambda,\mu)_{a_1+1, 1}^{a_1+1, 1}}
{\cal T}_{2, \bar{e}}(\mu)
{\cal T}_{a_1,a_1+2-\bar{e}}(\lambda)
~~~~ \mathrm{for}~~ a_1 \le N-1.
\ear

For the value $d_1=1$ we can now substitute Eq.(\ref{extraeq}) in Eq.(\ref{saniq131}). This cancels out the
undesirable product, providing  us the suitable commutation rule between   the operators
${\cal T}_{a_1+1,a_1-1}(\lambda)$ and ${\cal T}_{1,3}(\mu)$,
\bear
\lefteqn{ {\cal T}_{a_1+1, a_1-1}(\lambda)
{\cal T}_{1, 3}(\mu)
=
-\sum_{\bar{e}=2}^{a_1+1}
\frac{R(\lambda,\mu)_{a_1+1, 1}^{a_1, 2}}{R(\lambda,\mu)_{a_1+1, 1}^{a_1+1, 1}}
\frac{\left| \begin{array}{cc}
R(\lambda,\mu)_{a_1+1, 1}^{a_1-1, 3} & R(\lambda,\mu)_{a_1+1, 1}^{a_1+1, 1} \\
R(\lambda,\mu)_{a_1+2-\bar{e}, \bar{e}}^{a_1-1, 3} & R(\lambda,\mu)_{a_1+2-\bar{e}, \bar{e}}^{a_1+1, 1}
\end{array}\right|}{\left| \begin{array}{cc}
R(\lambda,\mu)_{a_1+1, 1}^{a_1, 2} & R(\lambda,\mu)_{a_1+1, 1}^{a_1+1, 1} \\
R(\lambda,\mu)_{a_1, 2}^{a_1, 2} & R(\lambda,\mu)_{a_1, 2}^{a_1+1, 1}
\end{array}\right|} {\cal T}_{2, \bar{e}}(\mu) }
\nonumber \\
& \times &
{\cal T}_{a_1, a_1+2-\bar{e}}(\lambda)
-
\sum_{\bar{e}=1}^{a_1+1}
\frac{R(\lambda,\mu)_{a_1+2-\bar{e}, \bar{e}}^{a_1-1,3} R(\lambda,\mu)_{a_1, 2}^{a_1, 2}}
{\left| \begin{array}{cc}
R(\lambda,\mu)_{a_1+1, 1}^{a_1, 2} & R(\lambda,\mu)_{a_1+1, 1}^{a_1+1, 1} \\
R(\lambda,\mu)_{a_1, 2}^{a_1, 2} & R(\lambda,\mu)_{a_1, 2}^{a_1+1, 1}
\end{array}\right|}
{\cal T}_{1, \bar{e}}(\mu)
{\cal T}_{a_1+1, a_1+2-\bar{e}}(\lambda)
\nonumber \\
&+&
\sum_{\bar{e}=1}^{a_1+1}
\frac{R(\lambda,\mu)_{a_1, 2}^{\bar{e}, a_1+2-\bar{e}}}{R(\lambda,\mu)_{a_1+1, 1}^{a_1+1, 1}}
\frac{R(\lambda,\mu)_{a_1+1, 1}^{a_1, 2} R(\lambda,\mu)_{a_1+1, 1}^{a_1-1, 3}}{\left| \begin{array}{cc}
R(\lambda,\mu)_{a_1+1, 1}^{a_1, 2} & R(\lambda,\mu)_{a_1+1, 1}^{a_1+1, 1} \\
R(\lambda,\mu)_{a_1, 2}^{a_1, 2} & R(\lambda,\mu)_{a_1, 2}^{a_1+1, 1}
\end{array}\right|}
{\cal T}_{\bar{e}, a_1+1}(\lambda)
{\cal T}_{a_1+2-\bar{e}, 1}(\mu)
\nonumber \\
&-&
\sum_{\bar{e}=1}^{a_1-1}
\frac{\left| \begin{array}{cc}
R(\lambda,\mu)_{a_1+1, 1}^{a_1, 2} & R(\lambda,\mu)_{a_1+1, 1}^{\bar{e}, a_1+2-\bar{e}} \\
R(\lambda,\mu)_{a_1, 2}^{a_1, 2} & R(\lambda,\mu)_{a_1, 2}^{\bar{e}, a_1+2-\bar{e}}
\end{array}\right|}
{\left| \begin{array}{cc}
R(\lambda,\mu)_{a_1+1, 1}^{a_1, 2} & R(\lambda,\mu)_{a_1+1, 1}^{a_1+1, 1} \\
R(\lambda,\mu)_{a_1, 2}^{a_1, 2} & R(\lambda,\mu)_{a_1, 2}^{a_1+1, 1}
\end{array}\right|}
{\cal T}_{\bar{e}, a_1-1}(\lambda)
{\cal T}_{a_1+2-\bar{e}, 3}(\mu)
\label{Ta+1a-1T13}
\ear

The case $d_1=0$ is more laborious due to the presence of three different types of undesirable products.
They come from the first ($\bar{e}=1,2$) and the second ($\bar{e}=1$) terms of Eq.(\ref{saniq132})
and their respective forms are
${\cal T}_{2, 1}(\mu) {\cal T}_{a_1-1, a_1+1}(\lambda)$, ${\cal T}_{2, 2}(\mu) {\cal T}_{a_1-1, a_1}(\lambda)$ and
${\cal T}_{1, 1}(\mu) {\cal T}_{a_1, a_1+1}(\lambda)$. All these terms can be eliminated  with
the help of the second linear system of equations  as follows. The last undesirable term
${\cal T}_{1, 1}(\mu) {\cal T}_{a_1, a_1+1}(\lambda)$ is the simplest one to be computed.
Its expression comes directly from Eq.(\ref{aniq3}) by taking the
values $c_1=d_1=0$,
\bear
{\cal T}_{1, 1}(\mu)
{\cal T}_{a_1, a_1+1}(\lambda)
&=&
\sum_{\bar{e}=1}^{a_1}
\frac{R(\lambda,\mu)_{a_1, 1}^{\bar{e},a_1+1-\bar{e}}}{R(\lambda,\mu)_{a_1+1, 1}^{a_1+1, 1}}
{\cal T}_{\bar{e}, a_1+1}(\lambda)
{\cal T}_{a_1+1-\bar{e}, 1}(\mu)
\nonumber \\
&-&
\sum_{\bar{e}=2}^{a_1+1}
\frac{R(\lambda,\mu)_{a_1+2-\bar{e}, \bar{e}}^{a_1+1, 1}}{R(\lambda,\mu)_{a_1+1, 1}^{a_1+1, 1}}
{\cal T}_{1, \bar{e}}(\mu)
{\cal T}_{a_1,a_1+2-\bar{e}}(\lambda)
\label{comT11Taa+1}
\ear

The remaining products
${\cal T}_{2, 1}(\mu) {\cal T}_{a_1-1, a_1+1}(\lambda)$ and ${\cal T}_{2, 2}(\mu) {\cal T}_{a_1-1, a_1}(\lambda)$ are
obtained as a solution of a $2 \times 2$ linear system of equations. It is derived from Eqs.(\ref{aniq3},\ref{aniq4})
for the values $c_1=1$ and $d_1=0$ and the final result is,
\bear
A^{(a_1-1,2)}_7(\lambda,\mu)
\left(\begin{array}{c}
{\cal T}_{2, 1}(\mu)
{\cal T}_{a_1-1, a_1+1}(\lambda) \\
{\cal T}_{2, 2}(\mu)
{\cal T}_{a_1-1, a_1}(\lambda) \\
\end{array}\right)
=
\sum_{\bar{e}=1}^{a_1} R(\lambda,\mu)_{a_1-1, 2}^{\bar{e},a_1+1-\bar{e}}
\left(\begin{array}{c}
{\cal T}_{\bar{e}, a_1}(\lambda)
{\cal T}_{a_1+1-\bar{e}, 2}(\mu) \\
{\cal T}_{\bar{e}, a_1+1}(\lambda)
{\cal T}_{a_1+1-\bar{e}, 1}(\mu)
\end{array}\right)
\nonumber \\
-
\sum_{\bar{e}=3}^{a_1+1}
v^{(a_1-1,2)}_7(\lambda,\mu)
{\cal T}_{2, \bar{e}}(\mu)
{\cal T}_{a_1-1,a_1+2-\bar{e}}(\lambda)
~~~~~~~~~~ \mathrm{for}~~~ a_1 \le N-1,
\label{sist2x2T21Ta-1a+1}
\ear
\bear
R(\lambda,\mu)_{N, 2}^{N, 2}
{\cal T}_{2, 2}(\mu) {\cal T}_{N-1, N}(\lambda)
&=&
\sum_{\bar{e}=1}^{N} R(\lambda,\mu)_{N-1, 2}^{\bar{e},N+1-\bar{e}}
{\cal T}_{\bar{e}, N}(\lambda) {\cal T}_{N+1-\bar{e}, 2}(\mu)
\nonumber \\
&-&
\sum_{\bar{e}=3}^{N}
R(\lambda,\mu)_{N+2-\bar{e}, \bar{e}}^{N, 2}
{\cal T}_{2, \bar{e}}(\mu)
{\cal T}_{N-1,N+2-\bar{e}}(\lambda),
\label{sist2x2T21TN-1N+1}
\ear
where
\EQ
A^{(a_1-1,2)}_7(\lambda,\mu)=
\left( \begin{array}{cc}
R(\lambda,\mu)_{a_1+1, 1}^{a_1, 2} & R(\lambda,\mu)_{a_1, 2}^{a_1, 2} \\
R(\lambda,\mu)_{a_1+1, 1}^{a_1+1, 1} & R(\lambda,\mu)_{a_1, 2}^{a_1+1, 1}
\end{array}\right)
~~~\mbox{and}~~
v^{(a_1-1,2)}_7(\lambda,\mu)=\left( \begin{array}{c}
R(\lambda,\mu)_{a_1+2-\bar{e}, \bar{e}}^{a_1, 2}
\\
R(\lambda,\mu)_{a_1+2-\bar{e}, \bar{e}}^{a_1+1, 1}
\end{array}\right).
\EN

By applying Cramer's rule  in Eq.(\ref{sist2x2T21Ta-1a+1}) we can easily compute the undesirable terms
\newline ${\cal T}_{2, 1}(\mu) {\cal T}_{a_1-1, a_1+1}(\lambda)$ and
${\cal T}_{2, 2}(\mu) {\cal T}_{a_1-1, a_1}(\lambda)$ for $a_1 \le N-1$. We then substitute
this result as well as the expression
(\ref{comT11Taa+1}) for
${\cal T}_{1, 1}(\mu)
{\cal T}_{a_1, a_1+1}(\lambda)$ in Eq.(\ref{saniq132}).
In this way we are able to obtain the commutation rule for the fields
${\cal T}_{a_1, a_1-1}(\lambda)$ and
${\cal T}_{1, 3}(\mu) $ for $a_1 \le N$.
For $a_1=N$  we just have to eliminate the term
${\cal T}_{2, 2}(\mu) {\cal T}_{N-1, N}(\lambda)$  of Eq.(\ref{saniq132}) with the assistance
of the relation
(\ref{sist2x2T21TN-1N+1}).
The final step to obtain suitable commutation relations concerns with the
reordering of the  products
${\cal T}_{a_1, a_1}(\lambda) {\cal T}_{1, 2}(\mu)$ with the help of
Eqs.(\ref{1parta},\ref{1partN}).

\section{Identities among the  weights}
\label{secI}

In this section we describe a procedure to derive identities between the amplitudes $R(\lambda,\mu)_{a,b}^{c,d}$
from both the unitarity relation and the Yang-Baxter equation. These identities are essential to
carry out simplifications on the eigenvalue problem without the necessity of referring to specific weights.
We shall start by considering the consequences of the unitarity relation (\ref{uni}).

\subsection{Unitarity relation}
\label{subI1}

A systematic study of the unitary relation (\ref{uni}) revealed us the existence
of two key sets of independent weights. They will play a relevant role in
the analysis of Eq.(\ref{uni}) and are defined as,
\EQ
\bullet ~~a^{j,q_1}_{b,c} = \begin{cases}
\displaystyle R(\lambda,\mu)_{q_1+1-b,b}^{c,q_1+1-c},~~~~\mbox{for} ~~
j=1 \cr \displaystyle R(\lambda,\mu)_{N+1-b,N-q_1+b}^{N-q_1+c,N+1-c},
~~~~\mbox{for} ~~ j=2, \label{nomea1}
\end{cases}
\EN
\EQ
\bullet ~~\bar{a}^{j,q_1}_{b,c} = \begin{cases} \displaystyle
R(\mu,\lambda)_{q_1+1-b,b}^{c,q_1+1-c},~~~~\mbox{for} ~~ j=1 \cr
\displaystyle R(\mu,\lambda)_{N+1-b,N-q_1+b}^{N-q_1+c,N+1-c},
~~~~\mbox{for} ~~ j=2, \label{nomea2}
\end{cases}
\EN
where $q_1=1,\dots,N$ and $b,c=1,\dots,q_1$.

The above definition explores the block
form of the $R$-matrix on the basis of the $U(1)$ operator
$S^{z} \otimes I_N + I_N \otimes S^{z}$. The index $j=1,2$
is used to split the charges with $q=1,\dots,N$ ($j=1$)  from
those with $q=N+1,\dots, 2N-1$ ($j=2$). Considering
Eqs.(\ref{nomea1},\ref{nomea2})   we are able to write
the matrix elements of unitarity relation (\ref{uni}) as,
\EQ
U[b,c]_j^{q_1} \equiv \sum_{k=1}^{q_1}
a_{b,k}^{j,q_1} \bar{a}_{k,c}^{j,q_1}-
\delta_{b,c}=0~~\mbox{for}~q_1=1,\dots,N;~~b,c=1,\dots,q_1;~~j=1,2.
\label{uniai}
\EN

A non-trivial identity between the weights is already relevant
for the analysis of the one-particle problem. It is obtained
from the
$U[1,2]_1^2$ component
and by substituting the respective indices in Eq.(\ref{uniai}) we
find,
\EQ \frac{R(\lambda,\mu)_{2,1}^{1,2}}
{R(\lambda,\mu)_{2,1}^{2,1}} = - \frac{R(\mu,\lambda)_{1,2}^{2,1}}
{R(\mu,\lambda)_{2,1}^{2,1}}. \label{apAid1}
\EN

The type of relations are necessary because the combined use of
the commutation rules, discussed in previous section, forces
us to reorder the spectral parameters of some $R$-matrix
elements.  It is fortunate that we can solve this problem
with the assistance of the unitary property.
In general,  more complex identities are needed and they
are derived by  analyzing
linear systems of
equations  based on Eq.(\ref{uniai}). The first such
system  is obtained by means of the linear combination
$U[b;i+1]^{i+2}_{j}\bar{a}_{1,i+2}^{j,i+2}-U[b;i+2]_{j}^{i+2}
\bar{a}_{1,i+1}^{j,i+2}$ for $b=1,\dots,i+1$.
These equations make it possible to write the
following linear system for variables
$\bar{a}_{b+1,i+1}^{j,i+2}\bar{a}_{1,i+2}^{j,i+2}-\bar{a}_{b+1,i+2}^{j,i+2}\bar{a}_{1,i+1}^{j,i+2}$,
\EQ \left(\begin{array}{cccc}
a_{1,2}^{j,i+2} & a_{1,3}^{j,i+2} & \dots & a_{1,i+2}^{j,i+2} \\
a_{2,2}^{j,i+2} & a_{2,3}^{j,i+2} & \dots & a_{2,i+2}^{j,i+2} \\
\vdots & \vdots & \ddots & \vdots \\
a_{i+1,2}^{j,i+2} & a_{i+1,3}^{j,i+2} & \dots &
a_{i+1,i+2}^{j,i+2}
\end{array}\right)
\left(\begin{array}{c}
\bar{a}_{2,i+1}^{j,i+2}\bar{a}_{1,i+2}^{j,i+2}-
\bar{a}_{2,i+2}^{j,i+2}\bar{a}_{1,i+1}^{j,i+2} \\
\bar{a}_{3,i+1}^{j,i+2}\bar{a}_{1,i+2}^{j,i+2}-
\bar{a}_{3,i+2}^{j,i+2}\bar{a}_{1,i+1}^{j,i+2} \\
\vdots \\
\bar{a}_{i+2,i+1}^{j,i+2}\bar{a}_{1,i+2}^{j,i+2}-
\bar{a}_{i+2,i+2}^{j,i+2}\bar{a}_{1,i+1}^{j,i+2}
\end{array} \right) = \left(
\begin{array}{c} 0 \\
\vdots \\
0 \\
\bar{a}_{1,i+2}^{j,i+2} \end{array}\right)
\label{unisl1}
\EN

The solution of this system of equations for three particular components is able
to generate classes of important identities. The first family is found by solving
the system (\ref{unisl1}) for the first component.
By employing  Cramer's rule we found,
\EQ
-\left|\begin{array}{ccc}
a_{1,2}^{j,i+2} & \dots & a_{1,i+2}^{j,i+2} \\
\vdots & \ddots & \vdots \\
a_{i+1,2}^{j,i+2} & \dots & a_{i+1,i+2}^{j,i+2}
\end{array}\right|
\left|\begin{array}{cc}
\bar{a}_{1,i+1}^{j,i+2} & \bar{a}_{1,i+2}^{j,i+2} \\
\bar{a}_{2,i+1}^{j,i+2} & \bar{a}_{2,i+2}^{j,i+2}
\end{array} \right| =
\bar{a}_{1,i+2}^{j,i+2} (-1)^{i} \left|\begin{array}{ccc}
a_{1,3}^{j,i+2} & \dots & a_{1,i+2}^{j,i+2} \\
\vdots & \ddots & \vdots \\
a_{i,3}^{j,i+2} & \dots & a_{i,i+2}^{j,i+2}
\end{array}\right|
\label{apA1i}
\EN

The remaining families of identities follow from the solution of Eq.(\ref{unisl1}) for
the second and last components. After making the ratio of these solutions with Eq.(\ref{apA1i})
we have,
\EQ
\frac{\left|\begin{array}{cc}
\bar{a}_{1,i+1}^{j,i+2} & \bar{a}_{1,i+2}^{j,i+2} \\
\bar{a}_{3,i+1}^{j,i+2} & \bar{a}_{3,i+2}^{j,i+2}
\end{array} \right|}{\left|\begin{array}{cc}
\bar{a}_{1,i+1}^{j,i+2} & \bar{a}_{1,i+2}^{j,i+2} \\
\bar{a}_{2,i+1}^{j,i+2} & \bar{a}_{2,i+2}^{j,i+2}
\end{array} \right|} =-
\frac{\left|\begin{array}{cccc}
a_{1,2}^{j,i+2} & a_{1,4}^{j,i+2} & \dots & a_{1,i+2}^{j,i+2} \\
\vdots & \vdots & \ddots & \vdots \\
a_{i,2}^{j,i+2} & a_{i,4}^{j,i+2} & \dots & a_{i,i+2}^{j,i+2}
\end{array}\right|}{\left|\begin{array}{ccc}
a_{1,3}^{j,i+2} & \dots & a_{1,i+2}^{j,i+2} \\
\vdots & \ddots & \vdots \\
a_{i,3}^{j,i+2} & \dots & a_{i,i+2}^{j,i+2}
\end{array}\right|}.
\label{apAid2i} \EN
and
\EQ \frac{\left|\begin{array}{cc}
\bar{a}_{1,i+1}^{j,i+2} & \bar{a}_{1,i+2}^{j,i+2} \\
\bar{a}_{i+2,i+1}^{j,i+2} & \bar{a}_{i+2,i+2}^{j,i+2}
\end{array} \right|}{\left|\begin{array}{cc}
\bar{a}_{1,i+1}^{j,i+2} & \bar{a}_{1,i+2}^{j,i+2} \\
\bar{a}_{2,i+1}^{j,i+2} & \bar{a}_{2,i+2}^{j,i+2}
\end{array} \right|} =(-1)^{i}
\frac{\left|\begin{array}{ccc}
a_{1,2}^{j,i+2} & \dots & a_{1,i+1}^{j,i+2} \\
\vdots & \ddots & \vdots \\
a_{i,2}^{j,i+1} & \dots & a_{i,i+1}^{j,i+2}
\end{array}\right|}{\left|\begin{array}{ccc}
a_{1,3}^{j,i+2} & \dots & a_{1,i+2}^{j,i+2} \\
\vdots & \ddots & \vdots \\
a_{i,3}^{j,i+2} & \dots & a_{i,i+2}^{j,i+2}
\end{array}\right|}.
\label{apAid3parti} \EN

We shall now discuss four
specific identities that are going to be very useful in next section.
The first two identities are relevant to verify the exchange property of the two-particle state
under the respective rapidities.
Both of them are derived by selecting the same indices $i=j=1$ in Eqs.(\ref{apA1i},\ref{apAid2i}).
Considering that
$R(\lambda,\mu)_{1,1}^{1,1} R(\mu,\lambda)_{1,1}^{1,1}=1$ it is possible to write
Eq.(\ref{apA1i}) with $i=j=1$ as,
\EQ
D_{2}^{(2,0)}(\lambda,\mu)
D_{2}^{(2,0)}(\mu,\lambda)=\frac{R(\lambda,\mu)_{1,1}^{1,1}}{R(\lambda,\mu)_{2,1}^{2,1}}
\frac{R(\mu,\lambda)_{1,1}^{1,1}}{R(\mu,\lambda)_{2,1}^{2,1}},
\label{uni2ans2} \EN
while Eq.(\ref{apAid2i}) with $i=j=1$ gives us,
\EQ
D_{2}^{(2,1)}(\lambda,\mu)=-\frac{R(\mu,\lambda)_{3,1}^{2,2}}{R(\mu,\lambda)_{3,1}^{3,1}}
D_{2}^{(2,0)}(\lambda,\mu). \label{uni1ans2} \EN

The two remaining examples are going to be used in the three-particle problem.
As before they are important to demonstrate the rapidities symmetrization properties of this state.
They are derived from Eqs.(\ref{apAid2i},\ref{apAid3parti}) by choosing the indices $i=2$ and $j=1$, namely
\bear
\frac{D_{2}^{(3,1)}(\lambda_1,\lambda)}{D_{2}^{(3,0)}(\lambda_1,\lambda)}
&=& \frac{\left|\begin{array}{cc}
R(\lambda_1,\lambda)_{4,1}^{3,2} & R(\lambda_1,\lambda)_{2,3}^{3,2} \\
R(\lambda_1,\lambda)_{4,1}^{4,1} & R(\lambda_1,\lambda)_{2,3}^{4,1}
\end{array} \right|}{\left|\begin{array}{cc}
R(\lambda_1,\lambda)_{4,1}^{3,2} & R(\lambda_1,\lambda)_{3,2}^{3,2} \\
R(\lambda_1,\lambda)_{4,1}^{4,1} & R(\lambda_1,\lambda)_{3,2}^{4,1}
\end{array} \right|} = -
\frac{\left|\begin{array}{cc}
R(\lambda,\lambda_1)_{4,1}^{2,3} & R(\lambda,\lambda_1)_{3,2}^{2,3} \\
R(\lambda,\lambda_1)_{4,1}^{4,1} & R(\lambda,\lambda_1)_{3,2}^{4,1}
\end{array} \right|}{\left|\begin{array}{cc}
R(\lambda,\lambda_1)_{4,1}^{3,2} & R(\lambda,\lambda_1)_{3,2}^{3,2} \\
R(\lambda,\lambda_1)_{4,1}^{4,1} & R(\lambda,\lambda_1)_{3,2}^{4,1}
\end{array} \right|} \label{D231/D230}
\\
\frac{D_{2}^{(3,2)}(\lambda_1,\lambda)}{D_{2}^{(3,0)}(\lambda_1,\lambda)}
&=& \frac{\left|\begin{array}{cc}
R(\lambda_1,\lambda)_{4,1}^{3,2} & R(\lambda_1,\lambda)_{1,4}^{3,2} \\
R(\lambda_1,\lambda)_{4,1}^{4,1} & R(\lambda_1,\lambda)_{1,4}^{4,1}
\end{array} \right|}{\left|\begin{array}{cc}
R(\lambda_1,\lambda)_{4,1}^{3,2} & R(\lambda_1,\lambda)_{3,2}^{3,2} \\
R(\lambda_1,\lambda)_{4,1}^{4,1} & R(\lambda_1,\lambda)_{3,2}^{4,1}
\end{array} \right|} =
\frac{\left|\begin{array}{cc}
R(\lambda,\lambda_1)_{4,1}^{2,3} & R(\lambda,\lambda_1)_{3,2}^{2,3} \\
R(\lambda,\lambda_1)_{4,1}^{3,2} & R(\lambda,\lambda_1)_{3,2}^{3,2}
\end{array} \right|}{\left|\begin{array}{cc}
R(\lambda,\lambda_1)_{4,1}^{3,2} & R(\lambda,\lambda_1)_{3,2}^{3,2} \\
R(\lambda,\lambda_1)_{4,1}^{4,1} & R(\lambda,\lambda_1)_{3,2}^{4,1}
\end{array} \right|} \label{D232/D230}. \ear

In order to construct further relevant identities we found the necessity
of combining the results of two systems of linear equations. The first of
them is obtained by means of the linear combination
$U[b;k+2]^{i+2}_j \bar{a}^{j,i+2}_{1,i+2}-U[b;i+2]^{i+2}_j
\bar{a}^{j,i+2}_{1,k+2}$ for $b=1,2,3$ and $k=1,\dots,i-1$.
From these equations we are able to write a $3 \times 3$ system of equations for
variables $\bar{a}_{\bar{e},k+2}^{j,i+2} \bar{a}_{1,i+2}^{j,i+2}-
\bar{a}_{\bar{e},i+2}^{j,i+2} \bar{a}_{1,k+2}^{j,i+2}$ for
$\bar{e}=i,i+1,i+2$, namely
\bear
\left(\begin{array}{ccc}
a_{1,i}^{j,i+2} & a_{1,i+1}^{j,i+2} & a_{1,i+2}^{j,i+2} \\
a_{2,i}^{j,i+2} & a_{2,i+1}^{j,i+2} & a_{2,i+2}^{j,i+2} \\
a_{3,i}^{j,i+2} & a_{3,i+1}^{j,i+2} & a_{3,i+2}^{j,i+2}
\end{array}\right)
\left(\begin{array}{c}
\bar{a}_{i,k+2}^{j,i+2}\bar{a}_{1,i+2}^{j,i+2}-
\bar{a}_{i,i+2}^{j,i+2}\bar{a}_{1,k+2}^{j,i+2} \\
\bar{a}_{i+1,k+2}^{j,i+2}\bar{a}_{1,i+2}^{j,i+2}-
\bar{a}_{i+1,i+2}^{j,i+2}\bar{a}_{1,k+2}^{j,i+2} \\
\bar{a}_{i+2,k+2}^{j,i+2}\bar{a}_{1,i+2}^{j,i+2}-
\bar{a}_{i+2,i+2}^{j,i+2}\bar{a}_{1,k+2}^{j,i+2}
\end{array} \right)
=
-\sum_{\bar{e}=2}^{i-1} \left(\begin{array}{c}
a_{1,i+1-\bar{e}}^{j,i+2}\\
a_{2,i+1-\bar{e}}^{j,i+2} \\
a_{3,i+1-\bar{e}}^{j,i+2}
\end{array}\right)
\nonumber \\
\times
\left|\begin{array}{cc}
\bar{a}_{1,i+2}^{j,i+2} & \bar{a}_{1,k+2}^{j,i+2} \\
\bar{a}_{i+1-\bar{e},i+2}^{j,i+2} & \bar{a}_{i+1-\bar{e},k+2}^{j,i+2}
\end{array}\right|
+
\delta_{k,1}
\left(
\begin{array}{c} 0 \\
0 \\
\bar{a}_{1,i+2}^{j,i+2} \end{array}\right) \label{apA3} \ear

The next system of equations is generated by solving Eq.(\ref{apA3})
for the first variable
$\bar{a}_{i,k+2}^{j,i+2}
\bar{a}_{1,i+2}^{j,i+2}- \bar{a}_{i,i+2}^{j,i+2}
\bar{a}_{1,k+2}^{j,i+2}$. By using Cramer's rule we find,
\bear
\sum_{\bar{e}=1}^{i-1}
\left|\begin{array}{ccc}
a_{1,i+1-\bar{e}}^{j,i+2} & a_{1,i+1}^{j,i+2} & a_{1,i+2}^{j,i+2} \\
a_{2,i+1-\bar{e}}^{j,i+2} & a_{2,i+1}^{j,i+2} & a_{2,i+2}^{j,i+2} \\
a_{3,i+1-\bar{e}}^{j,i+2} & a_{3,i+1}^{j,i+2} & a_{3,i+2}^{j,i+2}
\end{array}\right|
\left|\begin{array}{cc}
\bar{a}_{1,i+2}^{j,i+2} & \bar{a}_{1,k+2}^{j,i+2} \\
\bar{a}_{i+1-\bar{e},i+2}^{j,i+2} & \bar{a}_{i+1-\bar{e},k+2}^{j,i+2}
\end{array}\right|
&= &
\delta_{k,1}
\bar{a}_{1,i+2}^{j,i+2} \left|\begin{array}{cc}
a_{1,i+1}^{j,i+2} & a_{1,i+2}^{j,i+2} \\
a_{2,i+1}^{j,i+2} & a_{2,i+2}^{j,i+2}
\end{array}\right|,
\nonumber \\
&&\mbox{for}~~k=1,\dots,i-1,
\label{apA4i}
\ear

Note that Eqs.(\ref{apA4i}) can be interpreted as a system of linear equations
where the variables are the
$ 3 \times 3$ determinants. In order to obtain our final identity we
first solve Eqs.(\ref{apA4i}) for the $3 \times 3$ determinant indexed by
$\bar{e}=1$.  Next we make the ratio of this solution with
Eq.(\ref{apA1i}) considering the replacements $i \rightarrow i+1$ and $\lambda \leftrightarrow
\mu$.  After some cumbersome simplifications we find the following identity,
\EQ
\frac{\left|\begin{array}{ccc}
a_{1,i}^{j,i+2} & a_{1,i+1}^{j,i+2} & a_{1,i+2}^{j,i+2} \\
a_{2,i}^{j,i+2} & a_{2,i+1}^{j,i+2} & a_{2,i+2}^{j,i+2} \\
a_{3,i}^{j,i+2} & a_{3,i+1}^{j,i+2} & a_{3,i+2}^{j,i+2}
\end{array}\right| a_{1,i+1}^{j,i+1}}{\left|\begin{array}{cc}
a_{1,i}^{j,i+1} & a_{1,i+1}^{j,i+1} \\
a_{2,i}^{j,i+1} & a_{2,i+1}^{j,i+1}
\end{array} \right|
\left|\begin{array}{cc}
a_{1,i+1}^{j,i+2} & a_{1,i+2}^{j,i+2} \\
a_{2,i+1}^{j,i+2} & a_{2,i+2}^{j,i+2}
\end{array}\right|}=
(-1)^i \frac{\left|\begin{array}{ccc}
\bar{a}_{1,2}^{j,i+1} & \dots & \bar{a}_{1,i+1}^{j,i+1} \\
\vdots & \ddots & \vdots \\
\bar{a}_{i,2}^{j,i+1} & \dots & \bar{a}_{i,i+1}^{j,i+1}
\end{array}\right|}{\left|\begin{array}{ccc}
\bar{a}_{1,3}^{j,i+1} & \dots & \bar{a}_{1,i+1}^{j,i+1} \\
\vdots & \ddots & \vdots \\
\bar{a}_{i-1,3}^{j,i+1} & \dots & \bar{a}_{i-1,i+1}^{j,i+1}
\end{array}\right|}
\frac{\left|\begin{array}{cccc}
\bar{a}_{1,i+2}^{j,i+2} & \bar{a}_{1,i+1}^{j,i+2} & \dots & \bar{a}_{1,4}^{j,i+2} \\
\bar{a}_{2,i+2}^{j,i+2} & \bar{a}_{2,i+1}^{j,i+2} & \dots & \bar{a}_{2,4}^{j,i+2} \\
\vdots & \vdots & \ddots & \vdots \\
\bar{a}_{i-1,i+2}^{j,i+2} & \bar{a}_{i-1,i+1}^{j,i+2} & \dots &
\bar{a}_{i-1,4}^{j,i+2}
\end{array}\right|}{\left|\begin{array}{cccc}
\bar{a}_{1,i+2}^{j,i+2} & \bar{a}_{1,i+1}^{j,i+2} & \dots & \bar{a}_{1,3}^{j,i+2} \\
\bar{a}_{2,i+2}^{j,i+2} & \bar{a}_{2,i+1}^{j,i+2} & \dots & \bar{a}_{2,3}^{j,i+2} \\
\vdots & \vdots & \ddots & \vdots \\
\bar{a}_{i,i+2}^{j,i+2} & \bar{a}_{i,i+1}^{j,i+2} & \dots &
\bar{a}_{i,3}^{j,i+2}
\end{array}\right|}
\label{apAid6}
\EN

We close this section by presenting the explicit expressions of
two other  identities  in terms of the $R$-matrix elements.  These
relations are going to work together with those coming from the
Yang-Baxter equation in order to simplify the eigenvalue problem.
In what follows we shall present them in a notation that will be useful
for next subsection. The first one follows from
Eq.(\ref{apAid6}) choosing $j=1$,
\EQ
\frac{D_3^{(i,0)}(\lambda,\mu)}{D_2^{(i,0)}(\lambda,\mu)}
=
\frac{D_4^{(i+1,2)}(\lambda,\mu)}{D_4^{(i+1,3)}(\lambda,\mu)}
\frac{D_4^{(i+2,4)}(\lambda,\mu)}{D_4^{(i+2,3)}(\lambda,\mu)},
\label{apAid8}
\EN
where the determinant
$D_4^{(i,b)}(\lambda,\mu)$ is given by,
\EQ
D_4^{(i,b)}(\lambda,\mu)= \left|\begin{array}{ccc}
R(\mu,\lambda)_{i,1}^{b,i+1-b} & \dots & R(\mu,\lambda)_{i,1}^{i,1} \\
\vdots & \ddots & \vdots \\
R(\mu,\lambda)_{b,i+1-b}^{b,i+1-b} & \dots &
R(\mu,\lambda)_{b,i+1-b}^{i,1}
\end{array}\right|~~\mbox{for}~i \le N
\label{D4}
\EN

The second one follows from
Eq.(\ref{apAid2i}) also with $j=1$ and by performing the
rapidity exchange $\lambda \leftrightarrow \mu$ we find,
\EQ
\frac{D_2^{(i+1,1)}(\lambda,\mu)}{D_2^{(i+1,0)}(\lambda,\mu)}=
-\frac{D_5^{(i+2,2)}(\lambda,\mu)}{D_4^{(i+2,3)}(\lambda,\mu)},
\label{apAid7}
\EN
such that the determinant
$D_5^{(i,2)}(\lambda,\mu)$ is,
\EQ
D_5^{(i+2,2)}(\lambda,\mu)=\left|\begin{array}{cccc}
R(\mu,\lambda)_{i+2,1}^{2,i+1} &
R(\mu,\lambda)_{i+2,1}^{4,i-1} & \dots &
R(\mu,\lambda)_{i+2,1}^{i+2,1} \\
\vdots & \vdots & \ddots & \vdots \\
R(\mu,\lambda)_{3,i}^{2,i+1} &
R(\mu,\lambda)_{3,i}^{4,i-1} & \dots &
R(\mu,\lambda)_{3,i}^{i+2,1}
\end{array}\right|~~\mbox{for}~i+2 \le N. \label{D5} \EN

\subsection{Yang-Baxter equation}
\label{subI2}

In practice to extract informations from the Yang-Baxter equation one has to project out the
$N^3 \times N^3$ matrix (\ref{ybr}) on the Weyl basis. Considering the projection on
row $[(a_1-1)N^2+(a_2-1)N+a_3]$ and column $[(c_1-1)N^2+(c_2-1)N+c_3]$ we find,
\EQ
R(\lambda_{1},\lambda_{2})_{a_1,a_2}^{b_1,b_2}
R(\lambda_{1},\lambda_{3})_{b_1,a_3}^{c_1,b_3}
R(\lambda_{2},\lambda_{3})_{b_2,b_3}^{c_2,c_3} =
R(\lambda_{2},\lambda_{3})_{a_2,a_3}^{b_2,b_3}
R(\lambda_{1},\lambda_{3})_{a_1,b_3}^{b_1,c_3}
R(\lambda_{1},\lambda_{2})_{b_1,b_2}^{c_1,c_2}, \label{ybr1i} \EN
where sum on the repeated indices $b_i$ is assumed.

In order to describe our approach  we shall denote  Eq.(\ref{ybr1i})  for a given
set of indices $a_i$ and $c_i$ by the symbol
$YB_{a_1,a_2,a_3}^{c_1,c_2,c_3}(\lambda_1,\lambda_2,\lambda_3)$.
In general, we are required to elaborated on Eq.(\ref{ybr1i}) to
obtain suitable weights identities that are directly useful in the eigenvalue problem.
In fact, there  exists only two exceptions that follow from Eq.(\ref{ybr1i}) without further manipulations.
These identities come from the entries $YB_{3,1,1}^{2,2,1}(\lambda_1,\lambda_2,\lambda)$ and
$YB_{N-1,3,1}^{N-1,2,2}(\lambda,\lambda_1,\lambda_2)$. After appropriate normalizations they are given by,
\EQ
\frac{R(\lambda_2,\lambda)_{1,1}^{1,1}}{R(\lambda_2,\lambda)_{2,1}^{2,1}}
\frac{R(\lambda_1,\lambda_2)_{3,1}^{2,2}}{R(\lambda_1,\lambda_2)_{3,1}^{3,1}}=
\frac{R(\lambda_2,\lambda)_{1,2}^{2,1}}{R(\lambda_2,\lambda)_{2,1}^{2,1}}
\frac{R(\lambda_1,\lambda)^{2,2}_{3,1}}{R(\lambda_1,\lambda)^{3,1}_{3,1}}
+
\frac{R(\lambda_1,\lambda_2)_{3,1}^{2,2}}{R(\lambda_1,\lambda_2)_{3,1}^{3,1}}
\frac{R(\lambda_1,\lambda)_{2,1}^{2,1}}{R(\lambda_1,\lambda)_{3,1}^{3,1}}
\label{apAid3i} \EN
and
\EQ
\frac{R(\lambda,\lambda_2)_{N,2}^{N,2}}{R(\lambda,\lambda_2)_{N,1}^{N,1}}
\frac{R(\lambda_1,\lambda_2)_{3,1}^{2,2}}{R(\lambda_1,\lambda_2)_{3,1}^{3,1}}
=
\frac{R(\lambda_1,\lambda_2)_{3,1}^{2,2}}{R(\lambda_1,\lambda_2)_{3,1}^{3,1}}
\frac{R(\lambda,\lambda_1)_{N,3}^{N,3}}{R(\lambda,\lambda_1)_{N,2}^{N,2}}
-
\frac{R(\lambda,\lambda_1)_{N-1,3}^{N,2}}{R(\lambda,\lambda_1)_{N,2}^{N,2}}
\frac{R(\lambda,\lambda_2)_{N,1}^{N-1,2}}{R(\lambda,\lambda_2)_{N,1}^{N,1}}.
\label{apAid4i} \EN

The main feature of Eqs.(\ref{apAid3i},\ref{apAid4i}) is that they factorized
two different types of weights products into a single product.
Identities of this sort are essential to carry out  the necessary simplifications on the transfer matrix
eigenvalue problem. This
task, however, will require more complex classes of products factorization which involve
at least three distinct terms.
In order to derive such complicated relations we have to make combinations
of a special set of equations which follows from Eq.(\ref{ybr1i}).
In table (\ref{tab7}) we have summarized two families of linear combinations that will lead to three terms factorization identities.
\begin{table}[ht]
\begin{center}
\begin{tabular}{|c|c|c|}
\hline
 Equation & Row index & Column index \\
\hline
 $YB_{3+k,a-k,1}^{2+j,a-j,2}(\lambda_1,\lambda,\lambda_2)$ & $0 \le k \le m \{a-1,N-3 \}$ & $ 0\le j \le a-1$ \\
\hline
 $YB_{3+k,a-k,1}^{4+j,a-1-j,1}(\lambda_1,\lambda,\lambda_2)$ & $0 \le k \le m \{a-1,N-3 \}$ & $j=-2$ \\
 & & $ 0\le j \le m \{a-2,N-4 \}$ \\
\hline
\end{tabular}
\end{center}
\caption{\footnotesize{Two families of linear combination used to derive factorization identities of three  weights products.}} \label{tab7}
\end{table}

Let us now explain how to use the linear combination
described in the first row of table (\ref{tab7}).
We start by first considering the situation when row index is $k=0$. In this case  we are left with equations
$YB_{3,a,1}^{2+j,a-j,2}(\lambda_1,\lambda,\lambda_2)$ ($k=0$) for $ 0\le j \le a-1$ which can be combined as follows,
\bear
&& \left( \begin{array}{ccc}
R(\lambda_1,\lambda)_{a+1, 1}^{2, a} & \dots & R(\lambda_1,\lambda)_{2, a}^{2, a} \\
\vdots & \ddots & \vdots \\
R(\lambda_1,\lambda)_{a+1, 1}^{a+1, 1} & \dots &
R(\lambda_1,\lambda)_{2, a}^{a+1, 1}
\end{array}\right)
\left(\begin{array}{c}
R(\lambda,\lambda_2)_{a, 1}^{1, a} R(\lambda_1,\lambda_2)_{3, a}^{a+1, 2} \\
\vdots \\
R(\lambda,\lambda_2)_{a, 1}^{a, 1} R(\lambda_1,\lambda_2)_{3, 1}^{2, 2} \\
\end{array}\right)
= \left(\begin{array}{c}
v_{2,1}^{0,0} \\
\vdots \\
v_{2,1}^{0,a-1}\end{array}\right)
\nonumber \\
&& ~~~~~~~~~~~~~~~~~~~~~~~~~~~~~~~~~~~~~~~~~~~~~ \mathrm{for}~~~  2
\le a \le N-2, \label{sl1i}
\ear

The right-hand side components of the nonhomogenous  system (\ref{sl1i}) are particular
cases of more general elements
$v_{2,1}^{k,j}$,
\EQ
v_{2,1}^{k,j} = \sum_{b_2=1}^{a-j+1}
R(\lambda_{1},\lambda)_{3+k,a-k}^{a+3-b_2,b_2}
R(\lambda_{1},\lambda_{2})_{a+3-b_2,1}^{2+j,a-j+2-b_2}
R(\lambda,\lambda_{2})_{b_2,a-j+2-b_2}^{a-j,2},~~~\mbox{for}~~k=0, \dots, a-1
\label{vk21i}
\EN

The main purpose  of the linear combination (\ref{sl1i})
is that it provides us the means to get rid of the right-hand side of
equation
$YB_{3,a,1}^{2+j,a-j,2}(\lambda_1,\lambda,\lambda_2)$. For the
remaining values $k=1, \dots, a-1$  we are therefore
asked to eliminate the following terms,
\EQ
\sum_{b_2=1}^{a-k}
R(\lambda,\lambda_{2})_{a-k,1}^{b_2,a-k+1-b_2}
R(\lambda_{1},\lambda_{2})_{3+k, a-k+1-b_2}^{a+2-b_2,2}
R(\lambda_{1},\lambda)_{a+2-b_2,b_2}^{2+j,a-j}. \label{var21}
\EN

In order to do that we first rewrite the sum (\ref{var21}) in an equivalent
matrix form, namely
\EQ \sum_{b_2=1}^{a-k}
R(\lambda,\lambda_{2})_{a-k,1}^{b_2,a-k+1-b_2}
R(\lambda_{1},\lambda_{2})_{3+k, a-k+1-b_2}^{a+2-b_2,2}
\left( \begin{array}{c} R(\lambda_{1},\lambda)_{a+2-b_2,b_2}^{2,a} \\
\vdots \\
R(\lambda_{1},\lambda)_{a+2-b_2,b_2}^{a+1,1}
\end{array}\right) ~~\mbox{for}~1 \le k \le a-1. \label{var21M}
\EN

We next note that the right-hand side column of Eq.(\ref{var21M})
is exactly the
$b_2$ column of the matrix defining the linear system (\ref{sl1i}).
If now we substitute the last column of the $a \times a$  matrix of Eq.(\ref{sl1i})
by the sum (\ref{var21M}) we will end up with a new matrix whose determinant
is null since we are computing the sum of $a-k$ determinants with two
equal columns. Therefore, to eliminate the terms (\ref{var21}) for $k=1,\dots, a-1$
we just have to replace the last column of the matrix defining the linear
system (\ref{sl1i}) by the left-hand side of equations
$YB_{3+k,a-k,1}^{2+j,a-j,2}(\lambda_1,\lambda,\lambda_2)$. This leads us
to the following condition,
\EQ
\left|
\begin{array}{cccc}
R(\lambda_1,\lambda)_{a+1, 1}^{2, a} & \dots & R(\lambda_1,\lambda)_{3, a-1}^{2, a} & v_{2,1}^{k,0} \\
R(\lambda_1,\lambda)_{a+1, 1}^{3, a-1} & \dots &
R(\lambda_1,\lambda)_{3, a-1}^{3, a-1} &
v_{2,1}^{k,1} \\
\vdots & \ddots & \vdots & \vdots \\
R(\lambda_1,\lambda)_{a+1, 1}^{a+1, 1} & \dots &
R(\lambda_1,\lambda)_{3, a-1}^{a+1, 1} & v_{2,1}^{k,a-1}
\end{array}\right|=0
~~~~~~~ \mbox{for} ~~~ 1 \le k \le a-1,
\label{sl1geral}
\EN

In order to have a single relation for all values of $k$ we proceed as follows.
For $k=0$
we solve the linear system (\ref{sl1i}) for the last component
$R(\lambda,\lambda_2)_{a, 1}^{a, 1} R(\lambda_1,\lambda_2)_{3, 1}^{2, 2}$ while for $ 1 \le k \le a-1$ we single out
the element
$v_{2,1}^{k,0}$ from Eq.(\ref{sl1geral}). After considering these steps we find,
\bear
\sum_{b_2=1}^{a+1}
R(\lambda_{1},\lambda)_{3+k,a-k}^{a+3-b_2,b_2}
R(\lambda_{1},\lambda_{2})_{a+3-b_2,1}^{2,a+2-b_2}
R(\lambda,\lambda_{2})_{b_2,a+2-b_2}^{a,2} = \delta_{k,0}~ S^0_{2,0}+S^k_{2,1},
\nonumber \\
\mbox{for} ~~k=0,\dots,a-1~~
\mbox{and}~~ 2 \le a \le N-2.
\label{sl1si}
\ear

The dependence of functions $S^0_{2,0}$ and $S^k_{2,1}$ on the $R$-matrix amplitudes are
\bear
S^0_{2,0} = (-1)^{a+1} R(\lambda,\lambda_2)_{a, 1}^{a, 1}
R(\lambda_1,\lambda_2)_{3, 1}^{2, 2}
\frac{D_4^{(a+1,2)}(\lambda,\lambda_1)}{D_4^{(a+1,3)}(\lambda,\lambda_1)}
\label{s020i}
\ear
and
\EQ
S^k_{2,1} =
(-1)^a \frac{ \left| \begin{array}{cccc}
R(\lambda_1,\lambda)_{a+1, 1}^{2, a} & \dots & R(\lambda_1,\lambda)_{3, a-1}^{2, a} & 0 \\
R(\lambda_1,\lambda)_{a+1, 1}^{3, a-1} & \dots &
R(\lambda_1,\lambda)_{3, a-1}^{3, a-1} &
v_{2,1}^{k,1} \\
\vdots & \ddots & \vdots & \vdots \\
R(\lambda_1,\lambda)_{a+1, 1}^{a+1, 1} & \dots &
R(\lambda_1,\lambda)_{3, a-1}^{a+1, 1} & v_{2,1}^{k,a-1}
\end{array}\right|}
{D_4^{(a+1,3)}(\lambda,\lambda_1)}, \label{sk21i} \EN such that the
determinant $D_4^{(i,b)}(\lambda,\lambda_1)$ is defined by
Eq.(\ref{D4}).

We now turn our attention to the linear combination associated to the second row of table (\ref{tab7}).
As before we have to make a distinction  between the cases  $k=0$ and that involving the
remaining values $k=1, \dots, a-1$. In the former situation we have to combine equations
$YB_{3,a,1}^{4+j,a-1-j,1}(\lambda_1,\lambda,\lambda_2)$ ($k=0$) with $j=-2$ and $ 0\le j \le m \{a-2,N-4 \}$
which can be arranged as,
\bear
&& \left( \begin{array}{ccc}
R(\lambda_1,\lambda)_{a+2, 1}^{2, a+1} & \dots & R(\lambda_1,\lambda)_{3, a}^{2, a+1} \\
R(\lambda_1,\lambda)_{a+2, 1}^{4, a-1} & \dots & R(\lambda_1,\lambda)_{3, a}^{4, a-1} \\
\vdots & \ddots & \vdots \\
R(\lambda_1,\lambda)_{a+2, 1}^{a+2, 1} & \dots &
R(\lambda_1,\lambda)_{3, a}^{a+2, 1}
\end{array}\right)
\left(\begin{array}{c}
R(\lambda,\lambda_2)_{a, 1}^{1, a} R(\lambda_1,\lambda_2)_{3, a}^{a+2, 1} \\
R(\lambda,\lambda_2)_{a, 1}^{2, a-1} R(\lambda_1,\lambda_2)_{3, a-1}^{a+1, 1} \\
\vdots \\
R(\lambda,\lambda_2)_{a, 1}^{a, 1} R(\lambda_1,\lambda_2)_{3, 1}^{3, 1} \\
\end{array}\right)
= \left(\begin{array}{c}
v_{1,1}^{0,-2} \\
v_{1,1}^{0,0} \\
\vdots \\
v_{1,1}^{0,a-2}\end{array}\right)
\nonumber \\
&& ~~~~~~~~~~~~~~~~~~~~~~~~~~~~~~~~~~~~~~~~~~~~~ \mathrm{for}~~~  2
\le a \le N-2, \label{sl2i}
\ear
where the components $v_{1,1}^{0,j}$ are obtained from the following more general
elements,
\EQ
v_{1,1}^{k,j} = \sum_{b_2=1}^{a-j-1}
R(\lambda_{1},\lambda)_{3+k,a-k}^{a+3-b_2,b_2}
R(\lambda_{1},\lambda_{2})_{a+3-b_2,1}^{4+j,a-j-b_2}
R(\lambda,\lambda_{2})_{b_2,a-j-b_2}^{a-j-1,1}~~~\mbox{for}~~k=0,\dots,a-1.
\label{vk11i}
\EN

By the same token, the corresponding algebraic manipulations
for the values $k=1,\dots,a-1$ consist in the substitution of the last
column of the matrix (\ref{sl2i}) by left-hand side of equations
$YB_{3+k,a-k,1}^{4+j,a-1-j,1}(\lambda_1,\lambda,\lambda_2)$ with
$j=-2$ and $0 \le j \le a-2$. This leads us to the condition,
\EQ
\left| \begin{array}{cccc}
R(\lambda_1,\lambda)_{a+2, 1}^{2, a+1} & \dots & R(\lambda_1,\lambda)_{4, a-1}^{2, a+1} & v_{1,1}^{k,-2} \\
R(\lambda_1,\lambda)_{a+2, 1}^{4, a-1} & \dots & R(\lambda_1,\lambda)_{4, a-1}^{4, a-1} & v_{1,1}^{k,0} \\
\vdots & \ddots & \vdots & \vdots \\
R(\lambda_1,\lambda)_{a+2, 1}^{a+2, 1} & \dots &
R(\lambda_1,\lambda)_{4, a-1}^{a+2, 1} & v_{1,1}^{k,a-2}
\end{array}\right|
= 0
~~~~~~~~ \mbox{for} ~~~ 1 \le k \le a-1,
\label{sl2geral}
\EN

Now by solving Eq.(\ref{sl2i}) for the component
$R(\lambda,\lambda_2)_{a, 1}^{a, 1} R(\lambda_1,\lambda_2)_{3, 1}^{3, 1}$ and by calculating
$v_{1,1}^{k,0}$ from Eq.(\ref{sl2geral}) we find,
\bear
\sum_{b_2=1}^{a+1}
R(\lambda_{1},\lambda)_{3+k,a-k}^{a+3-b_2,b_2}
R(\lambda_{1},\lambda_{2})_{a+3-b_2,1}^{2,a+2-b_2}
R(\lambda,\lambda_{2})_{b_2,a+2-b_2}^{a+1,1}
=
\delta_{k,0}~ S^0_{1,0}+S^k_{1,1},
\nonumber \\
\mathrm{for}~~k=0, \dots, a-1~~
\mbox{and}~~  2 \le a \le N-2,
\label{sl2si}
\ear
where functions $S^0_{1,0}$ and $S^k_{1,1}$ are given by
\EQ
S^0_{1,0} = (-1)^{a+1} R(\lambda,\lambda_2)_{a, 1}^{a, 1}
R(\lambda_1,\lambda_2)_{3, 1}^{3, 1}
\frac{D_5^{(a+2,2)}(\lambda,\lambda_1)}{D_4^{(a+2,4)}(\lambda,\lambda_1)}
\label{s010i} \EN
and
\EQ
S^k_{1,1}
= (-1)^a \frac{ \left| \begin{array}{cccc}
R(\lambda_1,\lambda)_{a+2, 1}^{2, a+1} & \dots & R(\lambda_1,\lambda)_{4, a-1}^{2, a+1} & 0 \\
R(\lambda_1,\lambda)_{a+2, 1}^{4, a-1} & \dots & R(\lambda_1,\lambda)_{4, a-1}^{4, a-1} & v_{1,1}^{k,0} \\
\vdots & \ddots & \vdots & \vdots \\
R(\lambda_1,\lambda)_{a+2, 1}^{a+2, 1} & \dots &
R(\lambda_1,\lambda)_{4, a-1}^{a+2, 1} & v_{1,1}^{k,a-2}
\end{array}\right|}
{D_4^{(a+2,4)}(\lambda,\lambda_1)}, \label{sk11i} \EN such that the
determinant $D_5^{(i+2,2)}(\lambda,\lambda_1)$ is defined by
Eq.(\ref{D5}).

At this point we have at hand the basic ingredients that are necessary to obtain useful identities for the eigenvalue problem.
We just have to construct the total number of `$a$' $2 \times 2$ linear systems of equations by considering the expressions (\ref{sl1si},\ref{sl2si}). These $a$ systems are composed by Eqs.(\ref{sl1si},\ref{sl2si}) for $0 \le k \le a-1$, namely
\bear
\left( \begin{array}{cc}
R(\lambda,\lambda_2)_{a+1, 1}^{a, 2} & R(\lambda,\lambda_2)_{a, 2}^{a, 2} \\
R(\lambda,\lambda_2)_{a+1, 1}^{a+1, 1} & R(\lambda,\lambda_2)_{a,
2}^{a+1, 1}
\end{array}\right)
\left( \begin{array}{c}
R(\lambda_1,\lambda_2)_{2, 1}^{2, 1} R(\lambda_1,\lambda)_{3+k, a-k}^{2, a+1} \\
R(\lambda_1,\lambda_2)_{3, 1}^{2, 2} R(\lambda_1,\lambda)_{3+k, a-k}^{3, a}
\end{array}\right)
+
\sum_{b_2=1}^{a-1} R(\lambda_1,\lambda_2)_{a+3-b_2, 1}^{2, a+2-b_2}
\nonumber \\
\times  \left( \begin{array}{c}
R(\lambda,\lambda_2)_{b_2, a+2-b_2}^{a, 2} \\
R(\lambda,\lambda_2)_{b_2, a+2-b_2}^{a+1, 1}
\end{array}\right)
%\nonumber \\
%&\times&
R(\lambda_1,\lambda)_{3+k, a-k}^{a+3-b_2, b_2} =\left(
\begin{array}{c}
S^k_{2,1} \\
S^k_{1,1}
\end{array}\right)
+ \delta_{k,0} \left( \begin{array}{c}
S^0_{2,0} \\
S^0_{1,0}
\end{array}\right),
\nonumber \\
  \mbox{for} ~~0
\le k \le a-1 ~~\mbox{and}~~ 2 \le a \le N-2. \label{sl5i}
\ear

To obtain the desirable identity we proceed as follows.
We first compute the products \newline $R(\lambda_1,\lambda_2)_{3, 1}^{2, 2} R(\lambda_1,\lambda)_{3+k, a-k}^{3, a}$
by applying Cramer's rule in the system (\ref{sl5i}).
This will lead us to the following relations
\bear
 \sum_{b_2=1}^{a} \left|
\begin{array}{cc}
R(\lambda,\lambda_2)_{a+1, 1}^{a, 2} & R(\lambda,\lambda_2)_{b_2, a+2-b_2}^{a, 2} \\
R(\lambda,\lambda_2)_{a+1, 1}^{a+1, 1} & R(\lambda,\lambda_2)_{b_2,
a+2-b_2}^{a+1, 1}
\end{array}\right|
R(\lambda_1,\lambda_2)_{a+3-b_2, 1}^{2, a+2-b_2}
R(\lambda_1,\lambda)_{3+k, a-k}^{a+3-b_2, b_2}
\nonumber \\
= \left|
\begin{array}{cc}
R(\lambda,\lambda_2)_{a+1, 1}^{a, 2} & S^k_{2,1} \\
R(\lambda,\lambda_2)_{a+1, 1}^{a+1, 1} & S^k_{1,1}
\end{array}\right| + \delta_{k,0} \left|
\begin{array}{cc}
R(\lambda,\lambda_2)_{a+1, 1}^{a, 2} & S^0_{2,0} \\
R(\lambda,\lambda_2)_{a+1, 1}^{a+1, 1} & S^0_{1,0}
\end{array}\right|,
\nonumber \\
 ~~ \mbox{for} ~~0 \le k \le a-1~~\mbox{and}~~ 2
\le a \le N-2. \label{sl5si}
\ear

The next step is to solve the linear system (\ref{sl5si}) for the variable $b_2=a$ which allows
us to eliminate function $D_2^{(a,0)}(\lambda,\lambda_2)
R(\lambda_1,\lambda_2)_{3, 1}^{2, 2}$. By  normalizing the result by
$D_4^{(a+2,3)}(\lambda,\lambda_1) R(\lambda,\lambda_2)_{a,1}^{a,1}$ \newline $R(\lambda,\lambda_2)_{a+1,1}^{a+1,1}
R(\lambda_1,\lambda_2)_{3,1}^{3,1}$ and after few manipulations we find that,
\bear
D_2^{(a,0)}(\lambda,\lambda_2)
\frac{R(\lambda_1,\lambda_2)_{3,1}^{2,2}}{R(\lambda_1,\lambda_2)_{3,1}^{3,1}}
&=&
-\frac{R(\lambda,\lambda_2)_{a+1,1}^{a,2}}{R(\lambda,\lambda_2)_{a+1,1}^{a+1,1}}
\frac{D_{5}^{(a+2,2)}(\lambda,\lambda_1)}{D_{4}^{(a+2,3)}(\lambda,\lambda_1)}
+\frac{D_{5}^{(a+1,2)}(\lambda,\lambda_1)}{D_{4}^{(a+1,3)}(\lambda,\lambda_1)}
\frac{R(\lambda,\lambda_2)_{a,1}^{a-1,2}}{R(\lambda,\lambda_2)_{a,1}^{a,1}}
\nonumber \\
&+&
\frac{R(\lambda_1,\lambda_2)_{3,1}^{2,2}}{R(\lambda_1,\lambda_2)_{3,1}^{3,1}}
\frac{D_4^{(a+1,2)}(\lambda,\lambda_1)}{D_4^{(a+1,3)}(\lambda,\lambda_1)}
\frac{D_4^{(a+2,4)}(\lambda,\lambda_1)}{D_4^{(a+2,3)}(\lambda,\lambda_1)}
~~~\mbox{for}~~2 \le a \le N-2.
\nonumber \\
\label{apAid9i}
\ear

We finally discuss the special situation when $a=N-1$. It turns
out that the steps in this case are fairly parallel to those
already described for $2 \le a \le N-2$. The respective identity
is obtained from
Eq.(\ref{apAid9i}) by choosing $a=N-1$ except
for the terms involving the determinants
$D_4^{(N+1,b)}(\lambda,\lambda_1)$ and
$D_5^{(N+1,2)}(\lambda,\lambda_1)$. These determinants are then
replaced by the following analytic continuations,
\EQ
D_4^{(N+1,b)}(\lambda,\lambda_1)
\equiv \lim_{a\rightarrow N-1}
\frac{D_4^{(a+2,b)}(\lambda,\lambda_1)}{R(\lambda_1,\lambda)_{a+2,1}^{a+2,1}}=
(-1)^{N+1-b} \left|\begin{array}{ccc}
R(\lambda_1,\lambda)_{N,2}^{b,N+2-b} & \dots & R(\lambda_1,\lambda)_{N,2}^{N,2} \\
\vdots & \ddots & \vdots \\
R(\lambda_1,\lambda)_{b,N+2-b}^{b,N+2-b} & \dots &
R(\lambda_1,\lambda)_{b,N+2-b}^{N,2}
\end{array}\right|
\label{D4CAi}
\EN
and
\bear
D_5^{(N+1,2)}(\lambda,\lambda_1)
&\equiv &
\lim_{a\rightarrow N-1}
\frac{D_5^{(a+2,2)}(\lambda,\lambda_1)}{R(\lambda_1,\lambda)_{a+2,1}^{a+2,1}}
\nonumber \\
&=&
(-1)^{N}
\left|\begin{array}{cccc} R(\lambda_1,\lambda)_{N,2}^{2,N} &
R(\lambda_1,\lambda)_{N,2}^{4,N-2} & \dots &
R(\lambda_1,\lambda)_{N,2}^{N,2} \\
\vdots & \vdots & \ddots & \vdots \\
R(\lambda_1,\lambda)_{3,N-1}^{2,N} &
R(\lambda_1,\lambda)_{3,N-1}^{4,N-2} & \dots &
R(\lambda_1,\lambda)_{3,N-1}^{N,2}
\end{array}\right|.
\label{D5CAi} \ear

Though the element
$R(\lambda_1,\lambda)_{N+1,1}^{N+1,1}$ in Eqs.(\ref{D4CAi},\ref{D5CAi}) are not defined we observe that
in the identity (\ref{apAid9i}) such determinants always appear as ratios
$\frac{D_4^{(N+1,4)}(\lambda,\lambda_1)}{D_4^{(N+1,3)}(\lambda,\lambda_1)}$ or
$\frac{D_5^{(N+1,2)}(\lambda,\lambda_1)}{D_4^{(N+1,3)}(\lambda,\lambda_1)}$. This means that the common element
$R(\lambda_1,\lambda)_{N+1,1}^{N+1,1}$ will be canceled out in such ratios.
We end by mentioning
that the very particular case
$a=2$ e $N=3$ is obtained by setting $D_4^{(4,4)}(\lambda,\lambda_1)=1$.

\section{The eigenvalue problem}
\label{sec4}

In this section we shall consider the diagonalization of the transfer matrix (\ref{tran}) related to the inhomogeneous $U(1)$ vertex model defined in section \ref{sec2}.
The $U(1)$ invariance of the respective statistical weights implies that the transfer matrix (\ref{tran}) commutes with the total spin operator,
\EQ
\label{comutSz}
[T(\lambda), \sum_{i=1}^{L} S_{i}^{z}]=0.
\EN

A direct consequence of Eq.(\ref{comutSz}) is that the Hilbert space can be
separated in disjoint sectors labeled by the eigenvalues of the spin operator, namely
\EQ
\sum_{i=1}^{L} S_{i}^{z} \ket{\Phi_n}=[Ls-n] \ket{\Phi_n},
\EN
where $\ket{\Phi_n}$ denotes the eigenvector on the $n$-th sector.

In terms of the monodromy matrix elements the transfer matrix eigenvalue problem can then be defined by,
\EQ
\sum_{a=1}^{N} {\cal T}_{a, a}(\lambda) \ket{\Phi_n}=\Lambda_n(\lambda) \ket{\Phi_n}.
\label{tran1}
\EN

Here we are going to solve the eigenvalue problem (\ref{tran1}) by
solely relying on the three different families of commutation rules of section \ref{sec3}, the identities
between the $R$-matrix elements of section {\ref{secI}  as well as on the pseudovacuum state properties (\ref{mono1}).
To this end we will seek the eigenvectors $\ket{\Phi_n}$ in the form of multiparticle states which can formally be represented by the following expression,
\EQ
\ket{\Phi_n}=\phi_n(\lambda_1,\dots,\lambda_n) \ket{0},
\EN
where the rapidities $\lambda_1,\dots,\lambda_n$ parameterize the momenta of the particles.

For consistency the vector with null number of particles $\phi_0 \equiv 1$
is directly identified to the reference state $\ket{0}$.
The property (\ref{pro1}) concerning the spin of the monodromy matrix elements suggests us
that the mathematical structure of the vectors $\phi_n(\lambda_1,\dots,\lambda_n)$  for general $n$ can be searched with the help of linear combination of the product of the creation operators.
In what follows we are going to argue that such vectors are indeed given in terms of a recurrence relation 
involving the $N-1$ creation fields ${\cal T}_{1, b}(\lambda)$.

\subsection{The one-particle problem}
\label{sub41}

The one-particle state $n=1$ corresponds to an excitation of spin $s-1$ over the reference state $\ket{0}$.
From property (\ref{pro1}) we see that among the basis vectors ${\cal T}_{1,b}(\lambda)$ the operator responsible for this excitation is ${\cal T}_{1,2}(\lambda)$.
The one-particle vector $\phi_1(\lambda_1)$ is then given by,
\EQ
 \phi_1(\lambda_1)= {\cal T}_{1,2}(\lambda_{1})  .
\label{ans1}
\EN

The solution of the eigenvalue problem (\ref{tran1}) for the one-particle state
is achieved by using
the commutation rules between  the operators
${\cal T}_{a,a}(\lambda)$ and
${\cal T}_{1,2}(\lambda)$.
Taking into account the form of Eqs.(\ref{1part1}-\ref{1partN}) as well as the pseudovacuum
properties (\ref{mono1}) we find,
\bear
{\cal T}_{1,1}(\lambda) \ket{\Phi_1} &=&
w_{1}(\lambda)
\frac{R(\lambda_1,\lambda)_{1,1}^{1,1}}
{R(\lambda_1,\lambda)_{2,1}^{2,1}}
\ket{\Phi_1}
-w_{1}(\lambda_1) \frac{R(\lambda_1,\lambda)_{1,2}^{2,1}}{R(\lambda_1,\lambda)_{2, 1}^{2, 1}}
{\cal T}_{1,2}(\lambda) \ket{0},
\label{1partvac1}
\\
{\cal T}_{a,a}(\lambda) \ket{\Phi_1} &=&
w_{a}(\lambda) D_2^{(a,0)}(\lambda,\lambda_1)
\ket{\Phi_1}
+ w_{1}(\lambda_1) \frac{R(\lambda,\lambda_1)_{a+1, 1}^{a, 2}}{R(\lambda,\lambda_1)_{a+1, 1}^{ a+1, 1}}
{\cal T}_{a,a+1}(\lambda) \ket{0}
\nonumber \\
&-& w_{2}(\lambda_1) \frac{R(\lambda,\lambda_1)_{a, 1}^{a-1, 2}}{R(\lambda,\lambda_1)_{a,1}^{a,1}}
{\cal T}_{a-1,a}(\lambda) \ket{0},~~~\mathrm{for}~~~2 \le a \le N-1,
\label{1partvaca}
\\
{\cal T}_{N,N}(\lambda)
 \ket{\Phi_1} &=&
w_{N}(\lambda)
\frac{R(\lambda,\lambda_1)_{N,2}^{N,2}}{R(\lambda,\lambda_1)_{N,1}^{N,1}}
\ket{\Phi_1}
- w_{2}(\lambda_1) \frac{R(\lambda,\lambda_1)_{N, 1}^{N-1, 2}}{R(\lambda,\lambda_1)_{N,1}^{N,1}}
{\cal T}_{N-1,N}(\lambda) \ket{0}.
\label{1partvacN}
\ear

It is possible to  gather
Eqs.(\ref{1partvac1}-\ref{1partvacN})  into a single expression for all values of
the diagonal index $a$. This requires first the definition of a discrete function that
is able to project out a set of indices. Keeping in mind the extension to multi-particle
states we shall define this function by,
\EQ
\bar{\delta}_{i}^{j_1,\dots,j_n}=1 -\sum_{k=1}^{n}\delta_{i,j_k}.
\label{disc}
\EN

By using identity (\ref{apAid1}) to reorder the rapidities of the last term in Eq.(\ref{1partvac1})
we find that
Eqs.(\ref{1partvac1}-\ref{1partvacN}) can be recasted as,
\bear
{\cal T}_{a,a}(\lambda) \ket{\Phi_1} &=& w_{a}(\lambda)
P_a(\lambda,\lambda_1) \ket{\Phi_1}
- \bar{\delta}_{a}^{N} {\cal
T}_{a,a+1}(\lambda) w_{1}(\lambda_1) {_1}{\cal
F}_{1}^{(a)}(\lambda,\lambda_1) \ket{0}
\nonumber \\
&-& \bar{\delta}_{a}^{1}
{\cal T}_{a-1,a}(\lambda) w_{2}(\lambda_1) {_0}{\cal F}_{1}^{(a-1)}(\lambda,\lambda_1) \ket{0},~~~\mathrm{for}~~~1 \le a \le N,
\label{1partsimp}
\ear

The polynomials $P_a(\lambda,\mu)$ are the terms proportional to the eigenstate
$\ket{\Phi_1}$ in
Eqs.(\ref{1partvac1}-\ref{1partvacN}), namely
\EQ
P_{a}(\lambda,\mu) = \begin{cases}
\displaystyle
\frac{R(\mu,\lambda)_{1,1}^{1,1}}{R(\mu,\lambda)_{2,1}^{2,1}},~~~~\mbox{for} ~~ a=1 \cr
\displaystyle
D_2^{(a,0)}(\lambda,\mu),
~~~~ \mbox{for} ~~ 2 \le a \le N-1 \cr
\displaystyle
\frac{R(\lambda,\mu)_{N,2}^{N,2}}{R(\lambda,\mu)_{N,1}^{N,1}}, ~~~~\mbox{for} ~~ a=N.
\label{pa}
\end{cases}
\EN

The auxiliary functions ${_0}{\cal F}_{1}^{(a)}(\lambda,\mu)$ and ${_1} {\cal F}_{1}^{(a)}(\lambda,\mu)$
are proportional to the terms not parallel to the one-particle state. This notation for these off-shell
amplitudes have been
defined for later convenience. They are given by,
\EQ
\label{psi1}
{_0}{\cal F}_{1}^{(a)}(\lambda,\mu) =
-{_1}{\cal F}_{1}^{(a)}(\lambda,\mu) =
\frac{R(\lambda,\mu)_{a+1, 1}^{a, 2}}{R(\lambda,\mu)_{a+1, 1}^{ a+1, 1}}
\EN

The solution of the eigenvalue problem is obtained by summing
Eq.(\ref{1partsimp}) over the index $a$.  This gives us the action of the
transfer matrix on the one-particle state,
\EQ
T(\lambda) \ket{\Phi_1} = \sum_{a=1}^{N} w_{a}(\lambda)
P_{a}(\lambda,\lambda_1)\ket{\Phi_1} +[w_{1}(\lambda_1) -
w_{2}(\lambda_1)] \sum_{a=1}^{N-1} {_0}{\cal
F}_{1}^{(a)}(\lambda,\lambda_1) {\cal T}_{a,a+1}(\lambda) \ket{0}.
\EN

We see that the terms
${\cal T}_{a,a+1}(\lambda) \ket{0}$ for $a=1,\cdots,N$ are undesirable states because
they are not proportional to the one-particle eigenstate
$\ket{\Phi_1}$. Note that these states are built up from the many possible creation fields
possessing azimuthal spin $s=1$. All of them are eliminated by imposing
that the rapidity $\lambda_1$ satisfies the one-particle Bethe equation,
\EQ
\frac{w_{1}(\lambda_1)}{w_{2}(\lambda_1)}=1.
\label{BEA1}
\EN
which implies that the corresponding one-particle eigenvalue is,
\EQ
\Lambda_1(\lambda) =
\sum_{a=1}^{N} w_{a}(\lambda) P_{a}(\lambda,\lambda_1),
\label{gama1}
\EN

We  would like to close this section by commenting on the meaning of the
notation we have introduced for the off-shell function ${_c}{\cal F}_{b-a}^{(a)}(\lambda,\mu)$.
The two right-hand side indices $a$ and $b-a$ are directly related to the companion
unwanted operator ${\cal T}_{a,b}(\lambda)$, making explicit its azimuthal spin value $s_{a,b}=b-a$.
The remaining index $c$ accounts for the number of weight $w_1(\lambda_i)$
that is present in the respective undesirable term of form
$\displaystyle{\prod_{i=1}^{c} w_1(\lambda_i) {\cal T}_{a,b}(\lambda) \ket{0}}$. We shall see that
this is indeed a suitable notation
to accommodate similar results for general multiparticle states.

\subsection{The two-particle problem}
\label{sub42}

The two-particle state lies on the $n=2$ sector and is constructed by considering the possible excitations with spin $s-2$.
From Eq.(\ref{pro1}) we conclude that all products of form ${\cal T}_{1, k_{1}}(\lambda_{m_1})
{\cal T}_{1, k_{2}}(\lambda_{m_2})$ whose indices is constrained by $k_1+k_2=4$ can in principle contribute to such state.
Therefore, the most general ansatz for vector $\phi_{2}(\lambda_1,\lambda_2)$ one could start with is,
\EQ
\phi_{2}(\lambda_1,\lambda_2)=\sum_{\stackrel{ m_1 ,m_2 =1}{m_1 \ne m_2}}^{2}
\sum_{\stackrel{
k_{1},k_{2} =1}{k_1+k_2=4}}^{N} \bar{c}_{k_1,k_2}^{(m_1,m_2)}(\lambda_{m_1},\lambda_{m_2})
{\cal T}_{1, k_{1}}(\lambda_{m_1})
{\cal T}_{1, k_{2}}(\lambda_{m_2}),
\label{ans20}
\EN
where $\bar{c}_{k_1,k_2}^{(m_1,m_2)}(\lambda_{m_1},\lambda_{m_2})$ are the coefficients of an arbitrary linear combination.

However, the commutation rules of the operators ${\cal T}_{1,1}(\lambda)$ and ${\cal T}_{1,3}(\mu)$ as well as between
${\cal T}_{1,2}(\lambda)$ and ${\cal T}_{1,2}(\mu)$ shows us that not all the terms entering in Eq.(\ref{ans20}) are linearly independent.
In fact, by using the commutation relations (\ref{2part1},\ref{b12ba-1a}) we note that the products  ${\cal T}_{1,1}(\lambda_1) {\cal T}_{1,3}(\lambda_2)$, ${\cal T}_{1,1}(\lambda_2) {\cal T}_{1,3}(\lambda_1)$ and ${\cal T}_{1,2}(\lambda_2) {\cal T}_{1,2}(\lambda_1)$ can be eliminated from the linear combination (\ref{ans20}).
Taking this observation into account we conclude that an educated ansatz for the two-particle vector is,
\EQ
\phi_{2}(\lambda_1,\lambda_2) = {\cal T}_{1, 2}(\lambda_{1}) {\cal
T}_{1, 2}(\lambda_{2}) +g^{(1)}_{2}(\lambda_1,\lambda_2) {\cal T}_{1, 3}(\lambda_{1}) {\cal
T}_{1, 1}(\lambda_{2}) + g^{(2)}_{2}(\lambda_1,\lambda_2) {\cal T}_{1, 3}(\lambda_{2})
{\cal T}_{1, 1}(\lambda_{1}),
\label{ans2} \EN
where $g^{(i)}_{2}(\lambda_1,\lambda_2)$ are arbitrary functions to be determined.

The next task is to investigate under which conditions the
ansatz (\ref{ans2}) is able to provide us the two-particle transfer matrix eigenstates.
In other to do that we have to commute the matrix elements ${\cal T}_{a,a}(\lambda)$ with
the creation operators present in the two-particle ansatz
$\phi_{2}(\lambda_1,\lambda_2)$.
Let us start by describing the steps necessary to disentangle the action of
${\cal T}_{a,a}(\lambda)$ on the first  term of the two-particle ansatz (\ref{ans2}).
In this case we have to use all the three classes of commutation rules presented in section 3.
The first step is to move the diagonal operators
through the creation fields ${\cal T}_{1, 2}(\lambda_{1})$ and ${\cal T}_{1, 2}(\lambda_{2})$
with the help of relations (\ref{1part1}-\ref{1partN}). This generates products between creation
fields
${\cal T}_{1,2}(\lambda)$ and
${\cal T}_{a-1,a}(\mu)$ that are not ordered in the unique
desirable form  ${\cal T}_{a-1,a}(\lambda) {\cal T}_{1,2}(\lambda_1)$ or
${\cal T}_{a-1,a}(\lambda) {\cal T}_{1,2}(\lambda_2)$ for $a=2, \dots,N$.
This common rapidity ordering are then  established by using the
commutation rules (\ref{eq34},\ref{eq37}) among the fields
${\cal T}_{1,2}(\lambda_1)$ and ${\cal T}_{a-1,a}(\lambda)$.
The final  step concerns with the presence of annihilation fields
on the middle of products involving three monodromy matrix elements that are needed
to be carried out to the further right in order to act on the  reference state $\ket{0}$.
This is sorted out by employing the commutation rules (\ref{aniq0}) between
the annihilation fields with the creation operator ${\cal T}_{1,2}(\lambda_2)$.
The action of
${\cal T}_{a,a}(\lambda)$ on the second  and third terms of the two-particle ansatz (\ref{ans2})
is however much simpler. In this case we should not worry about the right-hand side diagonal
field ${\cal T}_{1,1}(\lambda_i)$ since its action on the reference state $\ket{0}$ is known to be given
by Eq.(\ref{mono1}).  Therefore,
the respective computation  involves
only the set of commutation rules among the operators
${\cal T}_{a,a}(\lambda)$ and ${\cal T}_{1, 3}(\lambda_i)$ for $i=1,2$.
This task is easily accomplished with the help of Eqs.(\ref{2part1}-\ref{2partN}).

A direct consequence of the above described commutations is the generation of
several types of terms that are not proportional to none of the terms present
in the two-particle ansatz (\ref{ans2}). Among many different types of such
undesirable terms there exists a very special family  that need to be single out
first. This family of unwanted terms are
always produced by
the same diagonal operator ${\cal T}_{a,a}(\lambda)$ for $a=2, \dots, N$ and carries
a dependence either on the couplings $g_2^{(i)}(\lambda_1,\lambda_2)$ or on the
specific amplitude ratio
$R_{3,1}^{2,2}(\lambda_1,\lambda_2)/R_{3,1}^{3,1}(\lambda_1,\lambda_2)$.  Therefore, the
elimination
of such unwanted terms are only possible after an appropriate choice of the
coefficients $g_2^{(i)}(\lambda_1,\lambda_2)$.  From now on we shall refer to the class
of undesirable terms that are responsible for fixing the linear
combination as easy unwanted terms.
In the two-particle problem there are
four different terms of this sort and  their structures are,
\bear
&&
\bullet~
w_{1}(\lambda_2)
\frac{R(\lambda,\lambda_1)_{a,1}^{a-1,2}}{R(\lambda,\lambda_1)_{a,1}^{a,1}}
\left[
\frac{R(\lambda_1,\lambda_2)_{3,1}^{2,2}}{R(\lambda_1,\lambda_2)_{3,1}^{3,1}}
+g_{2}^{(1)}(\lambda_1,\lambda_2)
\right]
{\cal T}_{a-1,a}(\lambda) {\cal T}_{2,3}(\lambda_1)
 \nonumber \\
&&
\bullet~
w_{1}(\lambda_2) w_{3}(\lambda_1)
\frac{R(\lambda,\lambda_1)_{a,1}^{a-2,3}}{R(\lambda,\lambda_1)_{a,1}^{a,1}}
\left[
\frac{R(\lambda_1,\lambda_2)_{3,1}^{2,2}}{R(\lambda_1,\lambda_2)_{3,1}^{3,1}}
+g_{2}^{(1)}(\lambda_1,\lambda_2)
\right]
{\cal T}_{a-2,a}(\lambda)
\nonumber \\
&&
\bullet~
w_{1}(\lambda_1)
\frac{R(\lambda,\lambda_2)_{a,1}^{a-1,2}}{R(\lambda,\lambda_2)_{a,1}^{a,1}}
g_{2}^{(2)}(\lambda_1,\lambda_2)
{\cal T}_{a-1,a}(\lambda) {\cal T}_{2,3}(\lambda_2)
\nonumber \\
&&
\bullet~
w_{1}(\lambda_1) w_{3}(\lambda_2)
\frac{R(\lambda,\lambda_2)_{a,1}^{a-2,3}}{R(\lambda,\lambda_2)_{a,1}^{a,1}}
g_{2}^{(2)}(\lambda_1,\lambda_2)
{\cal T}_{a-2,a}(\lambda)
\label{uwan2}
\ear

From Eq.(\ref{uwan2}) we conclude that all
the above easy unwanted terms are canceled out provided we choose functions $g^{(i)}_{2}(\lambda_1,\lambda_2)$ as
\EQ
g^{(1)}(\lambda_1,\lambda_2) = -\frac{R(\lambda_1,\lambda_2)_{3,1}^{2,2}}{R(\lambda_1,\lambda_2)_{3,1}^{3,1}},
~~~
g^{(2)}(\lambda_1,\lambda_2) = 0.
\label{constra}
\EN

Now, by comparing Eq.(\ref{psi1}) with Eq.(\ref{constra}) we see that
the coupling $g^{(1)}(\lambda_1,\lambda_2)$  is exactly  the amplitude
${_1}{\cal F}_{1}^{(2)}(\lambda_1,\lambda_2)$ introduced in the one-particle
problem. This means that the two-particle vector can be written as,
\EQ
\phi_{2}(\lambda_1,\lambda_2)= {\cal T}_{1,2}(\lambda_1) {\cal T}_{1, 2}(\lambda_2)
+ {_1}{\cal F}_{1}^{(2)}(\lambda_1,\lambda_2)
{\cal T}_{1,3}(\lambda_1) {\cal T}_{1, 1}(\lambda_2).
\label{ans2simp}
\EN

The two-particle vector (\ref{ans2simp}) has the nice feature of being
ordered as far as the rapidity
$\lambda_1$ is concerned.  Another important property is its
symmetry under the
permutation of the rapidities $\lambda_1$ and $\lambda_2$.
This property is derived by substituting the
commutation rule (\ref{b12ba-1a}) for the operators
${\cal T}_{1,2}(\lambda_1)$ and $ {\cal T}_{1, 2}(\lambda_2)$
in Eq.(\ref{ans2simp}) and by considering
the  definition of
${_1}{\cal F}_{1}^{(2)}(\lambda_1,\lambda_2)$ given in Eq.(\ref{psi1}).
As a result we find,
\EQ
\phi_{2}(\lambda_1,\lambda_2)=\theta(\lambda_1, \lambda_2)
\left[ {\cal T}_{1,2}(\lambda_2) {\cal T}_{1, 2}(\lambda_1)
+\frac{D_{2}^{(2,1)}(\lambda_1,\lambda_2)}{D_{2}^{(2,0)}(\lambda_1,\lambda_2)}
{\cal T}_{1,3}(\lambda_2) {\cal T}_{1, 1}(\lambda_1) \right],
\label{sym2incomp}
\EN
where the exchange function $\theta(\lambda,\mu)$ is,
\EQ
\theta(\lambda, \mu)=D_{2}^{(2,0)}(\lambda,\mu)
\frac{R(\lambda,\mu)_{2,1}^{2,1}}{R(\lambda,\mu)_{1,1}^{1,1}}
=\frac{\left|
\begin{array}{cc}
R(\lambda,\mu)_{2,2}^{2,2} & R(\lambda,\mu)_{3,1}^{2,2} \\
R(\lambda,\mu)_{2,2}^{3,1} & R(\lambda,\mu)_{3,1}^{3,1}
\end{array} \right|}{R(\lambda,\mu)_{1,1}^{1,1}
R(\lambda,\mu)_{3,1}^{3,1}}.
\label{thetasym}
\EN

The proof of the permutation symmetry is completed by using two
identities coming from the unitarity property derived in section 4.
In fact, if we substitute the identity (\ref{uni1ans2}) in
Eq.(\ref{sym2incomp}) we find that $\phi_{2}(\lambda_1,\lambda_2)$
satisfies the exchange property, \EQ
\phi_{2}(\lambda_1,\lambda_2)=\theta(\lambda_1,\lambda_2)
\phi_{2}(\lambda_2,\lambda_1), \label{sym2} \EN while the second
identity (\ref{uni2ans2}) assures  us the consistency condition, \EQ
\theta(\lambda_1,\lambda_2) \theta(\lambda_2,\lambda_1)=1.
\label{exch} \EN

The exchange symmetry is very helpful to simplify the expressions for
the action of the diagonal operators ${\cal T}_{a, a}(\lambda)$ on the two-particle state $\ket{\Phi_{2}}$.
It makes possible to represent the various distinct contributions
to the unwanted terms, generated by the commutation
between ${\cal T}_{a, a}(\lambda)$ and $\phi_{2}(\lambda_1,\lambda_2)$, in a compact
and illuminating form.
The details concerning such simplifications are rather cumbersome and for this reason they have
been summarized in Appendix A. In particular, we show how the identities derived in section 4
are very useful to write a recursive formula for the two-particle off-shell amplitudes.
In what follows we shall present the main
conclusions that can be reached from the results of Appendix A.
Considering the discrete function (\ref{disc}) we find that
${\cal T}_{a, a}(\lambda)\ket{\Phi_{2}}$ can be written as,
\bear
 {\cal T}_{a, a}(\lambda) \ket{\Phi_{2}}
&=& w_a(\lambda) \prod_{i=1}^{2}P_a(\lambda,\lambda_i) \ket{\Phi_{2}}
\nonumber \\
&-&
\bar{\delta}_{a}^{N}
{\cal T}_{a, a+1}(\lambda)
\sum_{\stackrel{i,j=1}{j \neq i}}^{2}
\phi_1 (\lambda_j)
w_1(\lambda_i) {_1}{\cal F}_{1}^{(a)}(\lambda,\lambda_i)
\frac{R(\lambda_j,\lambda_i)_{1, 1}^{1, 1}}{R(\lambda_j,\lambda_i)_{2, 1}^{2, 1}}
\theta_<(\lambda_j, \lambda_i)
\ket{0}
 \nonumber \\
 &-&
\bar{\delta}_{a}^{1}
{\cal T}_{a-1, a}(\lambda)
\sum_{\stackrel{i,j=1}{j \neq i}}^{2}
\phi_1 (\lambda_j)
w_2(\lambda_i) {_0}{\cal F}_{1}^{(a-1)}(\lambda,\lambda_i)
\frac{R(\lambda_i,\lambda_j)_{1, 1}^{1, 1}}{R(\lambda_i,\lambda_j)_{2, 1}^{2, 1}}
\theta_<(\lambda_i, \lambda_j)
\ket{0}
\nonumber \\
&-&
\bar{\delta}_{a}^{N-1,N}
{\cal T}_{a,a+2}(\lambda)
w_{1}(\lambda_1) w_{1}(\lambda_2)
{_2}{\cal F}_{2}^{(a)}(\lambda,\lambda_1,\lambda_2)
\ket{0}
\nonumber \\
&-&
\bar{\delta}_{a}^{N,1}
{\cal T}_{a-1, a+1}(\lambda)
\sum_{\stackrel{i,j=1}{j \neq i}}^{2}
w_{1}(\lambda_i) w_{2}(\lambda_j)
{_1}{\cal F}_{2}^{(a-1)}(\lambda,\lambda_i,\lambda_j)
\theta_<(\lambda_j, \lambda_i)
\ket{0}
\nonumber \\
&-&
\bar{\delta}_{a}^{1,2}
{\cal T}_{a-2, a}(\lambda)
w_{2}(\lambda_1) w_{2}(\lambda_2)
{_0}{\cal F}_{2}^{(a-2)}(\lambda,\lambda_1,\lambda_2)
\ket{0}, ~~~~  \mbox{for} ~~ 1 \le a \le N.
\label{comAaaPhi2geral}
 \ear
where function
$\theta_<(\lambda_{i},\lambda_{j})$ is defined by,
\EQ
\theta_<(\lambda_{i},\lambda_{j})
= \begin{cases} \displaystyle
\theta(\lambda_{i},\lambda_{j}),~~~~\mbox{for} ~~ i < j \cr
\displaystyle 1, ~~~~\mbox{for} ~~ i \ge j.
\label{theta<}
\end{cases}
\EN

The off-shell functions
${_c}{\cal F}_{2}^{(a)}(\lambda,\lambda_1,\lambda_2)$ for $c=0,1,2$
represent the three distinct
contributions of unwanted operators carrying azimuthal spin $s=2$. In appendix A
we have shown that they are given in terms of the
following recursive relations,
\bear
\label{f20}
{_0}{\cal F}{_2^{(a)}}(\lambda,\lambda_1,\lambda_2)&=&
\frac{R(\lambda,\lambda_1)_{a+1,1}^{a,2}}{R(\lambda,\lambda_1)_{a+2,1}^{a+2,1}}
{_0}{\cal F}_{1}^{(a+1)}(\lambda,\lambda_2)+
\frac{R(\lambda,\lambda_1)_{a+2,1}^{a,3}}{R(\lambda,\lambda_1)_{a+2,1}^{a+2,1}}
{_1}{\cal F}_{1}^{(2)}(\lambda_1,\lambda_2)
\\
\label{f21}
{_1}{\cal F}_{2}^{(a)}(\lambda,\lambda_1,\lambda_2)&=&
{_0}{\cal F}_{1}^{(a)}(\lambda,\lambda_2)
{_1}{\cal F}_{1}^{(a+1)}(\lambda,\lambda_1)
\frac{R(\lambda_2,\lambda_1)_{1,1}^{1,1}}{R(\lambda_2,\lambda_1)_{2,1}^{2,1}}
\\
\label{f22}
{_2}{\cal F}{_2^{(a)}}(\lambda,\lambda_1,\lambda_2)
&= &
-{_0}{\cal F}{_2^{(a)}}(\lambda,\lambda_1,\lambda_2)
-
\sum_{\stackrel{i,j=1}{j \neq i}}^{2}
{_1}{\cal F}{_2^{(a)}}(\lambda,\lambda_i,\lambda_j) \frac{R(\lambda_j,\lambda_i)_{2,1}^{2,1}}{R(\lambda_j,\lambda_i)_{1,1}^{1,1}} \frac{R(\lambda_i,\lambda_j)_{1,1}^{1,1}}{R(\lambda_i,\lambda_j)_{2,1}^{2,1}}
\theta_<(\lambda_i, \lambda_j). \nonumber \\
\ear

The symmetry of the two-particle vector under the exchange of rapidities is
expected to be reflected in functions
${_0}{\cal F}{_2^{(a)}}(\lambda,\lambda_1,\lambda_2)$ and
${_2}{\cal F}{_2^{(a)}}(\lambda,\lambda_1,\lambda_2)$.  In fact, from
Eq.(\ref{comAaaPhi2geral}) we observe that such functions
are proportional to products that do not contain
function $\theta(\lambda_1,\lambda_2)$. Therefore, the invariance
of the two-particle state $\ket{\Phi_2}$ by the permutation
$\lambda_1 \leftrightarrow \lambda_2$ implies that
\EQ
{_c}{\cal F}{_2^{(a)}}(\lambda,\lambda_1,\lambda_2) = \theta(\lambda_1,\lambda_2)
{_c}{\cal F}{_2^{(a)}}(\lambda,\lambda_2,\lambda_1)~~~\mbox{for}~~c=0,2.
\label{sym2F}
\EN

We now pause to comment on the results  we have  so far
obtained for the two-particle problem.
The alert reader will observe that the role of the discrete
functions $\bar{\delta}_{a}^{j_1}$ and $\bar{\delta}_{a}^{j_1,j_2}$
are merely to project out undefined monodromy matrix elements. These
are operators having either the form
${\cal T}_{a-i, a}(\lambda)$ with $a-i<1$ or
${\cal T}_{a,
a+i}(\lambda)$ with $a+i>N$ such that $i=1,2$. This remark suggests
Eq.(\ref{comAaaPhi2geral}) can still be
written in a more compact manner. Indeed, we find that
Eq.(\ref{comAaaPhi2geral}) is equivalent to the following
expression,
\bear
{\cal T}_{a, a}(\lambda) \ket{\Phi_2} &=& w_a(\lambda)
\prod_{i=1}^{2}P_{a}(\lambda,\lambda_{i}) \ket{\Phi_2}-
\sum_{t=1}^{2} \sum_{p=M \lbrace 0,a+t-N
\rbrace}^{m \lbrace a-1,t \rbrace} {\cal T}_{a-p,
a+t-p}(\lambda)
\nonumber \\
& \times & \sideset{}{^*}\sum_{\stackrel{1 \le j_1< \dots <
j_{(t-p)} \le 2}{1 \le j_{(t-p+1)}  < \dots < j_{t} \le 2}}
\phi_{2-t}(\{ \lambda_{i} \}_{i \ne j_1,\dots,j_{t} }^{i=1,2})~
{_{t-p}}{\cal F}_{t}^{(a-p)}(\lambda,
\lambda_{j_1},\dots,\lambda_{j_{t}} )
\nonumber \\
& \times & \left(
\prod_{k=1}^{t-p} w_1(\lambda_{j_k})
\prod_{\stackrel{i=1}{i \ne j_1,\dots,j_{t} }}^{2}
\frac{R(\lambda_{i},\lambda_{j_k})_{1, 1}^{1,
1}}{R(\lambda_{i},\lambda_{j_k})_{2, 1}^{2, 1}}
\theta_<(\lambda_{i},\lambda_{j_k}) \right)
\nonumber \\
& \times & \left( \prod_{l=t-p+1}^{t}
w_2(\lambda_{j_l}) \prod_{\stackrel{i=1}{i \ne
j_1,\dots,j_{t} }}^{2}
\frac{R(\lambda_{j_l},\lambda_{i})_{1, 1}^{1,
1}}{R(\lambda_{j_l},\lambda_{i})_{2, 1}^{2, 1}}
\theta_<(\lambda_{j_l},\lambda_{i}) \right)
\prod_{k=1}^{t-p} \prod_{l=t-p+1}^{t}
\theta_<(\lambda_{j_l},\lambda_{j_k}) \ket{0},
\nonumber \\
\label{comAaaPhi2compacta}
\ear
where $*$ in the sum means that any terms with $j_k= j_l$ for
$ l \in \{t-p+1,\dots,t \}  $ and $k \in \{ 1,\dots,t-p \} $ are different.
Moreover, the
notation $\{\lambda_i \}^{i=1,\dots,n}_{ i \ne j_1, \cdots,
j_{p}}$ represents that out of the many possible rapidities
$\lambda_1, \cdots, \lambda_n$ those indexed by $\lambda_{j_1},
\cdots, \lambda_{j_{p}}$ are not allowed in the set.

The latter results  contains all the basic ingredients we need to solve the two-particle problem.
The solution of the transfer matrix eigenvalue problem consists basically in taking
the sum of either Eq.(\ref{comAaaPhi2geral}) or Eq.(\ref{comAaaPhi2compacta}) over the diagonal index $a$.
By performing few rearrangements in this sum we find that,
\bear
\lefteqn{T(\lambda) \ket{\Phi_{2}} =
w_a(\lambda) \prod_{i=1}^{2}P_{a}(\lambda,\lambda_{i}) \ket{\Phi_2}}
\nonumber \\
&-&
\sum_{a=1}^{N-1}
\sum_{\stackrel{i,j=1}{j \neq i}}^{2}
\left[
w_1(\lambda_i) {_1}{\cal F}_{1}^{(a)}(\lambda,\lambda_i)
\frac{R(\lambda_j,\lambda_i)_{1, 1}^{1, 1}}{R(\lambda_j,\lambda_i)_{2, 1}^{2,1}}
+
w_2(\lambda_i) {_0}{\cal F}_{1}^{(a)}(\lambda,\lambda_i)
\frac{R(\lambda_i,\lambda_j)_{1, 1}^{1, 1}}{R(\lambda_i,\lambda_j)_{2, 1}^{2, 1}}
\theta(\lambda_i, \lambda_j)
\right]
\nonumber \\
&\times &
\theta_<(\lambda_j, \lambda_i)
{\cal T}_{a, a+1}(\lambda)
\phi_1 (\lambda_j)
\ket{0}
 \nonumber \\
 &-&
\sum_{a=1}^{N-2}
\left[
w_{1}(\lambda_1) w_{1}(\lambda_2)
{_2}{\cal F}_{2}^{(a)}(\lambda,\lambda_1,\lambda_2)
+
\sum_{\stackrel{i,j=1}{j \neq i}}^{2}
w_{1}(\lambda_i) w_{2}(\lambda_j)
{_1}{\cal F}_{2}^{(a)}(\lambda,\lambda_i,\lambda_j)
\theta_<(\lambda_j, \lambda_i)
\right.
\nonumber \\
&+&
\left.
w_{2}(\lambda_1) w_{2}(\lambda_2)
{_0}{\cal F}_{2}^{(a)}(\lambda,\lambda_1,\lambda_2)
\right]
{\cal T}_{a,a+2}(\lambda)\ket{0},
\label{traPhi2}
\ear

From Eq.(\ref{traPhi2})  we can see the possible classes of unwanted terms that should be
canceled out. The first term come as direct extension of the structure present in
the off-shell one-particle state. The second term is inherent to the two-particle
state  and corresponds to the three possible manners of constructing unwanted terms with azimuthal
spin $s=2$ of type  ${\cal T}_{a, a+2}(\lambda) w_i(\lambda_1) w_j(\lambda_2) \ket{0}$ for $i,j=1,2$.
In order to cancel all such unwanted terms we have to use the properties of the auxiliary functions
${_1}{\cal F}_1^{(a)}(\lambda,\lambda_1)$ and ${_2}{\cal F}_2^{(a)}(\lambda,\lambda_1,\lambda_2)$
given by Eqs.(\ref{psi1}, \ref{f22}).
More specifically, considering that
${_1}{\cal F}_1^{(a)}(\lambda,\lambda_1)=-{_0}{\cal F}_1^{(a)}(\lambda,\lambda_1)$
as well as by substituting the amplitude ${_2}{\cal F}_2^{(a)}(\lambda,\lambda_1,\lambda_2)$
in Eq.(\ref{traPhi2}) we find,
\bear
T(\lambda) \ket{\Phi_{2}} &=&
w_a(\lambda) \prod_{i=1}^{2}P_{a}(\lambda,\lambda_{i}) \ket{\Phi_2}
\nonumber \\
&+&
\sum_{a=1}^{N-1}
\sum_{\stackrel{i,j=1}{j \neq i}}^{2}
\left[
w_1(\lambda_j)
\frac{R(\lambda_i,\lambda_j)_{1, 1}^{1, 1}}{R(\lambda_i,\lambda_j)_{2, 1}^{2, 1}}
-
w_2(\lambda_j)
\frac{R(\lambda_j,\lambda_i)_{1, 1}^{1, 1}}{R(\lambda_j,\lambda_i)_{2, 1}^{2, 1}}
\theta(\lambda_j, \lambda_i)
\right]{_0}{\cal F}_{1}^{(a)}(\lambda,\lambda_j)
\nonumber \\
&\times &
\theta_<(\lambda_i, \lambda_j)
{\cal T}_{a, a+1}(\lambda)
\phi_1 (\lambda_i)
\ket{0}
 \nonumber \\
 &+&
\sum_{a=1}^{N-2}
\left\{
\left[
w_{1}(\lambda_1) w_{1}(\lambda_2)
-
w_{2}(\lambda_1) w_{2}(\lambda_2)
\right]
{_0}{\cal F}_{2}^{(a)}(\lambda,\lambda_1,\lambda_2)
\right.
\nonumber \\
&+&
\sum_{\stackrel{i,j=1}{j \neq i}}^{2}
w_{1}(\lambda_i)
\left[
w_1(\lambda_j)
\frac{R(\lambda_i,\lambda_j)_{1, 1}^{1, 1}}{R(\lambda_i,\lambda_j)_{2, 1}^{2, 1}}
-
w_2(\lambda_j)
\frac{R(\lambda_j,\lambda_i)_{1, 1}^{1, 1}}{R(\lambda_j,\lambda_i)_{2, 1}^{2, 1}}
\theta(\lambda_j, \lambda_i)
\right]
\frac{R(\lambda_j,\lambda_i)_{2, 1}^{2, 1}}{R(\lambda_j,\lambda_i)_{1, 1}^{1, 1}}
\nonumber \\
& \times &
\left.
{_1}{\cal F}_{2}^{(a)}(\lambda,\lambda_i,\lambda_j)
\theta_<(\lambda_i, \lambda_j)
\right\}
{\cal T}_{a,a+2}(\lambda)\ket{0}.
\label{traPhi2ii}
\ear

From Eq.(\ref{traPhi2ii}) we see that the unwanted terms either
proportional to ${\cal T}_{a,a+1}(\lambda)\phi_1 (\lambda_j)\ket{0}$
or to ${\cal T}_{a,a+2}(\lambda)\ket{0}$ vanish provided that the
rapidities $\lambda_1$ and $\lambda_2$ satisfy the following Bethe
ansatz equation\footnote{Note that Eq.(\ref{BEA2}) implies the
relation $\displaystyle{\prod_{i=1}^{2}
\frac{\omega_1(\lambda_i)}{\omega_2(\lambda_i)}}=1$ thanks to the
property $\theta(\lambda_1,\lambda_2)
\theta(\lambda_2,\lambda_1)=1$}.
\EQ \frac{w_1(\lambda_{j})}{w_2(\lambda_{j})} =
\prod_{\stackrel{i=1}{i\neq j}}^{2}
\theta(\lambda_{j},\lambda_{i})
\frac{R(\lambda_{j},\lambda_{i})_{1,1}^{1,1}}{R(\lambda_{j},\lambda_{i})_{2,1}^{2,1}}
\frac{R(\lambda_{i},\lambda_{j})_{2,1}^{2,1}}{R(\lambda_{i},\lambda_{j})_{1,1}^{1,1}}
~~~~ \mbox{for} ~~ j=1,2.
\label{BEA2}
\EN
and consequently the two-particle eigenvalue is,
\EQ
\Lambda_{2}(\lambda)=\sum_{a=1}^{N} w_a(\lambda) \prod_{i=1}^{2} P_a(\lambda,\lambda_i).
\label{gama2}
\EN

The typical feature of integrable theories is that the two-particle results already contain the
main flavour about the structure of the spectrum. This means that expressions (\ref{gama2}) and (\ref{BEA2})
are believed to be valid for general $n$-particle states. To benefit from the knowledge of the eigenvectors
one has, however, to perform the complete analysis for few other multiparticle states with
$n >2$.

\subsection{ The three-particle problem}
\label{sub43}

The three-particle state is constructed in terms of a linear combination of products of
creation fields with spin $s-3$. Thus, the most general ansatz for the three-particle vector
is,
\EQ
\phi_{3}(\lambda_1,\lambda_2,\lambda_3)=
\sum_{\stackrel{ m_1 ,m_2,m_3 =1}{m_1 \ne m_2 \ne m_3 \ne m_1}}^{3}
\sum_{\stackrel{
k_{1},k_{2},k_3 =1}{k_1+k_2+k_3=6}}^{N} \bar{c}_{k_1,k_2,k_3}^{(m_1,m_2,m_3)}(\lambda_{m_1},\lambda_{m_2},\lambda_{m_3})
{\cal T}_{1, k_{1}}(\lambda_{m_1})
{\cal T}_{1, k_{2}}(\lambda_{m_2})
{\cal T}_{1, k_{3}}(\lambda_{m_3}),
\label{ans30}
\EN
where $\bar{c}_{k_1,k_2,k_3}^{(m_1,m_2,m_3)}(\lambda_{m_1},\lambda_{m_2},\lambda_{m_3})$ are arbitrary coefficients.

This general ansatz can be further simplified by considering the following two step procedure.
First we use the commutation rules between the fields
${\cal T}_{1,1}(\lambda)$ and ${\cal T}_{1,b}(\mu)$ for $b=2,\dots,4$ to eliminate from Eq.(\ref{ans30})
the linear dependent products
${\cal T}_{1,1}(\lambda_{m_1}) {\cal T}_{1,1}(\lambda_{m_2}) {\cal T}_{1,4}(\lambda_{m_3})$,
${\cal T}_{1,1}(\lambda_{m_1}) {\cal T}_{1,4}(\lambda_{m_2}) {\cal T}_{1,1}(\lambda_{m_3})$,
${\cal T}_{1,1}(\lambda_{m_1}) {\cal T}_{1,2}(\lambda_{m_2}) {\cal T}_{1,3}(\lambda_{m_3})$,
${\cal T}_{1,1}(\lambda_{m_1}) {\cal T}_{1,3}(\lambda_{m_2}) {\cal T}_{1,2}(\lambda_{m_3})$,
${\cal T}_{1,2}(\lambda_{m_1})$ \newline $\times {\cal T}_{1,1}(\lambda_{m_2}) {\cal T}_{1,3}(\lambda_{m_3})$ and
${\cal T}_{1,3}(\lambda_{m_1}) {\cal T}_{1,1}(\lambda_{m_2}) {\cal T}_{1,2}(\lambda_{m_3})$. The next step
is to reorder products of creation operators associated to the combinations
${\cal T}_{1, 2}(\lambda)$ with ${\cal T}_{1, 2}(\mu)$, ${\cal T}_{1, 2}(\lambda)$ with ${\cal T}_{1, 3}(\mu)$ and
 ${\cal T}_{1, 3}(\lambda)$ with ${\cal T}_{1, 2}(\mu)$. This task is performed with the
help of the commutation rules given by Eq.(\ref{eq37}). Taking into account these steps we find that the
ansatz for the three-particle vector  (\ref{ans30})
becomes,
\bear
\lefteqn{\phi_3(\lambda_1,\lambda_2,\lambda_3) = {\cal T}_{1, 2}(\lambda_1)
\phi_{2}(\lambda_2,\lambda_3)
+ g^{(1)}_{3}(\lambda_1,\lambda_2,\lambda_3)
{\cal T}_{1, 3}(\lambda_1) \phi_{1}(\lambda_2) {\cal T}_{1,1}(\lambda_3)}
\nonumber \\
&+& g^{(2)}_{3}(\lambda_1,\lambda_2,\lambda_3)
{\cal T}_{1, 3}(\lambda_1) \phi_{1}(\lambda_3) {\cal T}_{1,1}(\lambda_2)
+ g^{(3)}_{3}(\lambda_1,\lambda_2,\lambda_3)
{\cal T}_{1, 3}(\lambda_2) \phi_{1}(\lambda_3) {\cal T}_{1,1}(\lambda_1)
\nonumber \\
&+& g^{(4)}_{3}(\lambda_1,\lambda_2,\lambda_3)
{\cal T}_{1, 4}(\lambda_1) {\cal T}_{1,1}(\lambda_2) {\cal T}_{1,1}(\lambda_3)
+ g^{(5)}_{3}(\lambda_1,\lambda_2,\lambda_3)
{\cal T}_{1, 4}(\lambda_2) {\cal T}_{1,1}(\lambda_1) {\cal T}_{1,1}(\lambda_3)
\nonumber \\
&+& g^{(6)}_{3}(\lambda_1,\lambda_2,\lambda_3)
{\cal T}_{1, 4}(\lambda_3) {\cal T}_{1,1}(\lambda_1) {\cal T}_{1,1}(\lambda_2)
+ g^{(7)}_{3}(\lambda_1,\lambda_2,\lambda_3)
{\cal T}_{1, 2}(\lambda_1) {\cal T}_{1, 3}(\lambda_2) {\cal T}_{1,1}(\lambda_3)
\nonumber \\
&+& g^{(8)}_{3}(\lambda_1,\lambda_2,\lambda_3)
{\cal T}_{1, 2}(\lambda_1) {\cal T}_{1, 3}(\lambda_3) {\cal T}_{1,1}(\lambda_2)
+ g^{(9)}_{3}(\lambda_1,\lambda_2,\lambda_3)
{\cal T}_{1, 2}(\lambda_2) {\cal T}_{1, 3}(\lambda_3) {\cal T}_{1,1}(\lambda_1),
\nonumber \\
\label{ans3} \ear
where $g^{(i)}_{3}(\lambda_1,\lambda_2,\lambda_3)$ are going to be fixed bellow.

Note that the three-particle ansatz has been written in terms of the one-particle and two-particle
vectors. In this manner it will be able to use the previous results already established in sections
\ref{sub41} and \ref{sub42}. Further progress is made by investigating the result of the commutation of the
diagonal operators with the four distinct types of product of operators defining the three-particle
vector (\ref{ans3}). As in the two-particle problem there is no need to worry about the right-hand side
diagonal fields ${\cal T}_{1,1}(\lambda_i)$.
In what follows we describe the main steps one has to perform in order to solve this problem.

Let us  start by considering the commutation of ${\cal
T}_{a,a}(\lambda)$ with the most involved term ${\cal T}_{1,
2}(\lambda_1)$ $\times \phi_{2}(\lambda_2,\lambda_3)$.  The commutation of the
diagonal fields with ${\cal T}_{1, 2}(\lambda_1)$  is once again
done by using Eqs.(\ref{1part1}-\ref{1partN}). This step generates
besides the diagonal fields several annihilation operators  acting
on the two-particle vector $\phi_{2}(\lambda_2,\lambda_3)$. More specifically,
we have carried on operators of type
${\cal T}_{d+a,a}(\lambda)$ or ${\cal T}_{d+a,a}(\lambda_1)$
for $d=0, \cdots, N-a$ through the two-particle vector
$\phi_{2}(\lambda_2,\lambda_3)$. This task is immediate for $d \ge 3$ but it
requires a number of computations for $d=0,1,2$.
In particular, we have to reorder, as far as the rapidity $\lambda$ is concerned, the following
products of operators
${\cal T}_{1, 2}(\lambda_1)
{\cal T}_{a_1, a_1+1}(\lambda) {\cal T}_{1, 2}(\lambda_j)$ with
$j=2,3$ and  ${\cal T}_{1, 2}(\lambda_1) {\cal T}_{a_2, a_2+2}(\lambda)$
as well as  ${\cal T}_{1, 3}(\lambda_1) {\cal T}_{a_1, a_1+1}(\lambda)$ where
$a_i=a,\dots,a-i$. These computations
are rather cumbersome and have been summarized in Appendix B.

We now discuss the commutation
of the diagonal fields with the second term
${\cal T}_{1, 3}(\lambda_i) \phi_{1}(\lambda_j)$. In this case we start by commuting the operators
${\cal T}_{a, a}(\lambda)$
with ${\cal T}_{1, 3}(\lambda_1)$ by means of Eqs.(\ref{2part1}-\ref{2partN})  and as  before we
produce diagonal and annihilation fields. These operators are then carried out to the right with
the help of
Eqs.(\ref{aniq0},\ref{1partsimp}).  The final step consists of reordering
products of creation fields
such as  ${\cal T}_{1, 3}(\lambda_1) {\cal T}_{a, a+1}(\lambda) $
and ${\cal T}_{1, 3}(\lambda_1) {\cal T}_{a-1, a}(\lambda) $ having the rapidity $\lambda$ in the further right.
This problem is resolved by using the commutation rules given by Eq.(\ref{eq37}) as well as by
the linear system of equations
(\ref{creation1},\ref{creation2}) with $d_1=0,~b_1=3$.
The commutations  between
${\cal T}_{a, a}(\lambda)$ and the  third operator
${\cal T}_{1, 4}(\lambda_i)$ are obtained more directly. In the cases $a=1$ and $a=N$ they follow from
Eqs.(\ref{t11geral},\ref{tNNgeral}) with  $b=4$, respectively. For
$2 \le a \le N-1$ we have to solve the linear system of
equations (\ref{diagon1},\ref{diagon2}) with $b=4$ by applying
the standard Cramer's rule.

We finally discuss the commutation of the diagonal
fields with the last term ${\cal T}_{1, 2}(\lambda_i) {\cal T}_{1, 3}(\lambda_j)$ for $i,j=1,2,3$ such
that $i \neq j $  This turns out to be the simplest case in our analysis  due to the following reason.
The  commutation of
${\cal T}_{a,a}(\lambda)$  with
${\cal T}_{1, 2}(\lambda_i){\cal T}_{1, 3}(\lambda_j)$, by employing
Eqs.(\ref{1part1}-\ref{1partN},
\ref{2part1}-\ref{2partN}), produces a special class of easy unwanted  products that
are not  generated by any of the previous three terms
entering the three-particle ansatz. The form of such undesirable products is given by,
\bear
&\bullet &~~g^{(4+i+j)}_{3}(\lambda_1,\lambda_2,\lambda_3) {\cal T}_{a-3, a-1}(\lambda) {\cal T}_{2, 3}(\lambda_i) {\cal T}_{3, 3}(\lambda_j)~~ \mbox{for} ~a>3
\label{easy31}
\\
&\bullet &~~g^{(4+i+j)}_{3}(\lambda_1,\lambda_2,\lambda_3)  {\cal T}_{a-4, a-1}(\lambda) {\cal T}_{3, 3}(\lambda_i) {\cal T}_{3, 3}(\lambda_j)~~ \mbox{for} ~a>4.
\label{easy32}
\ear

Therefore  we must project out the last term
${\cal T}_{1, 2}(\lambda_i) {\cal T}_{1, 3}(\lambda_j)$  from the linear combination (\ref{ans3}) which is
achieved by setting
$g^{(7)}_{3}(\lambda_1,\lambda_2,\lambda_3)=g^{(8)}_{3}(\lambda_1,\lambda_2,\lambda_3)=g^{(9)}_{3}(\lambda_1,\lambda_2,\lambda_3)=0$.
This choice prompts a sequence of similar cancellations of other terms in the linear
combination (\ref{ans3}). The first one concerns
with the  product
${\cal T}_{1, 3}(\lambda_2)\phi_{1}(\lambda_3)$ which generates the following easy unwanted terms,
\bear &\bullet &~~g^{(3)}_{3}(\lambda_1,\lambda_2,\lambda_3) {\cal T}_{a-1, a}(\lambda) {\cal T}_{2,
3}(\lambda_2) \phi_{1}(\lambda_3)~~ \mbox{for} ~a>1 \label{easy33}
\\
&\bullet &~~g^{(3)}_{3}(\lambda_1,\lambda_2,\lambda_3)
{\cal T}_{a-2, a}(\lambda) {\cal T}_{2, 3}(\lambda_2) {\cal T}_{2, 2}(\lambda_3)~~ \mbox{for} ~a>2
\label{easy34}
\\
&\bullet &~~g^{(3)}_{3}(\lambda_1,\lambda_2,\lambda_3) {\cal T}_{a-3, a}(\lambda) {\cal T}_{3, 3}(\lambda_2)
{\cal T}_{2, 2}(\lambda_3)~~ \mbox{for} ~a>3. \label{easy35} \ear
which are eliminated provided that
$g^{(3)}_{3}(\lambda_1,\lambda_2,\lambda_3)=0$

The next case is associated  to the operator
${\cal T}_{1, 4}(\lambda_j)$ for $j=2,3$. Considering the above constraints, the
commutation of
${\cal T}_{a,a}(\lambda_j)$ with
${\cal T}_{1, 4}(\lambda_j)$ generates the following types of easy unwanted terms,
\bear
&\bullet & ~~g^{(3+j)}_{3}(\lambda_1,\lambda_2,\lambda_3) {\cal T}_{a-1, a}(\lambda) {\cal T}_{2, 4}(\lambda_j) ~~ \mbox{for} ~a>1
\label{easy36}
\\
&\bullet &~~g^{(3+j)}_{3}(\lambda_1,\lambda_2,\lambda_3) {\cal T}_{a-2, a}(\lambda) {\cal T}_{3, 4}(\lambda_j)~~ \mbox{for} ~a>2
\label{easy37}
\\
&\bullet &~~g^{(3+j)}_{3}(\lambda_1,\lambda_2,\lambda_3)  {\cal T}_{a-3, a}(\lambda) {\cal T}_{4, 4}(\lambda_j) ~~ \mbox{for} ~a>3,
\label{easy38}
\\
&\bullet &~~g^{(3+j)}_{3}(\lambda_1,\lambda_2,\lambda_3) {\cal T}_{a-1, a+1}(\lambda) {\cal T}_{2, 3}(\lambda_j) ~~ \mbox{for} ~1< a < N
\label{easy39}
\\
&\bullet &~~g^{(3+j)}_{3}(\lambda_1,\lambda_2,\lambda_3) {\cal T}_{a-2, a+1}(\lambda) {\cal T}_{3, 3}(\lambda_j) ~~ \mbox{for} ~2< a < N
\label{easy310}.
\ear
which are canceled out provided we choose
$g^{(5)}_{3}(\lambda_1,\lambda_2,\lambda_3)=g^{(6)}_{3}(\lambda_1,\lambda_2,\lambda_3)=0$.

The restrictions we have found so far bring a considerable simplification to the
structure of the three-particle ansatz (\ref{ans3}).
We are now left with
only four independent terms,
\bear
\phi_3(\lambda_1,\lambda_2,\lambda_3) &=& {\cal T}_{1, 2}(\lambda_1)
\phi_{2}(\lambda_2,\lambda_3)
\nonumber \\
&+& g^{(1)}_{3}(\lambda_1,\lambda_2,\lambda_3)
{\cal T}_{1, 3}(\lambda_1) \phi_{1}(\lambda_2) {\cal T}_{1,1}(\lambda_3)
+g^{(2)}_{3}(\lambda_1,\lambda_2,\lambda_3)
{\cal T}_{1, 3}(\lambda_1) \phi_{1}(\lambda_3) {\cal T}_{1,1}(\lambda_2)
\nonumber \\
&+& g^{(4)}_{3}(\lambda_1,\lambda_2,\lambda_3)
{\cal T}_{1, 4}(\lambda_1) {\cal T}_{1,1}(\lambda_2) {\cal T}_{1,1}(\lambda_3)
\label{ans3sim} \ear

The structure of the three-particle vector (\ref{ans3sim}) shows some remarkable
characteristics.
We first note that the left-hand side of the linear combination
is always commanded by the creation fields ${\cal T}_{1,b}(\lambda_1)$ where the index $b$
is limited by the maximum possible spin value. Next, each term
of the linear combination is multiplied
by the previous admissible (4-b)-particle state
with rapidity $\lambda_2$ and $\lambda_3$. Finally, diagonal
fields
${\cal T}_{1,1}(\lambda_i)$ are added to the further right to complete the
total number of three possible rapidities. Altogether, these are rather illuminating
features that will be of great help  to set up the ansatz for arbitrary
multiparticle states.
To determine completely the linear combination (\ref{ans3sim}) we have to return
to the analysis of the easy unwanted terms.  Now the same form of a given remaining
easy unwanted term is generated by two distinct types of operators present in
the three-particle vector (\ref{ans3sim}). In particular,
the  operators
${\cal T}_{1, 2}(\lambda_1)\phi_{2}(\lambda_2,\lambda_3)$ and  ${\cal T}_{1, 3}(\lambda_1) \phi_{1}(\lambda_j)$
generate the following class of easy unwanted terms,
\bear
\bullet~~
&& \frac{R(\lambda,\lambda_1)_{a, 1}^{a-1, 2}}{R(\lambda,\lambda_1)_{a, 1}^{a, 1}}
\left[
{_1}{\cal F}_{1}^{(2)}(\lambda_1,\lambda_j)
\frac{R(\lambda_i,\lambda_j)_{1, 1}^{1, 1}}{R(\lambda_i,\lambda_j)_{2, 1}^{2, 1}}
\prod_{\bar{e}=2}^{j-1}
\theta(\lambda_{\bar{e}}, \lambda_j)
-g_{3}^{(i-1)}(\lambda_1,\lambda_2,\lambda_3)
\right]
\nonumber \\
&& \times {\cal T}_{a-1, a}(\lambda) {\cal T}_{2, 3}(\lambda_1) \phi_{1}(\lambda_i) {\cal T}_{1, 1}(\lambda_j)
~~\mbox{for}~~ i,j=2,3; i \ne j.
\label{restr34}
\ear

On the other hand
operators
${\cal T}_{1, 2}(\lambda_1)\phi_{2}(\lambda_2,\lambda_3)$ and ${\cal T}_{1, 4}(\lambda_1)$ are able to produce yet
another family of easy unwanted terms,
\EQ
\bullet ~~
\frac{R(\lambda,\lambda_1)_{a, 1}^{a-1, 2}}{R(\lambda,\lambda_1)_{a, 1}^{a, 1}}
\left[
{_2}{\cal F}_{2}^{(2)}(\lambda_1,\lambda_2,\lambda_3)
-g_{3}^{(4)}(\lambda_1,\lambda_2,\lambda_3)
\right]
{\cal T}_{a-1, a}(\lambda) {\cal T}_{2, 4}(\lambda_1) {\cal T}_{1, 1}(\lambda_2) {\cal T}_{1, 1}(\lambda_2)~~ \mbox{for} ~a>1.
\label{easy314}
\EN

The cancellation of these two families of easy unwanted terms is obtained by choosing functions
$g_3^{(1)}(\lambda_1,\lambda_2,\lambda_3)$,
$g_3^{(2)}(\lambda_1,\lambda_2,\lambda_3)$,
and $g_3^{(4)}(\lambda_1,\lambda_2,\lambda_3)$ as,
\bear
g_{3}^{(1)}(\lambda_1,\lambda_2,\lambda_3) &=&
{_1}{\cal F}_{1}^{(2)}(\lambda_1,\lambda_3)
\frac{R(\lambda_2,\lambda_3)_{1, 1}^{1, 1}}{R(\lambda_2,\lambda_3)_{2, 1}^{2, 1}}
\theta(\lambda_2, \lambda_3)
\label{restr34a}
\\
g_{3}^{(2)}(\lambda_1,\lambda_2,\lambda_3) &=&
{_1}{\cal F}_{1}^{(2)}(\lambda_1,\lambda_2)
\frac{R(\lambda_3,\lambda_2)_{1, 1}^{1, 1}}{R(\lambda_3,\lambda_2)_{2, 1}^{2, 1}}.
\label{restr34b}
\\
g_{3}^{(4)}(\lambda_1,\lambda_2,\lambda_3)& = & {_2}{\cal F}_{2}^{(2)}(\lambda_1,\lambda_2,\lambda_3).
\label{restr35}
\ear

It is important to emphasize that the choices (\ref{restr34a}-\ref{restr35}) eliminate all possible
easy unwanted terms generated by the commutation of the diagonal field ${\cal T}_{a,a}(\lambda)$
with the simplified three-particle vector  (\ref{ans3sim}). By substituting the constraints
(\ref{restr34a}-\ref{restr35})  in Eq.(\ref{ans3sim}) we find the three-particle vector
is now fixed by the expression,
\bear
\phi_3(\lambda_1,\lambda_2,\lambda_3)  &=& {\cal T}_{1, 2}(\lambda_1)
\phi_{2}(\lambda_2,\lambda_3)
\nonumber \\
&+& {\cal T}_{1, 3}(\lambda_1) \sum_{\stackrel{i,j=2}{j \neq i}}^{n=3}
{_1}{\cal F}_{1}^{(2)}(\lambda_1,\lambda_j)
\frac{R(\lambda_i,\lambda_j)_{1, 1}^{1, 1}}{R(\lambda_i,\lambda_j)_{2, 1}^{2, 1}}
\theta_<(\lambda_{i}, \lambda_j)
 \phi_{1}(\lambda_i) {\cal T}_{1,1}(\lambda_j)
\nonumber \\
&+& {_2}{\cal F}_{2}^{(2)}(\lambda_1,\lambda_2,\lambda_3)
{\cal T}_{1, 4}(\lambda_1) {\cal T}_{1,1} (\lambda_2) {\cal T}_{1,1} (\lambda_3) .
\label{ans3s}
\ear

We now turn our attention to investigate the
exchange properties of the three-particle vector (\ref{ans3s}) concerning the
rapidity permutations $\lambda_2 \leftrightarrow \lambda_3$ and
$\lambda_1 \leftrightarrow \lambda_2$. In what follows we shall show that this vector
satisfies the relations,
\EQ
\phi_{3}(\lambda_1,\lambda_2,\lambda_3)  = \theta(\lambda_2,\lambda_3) \phi_{3}(\lambda_1,\lambda_3,\lambda_2)
= \theta(\lambda_1,\lambda_2) \phi_{3}(\lambda_2,\lambda_1,\lambda_3)
\label{sym32}
\EN

The permutation for the rapidities $\lambda_2$ and $\lambda_3$ follows immediately from the two-particle exchange
symmetry. In this case we just have to use
Eqs.(\ref{sym2},{\ref{sym2F}).  The
technical steps to show the permutation for the rapidities $\lambda_1$ and $\lambda_2$ are more intricate.
We start by commuting the operators ${\cal T}_{1, 2}(\lambda_1)$ and ${\cal T}_{1, 2}(\lambda_2)$ in the product
${\cal T}_{1, 2}(\lambda_1) \phi_{2}(\lambda_2,\lambda_3)$ by using the two-particle expression (\ref{ans2simp})
and its symmetry property (\ref{exch}).
We next carry on the operators ${\cal T}_{1, 1}(\lambda_1)$ and ${\cal T}_{1, 1}(\lambda_2)$
through the operator ${\cal T}_{1, 2}(\lambda_3)$ by using Eq.(\ref{1part1}).
By performing these two steps procedure we find the expression,
\bear
\lefteqn{{\cal T}_{1, 2}(\lambda_1) \phi_{2}(\lambda_2,\lambda_3)
=
\theta(\lambda_1,\lambda_2) {\cal T}_{1, 2}(\lambda_2) \phi_{2}(\lambda_1,\lambda_3)
+
\theta(\lambda_1,\lambda_2) {_1}{\cal F}_{1}^{(2)}(\lambda_2,\lambda_1) {\cal T}_{1, 3}(\lambda_2)}
\nonumber \\
&\times &
\left[
P_1(\lambda_1,\lambda_3) {\cal T}_{1, 2}(\lambda_3) {\cal T}_{1, 1}(\lambda_1) -
{_1}{\cal F}_{1}^{(1)}(\lambda_1,\lambda_3) {\cal T}_{1, 2}(\lambda_1) {\cal T}_{1, 1}(\lambda_3)
\right]
\nonumber \\
&-&
{_1}{\cal F}_{1}^{(2)}(\lambda_1,\lambda_2) {\cal T}_{1, 3}(\lambda_1)
\left[
P_1(\lambda_2,\lambda_3) {\cal T}_{1, 2}(\lambda_3) {\cal T}_{1, 1}(\lambda_2) -
{_1}{\cal F}_{1}^{(1)}(\lambda_2,\lambda_3) {\cal T}_{1, 2}(\lambda_2) {\cal T}_{1, 1}(\lambda_3)
\right]
\nonumber \\
&+&
{_1}{\cal F}_{1}^{(2)}(\lambda_2,\lambda_3) {\cal T}_{1, 2}(\lambda_1) {\cal T}_{1, 3}(\lambda_2) {\cal T}_{1, 1}(\lambda_3)
\nonumber \\
&-&
\theta(\lambda_1,\lambda_2)
{_1}{\cal F}_{1}^{(2)}(\lambda_1,\lambda_3) {\cal T}_{1, 2}(\lambda_2) {\cal T}_{1, 3}(\lambda_1) {\cal T}_{1, 1}(\lambda_3).
\label{demonsym31}
\ear

The next task is concerned with the elimination of
products ${\cal T}_{1, 2}(\lambda_2) {\cal T}_{1, 3}(\lambda_1) {\cal T}_{1, 1}(\lambda_3)$ and ${\cal T}_{1, 2}(\lambda_1) $ $\times {\cal T}_{1, 3}(\lambda_2) {\cal T}_{1, 1}(\lambda_3)$ from Eq.(\ref{demonsym31}).
This is accomplished by using the commutation rule between ${\cal T}_{1, 3}(\lambda)$
and ${\cal T}_{1, 2}(\mu)$ given by Eq.(\ref{eq37}) with $d_1=0$ and $b_1=3$.
After commuting these operators in Eq.(\ref{demonsym31}) we find,
\bear
\lefteqn{{\cal T}_{1, 2}(\lambda_1) \phi_{2}(\lambda_2,\lambda_3)
=
\theta(\lambda_1,\lambda_2) \Big[ {\cal T}_{1, 2}(\lambda_2) \phi_{2}(\lambda_1,\lambda_3)
+
{_1}{\cal F}_{1}^{(2)}(\lambda_2,\lambda_1) P_1(\lambda_1,\lambda_3) {\cal T}_{1, 3}(\lambda_2) {\cal T}_{1, 2}(\lambda_3)}
\nonumber \\
&\times&
{\cal T}_{1, 1}(\lambda_1)
+
{_1}{\cal \bar{H}}_{1}^{(2)}(\lambda_2,\lambda_1,\lambda_3|2) {\cal T}_{1, 3}(\lambda_2) {\cal T}_{1, 2}(\lambda_1) {\cal T}_{1, 1}(\lambda_3)
+
{_2}{\cal \bar{F}}_{2}^{(2)}(\lambda_2,\lambda_1,\lambda_3) {\cal T}_{1, 4}(\lambda_2)
\nonumber \\
&\times&
{\cal T}_{1, 1}(\lambda_1) {\cal T}_{1, 1}(\lambda_3) \Big]
-
\Big[ {_1}{\cal F}_{1}^{(2)}(\lambda_1,\lambda_2) P_1(\lambda_2,\lambda_3) {\cal T}_{1, 3}(\lambda_1) {\cal T}_{1, 2}(\lambda_3) {\cal T}_{1, 1}(\lambda_2)
\nonumber \\
&+&
{_1}{\cal \bar{H}}_{1}^{(2)}(\lambda_1,\lambda_2,\lambda_3|2) {\cal T}_{1, 3}(\lambda_1) {\cal T}_{1, 2}(\lambda_2) {\cal T}_{1, 1}(\lambda_3)
+
{_2}{\cal \bar{F}}_{2}^{(2)}(\lambda_1,\lambda_2,\lambda_3) {\cal T}_{1, 4}(\lambda_1) {\cal T}_{1, 1}(\lambda_2) {\cal T}_{1, 1}(\lambda_3) \Big],
\nonumber \\
\label{demonsym32}
\ear
where functions ${_1}{\cal \bar{H}}_{1}^{(2)}(\lambda,\lambda_1,\lambda_2|2)$
and ${_2}{\cal \bar{F}}_{2}^{(2)}(\lambda,\lambda_1,\lambda_2)$ are given by
\bear
{_1}{\cal \bar{H}}_{1}^{(2)}(\lambda,\lambda_1,\lambda_2|2)
&=&
-
\frac{D_{2}^{(2,0)}(\lambda,\lambda_1)}{D_{2}^{(3,0)}(\lambda,\lambda_1)}
\frac{R(\lambda,\lambda_2)_{3,1}^{2,2}}{R(\lambda,\lambda_2)_{3,1}^{3,1}}
\frac{R(\lambda,\lambda_1)_{2,1}^{2,1}}{R(\lambda,\lambda_1)_{3,1}^{3,1}}
-\frac{R(\lambda,\lambda_1)_{3,1}^{2,2}}{R(\lambda,\lambda_1)_{3,1}^{3,1}}
\frac{R(\lambda_1,\lambda_2)_{2,1}^{1,2}}{R(\lambda_1,\lambda_2)_{2,1}^{2,1}}
\nonumber \\
&-&
\frac{R(\lambda_1,\lambda_2)_{3,1}^{2,2}}{R(\lambda_1,\lambda_2)_{3,1}^{3,1}}
\frac{D_{2}^{(3,1)}(\lambda_1,\lambda)}{D_{2}^{(3,0)}(\lambda_1,\lambda)}
\label{barH211|2}
\\
{_2}{\cal \bar{F}}{_2^{(2)}}(\lambda,\lambda_1,\lambda_2)
&=&
\frac{D_{2}^{(2,0)}(\lambda,\lambda_1)}{D_{2}^{(3,0)}(\lambda,\lambda_1)}
\frac{R(\lambda,\lambda_2)_{3,1}^{2,2}}{R(\lambda,\lambda_2)_{3,1}^{3,1}}
\frac{R(\lambda,\lambda_1)_{2,1}^{2,1}}{R(\lambda,\lambda_1)_{3,1}^{3,1}}
\frac{R(\lambda,\lambda_1)_{4,1}^{3,2}}{R(\lambda,\lambda_1)_{4,1}^{4,1}}
-
\frac{R(\lambda_1,\lambda_2)_{3,1}^{2,2}}{R(\lambda_1,\lambda_2)_{3,1}^{3,1}}
\frac{D_{2}^{(3,2)}(\lambda_1,\lambda)}{D_{2}^{(3,0)}(\lambda_1,\lambda)}
\nonumber \\
\label{barFB222}
\ear

We now reached a point in which we have to implement simplifications
on the expressions (\ref{barH211|2},\ref{barFB222}). This is done by
taking into account identities coming from the unitarity relation.
We first use Eq.(\ref{apAid1}) to change the order of the rapidities
$\lambda_1$ and $\lambda_2$ on the second term of
Eq.(\ref{barH211|2}). We then consider the results
(\ref{D231/D230},\ref{D232/D230}) to reorder the rapidities
$\lambda$ and $\lambda_1$ on the determinant ratios
$\frac{D_{2}^{(3,1)}(\lambda_1,\lambda)}{D_{2}^{(3,0)}(\lambda_1,\lambda)}$
and
$\frac{D_{2}^{(3,2)}(\lambda_1,\lambda)}{D_{2}^{(3,0)}(\lambda_1,\lambda)}$
of Eqs.(\ref{barH211|2},\ref{barFB222}). After implementing these
steps  we find the following identities for functions ${_1}{\cal
\bar{H}}_{1}^{(2)}(\lambda,\lambda_1,\lambda_2|2)$ and ${_2}{\cal
\bar{F}}{_2^{(2)}}(\lambda,\lambda_1,\lambda_2)$, \bear {_1}{\cal
\bar{H}}_{1}^{(2)}(\lambda,\lambda_1,\lambda_2|2) &=& {_1}{\cal
H}_{1}^{(2)}(\lambda,\lambda_1,\lambda_2|2)
=\theta(\lambda_1,\lambda_2)
\frac{R(\lambda_1,\lambda_2)_{1,1}^{1,1}}{R(\lambda_1,\lambda_2)_{2,1}^{2,1}}
{_1}{\cal F}_{1}^{(2)}(\lambda,\lambda_2) \label{barH211}
\\
{_2}{\cal \bar{F}}_{2}^{(2)}(\lambda,\lambda_1,\lambda_2)
&=&
{_2}{\cal F}_{2}^{(2)}(\lambda,\lambda_1,\lambda_2)
\label{barF211i}
\ear
where
${_1}{\cal
{H}}_{1}^{(2)}(\lambda,\lambda_1,\lambda_2|2)$ has been defined in Eqs.(\ref{H11|2}).

Now if we substitute the above expressions in Eq.(\ref{demonsym32}) we find that the product
${\cal T}_{1, 2}(\lambda_1)$ $\phi_{2}(\lambda_2,\lambda_3)$ can be finally be written as,
\bear
&&
{\cal T}_{1, 2}(\lambda_1) \phi_{2}(\lambda_2,\lambda_3)
=
\theta(\lambda_1,\lambda_2) \left[ {\cal T}_{1, 2}(\lambda_2) \phi_{2}(\lambda_1,\lambda_3)
+
{_1}{\cal F}_{1}^{(2)}(\lambda_2,\lambda_1) P_1(\lambda_1,\lambda_3) {\cal T}_{1, 3}(\lambda_2) {\cal T}_{1, 2}(\lambda_3)
\right.
\nonumber \\
&&\times
{\cal T}_{1, 1}(\lambda_1)
+
\theta(\lambda_1,\lambda_3) {_1}{\cal F}_{1}^{(2)}(\lambda_2,\lambda_3) P_1(\lambda_3,\lambda_1)
{\cal T}_{1, 3}(\lambda_2) {\cal T}_{1, 2}(\lambda_1) {\cal T}_{1, 1}(\lambda_3)
+
{_2}{\cal F}_{2}^{(2)}(\lambda_2,\lambda_1,\lambda_3)
\nonumber \\
&&\times
\left.
{\cal T}_{1, 4}(\lambda_2) {\cal T}_{1, 1}(\lambda_1) {\cal T}_{1, 1}(\lambda_3) \right]
-
\left[ {_1}{\cal F}_{1}^{(2)}(\lambda_1,\lambda_2) P_1(\lambda_2,\lambda_3) {\cal T}_{1, 3}(\lambda_1) {\cal T}_{1, 2}(\lambda_3) {\cal T}_{1, 1}(\lambda_2)
\right.
\nonumber \\
&& +
\theta(\lambda_2,\lambda_3) {_1}{\cal F}_{1}^{(2)}(\lambda_1,\lambda_3) P_1(\lambda_3,\lambda_2)
{\cal T}_{1, 3}(\lambda_1) {\cal T}_{1, 2}(\lambda_2) {\cal T}_{1, 1}(\lambda_3)
+
{_2}{\cal F}_{2}^{(2)}(\lambda_1,\lambda_2,\lambda_3)
\nonumber \\
&&\times
\left.
{\cal T}_{1, 4}(\lambda_1) {\cal T}_{1, 1}(\lambda_2) {\cal T}_{1, 1}(\lambda_3) \right],
\ear
which is indeed equivalent to exchange property (\ref{sym32}).

We now turn our attention to the action of diagonal fields
${\cal T}_{a, a}(\lambda)$ on the three-particle state $\ket{\Phi_3}$.
The permutation property (\ref{sym32}) is once again of fundamental help
to gather the contributions of ${\cal T}_{a, a}(\lambda) \ket{\Phi_3}$ in closed forms.
In Appendix B we have described main technical details of these computations and our
final results are given by Eq.(\ref{comAaaPhi3ss}). It turns out that this expression can
also be rewritten in a neat compact form, namely
\bear
{\cal T}_{a, a}(\lambda) \ket{\Phi_3} &=& w_a(\lambda)
\prod_{i=1}^{3}P_{a}(\lambda,\lambda_{i}) \ket{\Phi_3}-
\sum_{t=1}^{3} \sum_{p=M \lbrace 0,a+t-N
\rbrace}^{m \lbrace a-1,t \rbrace} {\cal T}_{a-p,
a+t-p}(\lambda)
\nonumber \\
& \times & \sideset{}{^*}\sum_{\stackrel{1 \le j_1< \dots <
j_{(t-p)} \le 3}{1 \le j_{(t-p+1)}  < \dots < j_{t} \le 3}}
\phi_{3-t}(\{ \lambda_{i} \}_{i \ne j_1,\dots,j_{t} }^{i=1,2,3})~
{_{t-p}}{\cal F}_{t}^{(a-p)}(\lambda,
\lambda_{j_1},\dots,\lambda_{j_{t}} )
\nonumber \\
& \times & \left(
\prod_{k=1}^{t-p} w_1(\lambda_{j_k})
\prod_{\stackrel{i=1}{i \ne j_1,\dots,j_{t} }}^{3}
\frac{R(\lambda_{i},\lambda_{j_k})_{1, 1}^{1,
1}}{R(\lambda_{i},\lambda_{j_k})_{2, 1}^{2, 1}}
\theta_<(\lambda_{i},\lambda_{j_k}) \right)
\nonumber \\
& \times & \left( \prod_{l=t-p+1}^{t}
w_2(\lambda_{j_l}) \prod_{\stackrel{i=1}{i \ne
j_1,\dots,j_{t} }}^{3}
\frac{R(\lambda_{j_l},\lambda_{i})_{1, 1}^{1,
1}}{R(\lambda_{j_l},\lambda_{i})_{2, 1}^{2, 1}}
\theta_<(\lambda_{j_l},\lambda_{i}) \right)
\prod_{k=1}^{t-p} \prod_{l=t-p+1}^{t}
\theta_<(\lambda_{j_l},\lambda_{j_k}) \ket{0},
\nonumber \\
\label{comAaaPhi3compacta}
\ear

As expected the three-particle state generates new unwanted terms proportional
to the creation operators ${\cal T}_{a,a+3}(\lambda)$ carrying spin $s=3$.
The corresponding off-shell amplitudes
\newline ${_c}{\cal F}_{3}^{(a)}(\lambda,\lambda_1,\lambda_2,\lambda_3)$ for $c=0,1,2,3$
are obtained from the following relations,
\bear
{_0}{\cal
F}_{3}^{(a)}(\lambda,\lambda_1,\lambda_2,\lambda_3)& =&
\frac{R(\lambda,\lambda_1)_{a+1, 1}^{a,
2}}{R(\lambda,\lambda_1)_{a+3, 1}^{a+3, 1}} {_0}{\cal
F}_{2}^{(a+1)}(\lambda,\lambda_2,\lambda_3) +
\sum_{\stackrel{i,j=2}{j \neq i}}^{n=3}
\frac{R(\lambda,\lambda_1)_{a+2, 1}^{a,
3}}{R(\lambda,\lambda_1)_{a+3, 1}^{a+3, 1}}
{_0}{\cal
F}_{1}^{(a+2)}(\lambda,\lambda_i)
\nonumber \\
& \times &
{_1}{\cal
F}_{1}^{(2)}(\lambda_1,\lambda_j)
\frac{R(\lambda_i,\lambda_j)_{1, 1}^{1, 1}}{R(\lambda_i,\lambda_j)_{2, 1}^{2, 1}}
\theta_<(\lambda_i,\lambda_j)
+
\frac{R(\lambda,\lambda_1)_{a+3, 1}^{a, 4}}{R(\lambda,\lambda_1)_{a+3, 1}^{a+3, 1}}
{_2}{\cal F}_{2}^{(2)}(\lambda_1,\lambda_2,\lambda_3)
\nonumber \\
\label{f30i}
\ear
\bear
{_1}{\cal F}_{3}^{(a)}(\lambda,\lambda_1,\lambda_2,\lambda_3)
&=&
{_0}{\cal F}_{2}^{(a)}(\lambda,\lambda_2,\lambda_3)
{_1}{\cal F}_{1}^{(a+2)}(\lambda,\lambda_1)
\frac{R(\lambda_2,\lambda_1)_{1, 1}^{1, 1}}{R(\lambda_2,\lambda_1)_{2, 1}^{2, 1}}
\frac{R(\lambda_3,\lambda_1)_{1, 1}^{1, 1}}{R(\lambda_3,\lambda_1)_{2, 1}^{2, 1}}
\label{f31i}
\ear
\bear
{_2}{\cal F}_{3}^{(a)}(\lambda,\lambda_1,\lambda_2,\lambda_3)
&=&
{_0}{\cal F}_{1}^{(a)}(\lambda,\lambda_3)
{_2}{\cal F}_{2}^{(a+1)}(\lambda,\lambda_1,\lambda_2)
\frac{R(\lambda_3,\lambda_1)_{1, 1}^{1, 1}}{R(\lambda_3,\lambda_1)_{2, 1}^{2, 1}}
\frac{R(\lambda_3,\lambda_2)_{1, 1}^{1, 1}}{R(\lambda_3,\lambda_2)_{2, 1}^{2, 1}}
\label{f32i}
\ear
\bear
&&{_3}{\cal F}_{3}^{(a)}(\lambda,\lambda_1,\lambda_2,\lambda_3)
=
-{_0}{\cal F}_{3}^{(a)}(\lambda,\lambda_1,\lambda_2,\lambda_3)
\nonumber \\
&&-
\sum_{j_1=1}^{3}
\sum_{\stackrel{1 \le l_1 < l_2 \le 3}{l_1 \neq j_1 \neq l_2}}
{_1}{\cal F}_{3}^{(a)}(\lambda,\lambda_{j_1},\lambda_{l_1},\lambda_{l_2})
\prod_{k=1}^{2}
\theta_<(\lambda_{j_1},\lambda_{l_k})
\frac{R(\lambda_{j_1},\lambda_{l_k})_{1, 1}^{1, 1}}{R(\lambda_{j_1},\lambda_{l_k})_{2, 1}^{2, 1}}
\frac{R(\lambda_{l_k},\lambda_{j_1})_{2, 1}^{2, 1}}{R(\lambda_{l_k},\lambda_{j_1})_{1, 1}^{1, 1}}
\nonumber \\
&&-
\sum_{1 \le j_1 < j_2 \le 3}
\sum_{\stackrel{l_1=1}{j_1 \neq l_1 \neq j_2}}^{3}
{_2}{\cal F}_{3}^{(a)}(\lambda,\lambda_{j_1},\lambda_{j_2},\lambda_{l_1})
\prod_{k=1}^{2}
\theta_<(\lambda_{j_k},\lambda_{l_1})
\frac{R(\lambda_{j_k},\lambda_{l_1})_{1, 1}^{1, 1}}{R(\lambda_{j_k},\lambda_{l_1})_{2, 1}^{2, 1}}
\frac{R(\lambda_{l_1},\lambda_{j_k})_{2, 1}^{2, 1}}{R(\lambda_{l_1},\lambda_{j_k})_{1, 1}^{1, 1}}.
\nonumber \\
&&
\label{f33i}
\ear

Direct inspection of Eqs.(\ref{f30i}-\ref{f33i}) reveals us that all the three-particle
off-shell amplitudes are determined by a recurrence relation that depends on the previously
calculated
off-shell functions
associated to the one-particle and two-particle problems. From
Eq.(\ref{comAaaPhi3compacta}) we note that the only  unwanted terms not
carrying function $\theta(\lambda_i,\lambda_j)$  are those proportional
to
${_0}{\cal
F}_{3}^{(a)}(\lambda,\lambda_1,\lambda_2,\lambda_3)$ and ${_3}{\cal
F}_{3}^{(a)}(\lambda,\lambda_1,\lambda_2,\lambda_3)$. Therefore, the
three-particle vector permutation property implies the following relation,
\EQ
{_c}{\cal
F}_{3}^{(a)}(\lambda,\lambda_1,\lambda_2,\lambda_3) =
\theta(\lambda_1,\lambda_2) {_c}{\cal
F}_{3}^{(a)}(\lambda,\lambda_2,\lambda_1,\lambda_3)=
\theta(\lambda_2,\lambda_3) {_c}{\cal
F}_{3}^{(a)}(\lambda,\lambda_1,\lambda_3,\lambda_2),~~\mbox{for}~~c=0,3.
\label{sym3F}
\EN

In order to solve the three-particle problem we have first to perform the sum
over the diagonal indices $a=1,\dots,N$ in
Eq.(\ref{comAaaPhi3compacta}). After few rearrangements of the sums
entering in
Eq.(\ref{comAaaPhi3compacta}) we obtain,
\bear
&& T(\lambda) \ket{\Phi_3} =
w_a(\lambda) \prod_{i=1}^{3} P_{a}(\lambda,\lambda_{i}) \ket{\Phi_3}
- \sum_{t=1}^3 \sum_{a=1}^{N-t} {\cal T}_{a,
a+t}(\lambda) \sum_{p=0}^{t}
\sideset{}{^*}\sum_{\stackrel{1 \le j_1< \dots < j_{p} \le
3}{1 \le j_{(p+1)}  < \dots < j_{t} \le 3}}
\phi_{3-t}(\{ \lambda_{i} \}_{i \ne j_1,\dots,j_{t}
}^{i=1,2,3})
\nonumber \\
&& \times  {_{p}}{\cal
F}_{t}^{(a)}(\lambda,\lambda_{j_1},\dots,\lambda_{j_{t}})
\left( \prod_{s=1}^{p} w_1(\lambda_{j_s})
\prod_{\stackrel{i=1}{i \ne j_1,\dots,j_{t} }}^{3}
\frac{R(\lambda_{i},\lambda_{j_s})_{1, 1}^{1,
1}}{R(\lambda_{i},\lambda_{j_s})_{2, 1}^{2, 1}}
\theta_<(\lambda_{i},\lambda_{j_s}) \right)
\nonumber \\
&& \times \left( \prod_{r=p+1}^{t} w_2(\lambda_{j_r})
\prod_{\stackrel{i=1}{i \ne j_1,\dots,j_{t} }}^{3}
\frac{R(\lambda_{j_r},\lambda_{i})_{1, 1}^{1,
1}}{R(\lambda_{j_r},\lambda_{i})_{2, 1}^{2, 1}}
\theta_<(\lambda_{j_r},\lambda_{i}) \right) \prod_{s=1}^{p}
\prod_{r=p+1}^{t} \theta_<(\lambda_{j_r},\lambda_{j_s}).
\label{traPhi3} \ear

The next step consists of collecting together the contributions proportional to
the same type of unwanted terms ${\cal T}_{a,a+\bar{e}}(\lambda)$. This
task is performed by plugging in Eq.(\ref{traPhi3}) the off-shell amplitudes
${_{1}}{\cal
F}_{1}^{(a)}(\lambda,\lambda_{j_1})$, ${_{2}}{\cal
F}_{2}^{(a)}(\lambda,\lambda_{j_1},\lambda_{j_2})$ and ${_{3}}{\cal
F}_{3}^{(a)}(\lambda,\lambda_{j_1},\lambda_{j_2},\lambda_{j_3})$
defined by
Eqs.(\ref{psi1},\ref{f22},\ref{f33i}). After some simplifications
we find the following expression,
\bear && T(\lambda) \ket{\Phi_3} =
\sum_{a=1}^N w_a(\lambda) \prod_{i=1}^{3}P_{a}(\lambda,\lambda_{i}) \ket{\Phi_3}
- \sum_{t=1}^3 \sum_{a=1}^{N-t} {\cal T}_{a, a+t}(\lambda)
\sum_{p=0}^{t-1} \sideset{}{^*}\sum_{\stackrel{1 \le j_1 < \dots <
j_{p} \le 3}{1 \le j_{(p+1)} < \dots < j_{t} \le 3}}
\phi_{3-t}(\{ \lambda_{i} \}_{i \ne j_1,\dots,j_{t} }^{i=1,2,3})
\nonumber \\
&& \times  {_{p}}{\cal
F}_{t}^{(a)}(\lambda,\lambda_{j_1},\dots,\lambda_{t})
\left( \prod_{s=1}^{p} w_1(\lambda_{j_s})
\prod_{\stackrel{i=1}{i \ne  j_1,\dots,j_{t} }}^{3}
\frac{R(\lambda_{i},\lambda_{j_s})_{1, 1}^{1,
1}}{R(\lambda_{i},\lambda_{j_s})_{2, 1}^{2, 1}}
\theta_<(\lambda_{i},\lambda_{j_s}) \right)
\nonumber \\
&& \times \left( \prod_{r=p+1}^{t} \prod_{s=1}^{p}
\theta_<(\lambda_{j_s},\lambda_{j_r}) \prod_{\stackrel{i=1}{i \ne
j_1,\dots,j_{t} }}^{3} \theta_<(\lambda_{i},\lambda_{j_r})
\right)
 \left[ \prod_{r=p+1}^{t} w_2(\lambda_{j_r})
\prod_{\stackrel{i=1}{i \ne j_1,\dots,j_{t} }}^{3}
\frac{R(\lambda_{j_r},\lambda_{i})_{1, 1}^{1,
1}}{R(\lambda_{j_r},\lambda_{i})_{2, 1}^{2, 1}}
\theta(\lambda_{j_r},\lambda_{i}) \right.
\nonumber \\
&& \left. \times \prod_{s=1}^{p}
\theta(\lambda_{j_r},\lambda_{j_s}) - \prod_{r=p+1}^{t}
w_1(\lambda_{j_r}) \prod_{\stackrel{i=1}{i \ne
j_1,\dots,j_{t} }}^{3}
\frac{R(\lambda_{i},\lambda_{j_r})_{1, 1}^{1,
1}}{R(\lambda_{i},\lambda_{j_r})_{2, 1}^{2, 1}}
\prod_{s=1}^{p}
\frac{R(\lambda_{j_s},\lambda_{j_r})_{1, 1}^{1, 1}}{R(\lambda_{j_s},\lambda_{j_r})_{2, 1}^{2, 1}}
\frac{R(\lambda_{j_r},\lambda_{j_s})_{2, 1}^{2,
1}}{R(\lambda_{j_r},\lambda_{j_s})_{1, 1}^{1, 1}} \right].
\label{traPhi3compactaii}
\ear

From Eq.(\ref{traPhi3compactaii}) we conclude that all unwanted terms
proportional to
${\cal T}_{a,a+t}(\lambda)$ can be eliminated
by imposing that functions inside the brackets are null, namely
\bear \prod_{r=p+1}^{t}
\frac{w_1(\lambda_{j_r})}{w_2(\lambda_{j_r})}& =&
\prod_{r=p+1}^{t} \prod_{\stackrel{i=1}{i \ne j_1,\dots,j_{t} }}^{3} \theta(\lambda_{j_r},\lambda_{i})
\frac{R(\lambda_{j_r},\lambda_{i})_{1, 1}^{1,
1}}{R(\lambda_{j_r},\lambda_{i})_{2, 1}^{2, 1}}
\frac{R(\lambda_{i},\lambda_{j_r})_{2, 1}^{2,
1}}{R(\lambda_{i},\lambda_{j_r})_{1, 1}^{1, 1}}
\prod_{s=1}^{p} \theta(\lambda_{j_r},\lambda_{j_s})
\nonumber \\
& \times &
\frac{R(\lambda_{j_r},\lambda_{j_s})_{1, 1}^{1,
1}}{R(\lambda_{j_r},\lambda_{j_s})_{2, 1}^{2, 1}}
\frac{R(\lambda_{j_s},\lambda_{j_r})_{2, 1}^{2,
1}}{R(\lambda_{j_s},\lambda_{j_r})_{1, 1}^{1, 1}} ~~\mbox{for}
~~t=1,2,3 ~~\mbox{and} ~~p=0,\dots,t-1.
\label{almostbea3}
\ear

It turns out that this condition  is fulfilled provided that the rapidities
$\lambda_i$ satisfy the following Bethe ansatz equations,
\EQ
\frac{w_1(\lambda_j)}{w_2(\lambda_j)} =
\prod_{\stackrel{i=1}{i\neq j}}^{3} \theta(\lambda_{j},\lambda_{i})
\frac{R(\lambda_{j},\lambda_{i})_{11}^{11}}{R(\lambda_{j},\lambda_{i})_{21}^{21}}
\frac{R(\lambda_{i},\lambda_{j})_{21}^{21}}{R(\lambda_{i},\lambda_{j})_{11}^{11}}
~~~\mbox{for}~~ j=1,2,3.
\label{BEA3}
\EN

We observe that Eq.(\ref{BEA3}) follows directly from Eq.(\ref{almostbea3})
for $t=1$. To show
that
the other unwanted terms proportional to
${\cal T}_{a, a+2}(\lambda)$ and ${\cal
T}_{a, a+3}(\lambda)$ are also canceled out we just have to substitute
Eq.(\ref{BEA3}) in the left-hand side of Eq.(\ref{almostbea3}). As a result we obtain,
\bear
\prod_{r=p+1}^{t} \prod_{\stackrel{i=1}{i \ne j_r}}^{3}
\theta(\lambda_{j_r},\lambda_{i})
\frac{R(\lambda_{j_r},\lambda_{i})_{1, 1}^{1,
1}}{R(\lambda_{j_r},\lambda_{i})_{2, 1}^{2, 1}}
\frac{R(\lambda_{i},\lambda_{j_r})_{2, 1}^{2,
1}}{R(\lambda_{i},\lambda_{j_r})_{1, 1}^{1, 1}} &=&
\prod_{r=p+1}^{t} \prod_{\stackrel{i=1}{i \ne j_1,\dots,j_{t} }}^{3} \theta(\lambda_{j_r},\lambda_{i})
\frac{R(\lambda_{j_r},\lambda_{i})_{1, 1}^{1,
1}}{R(\lambda_{j_r},\lambda_{i})_{2, 1}^{2, 1}}
\frac{R(\lambda_{i},\lambda_{j_r})_{2, 1}^{2,
1}}{R(\lambda_{i},\lambda_{j_r})_{1, 1}^{1, 1}} \nonumber \\
& \times & \prod_{s=1}^{p} \theta(\lambda_{j_r},\lambda_{j_s})
\frac{R(\lambda_{j_r},\lambda_{j_s})_{1, 1}^{1,
1}}{R(\lambda_{j_r},\lambda_{j_s})_{2, 1}^{2, 1}}
\frac{R(\lambda_{j_s},\lambda_{j_r})_{2, 1}^{2,
1}}{R(\lambda_{j_s},\lambda_{j_r})_{1, 1}^{1, 1}}.
\label{besimp}
\ear

By carrying out few simplifications on Eq.(\ref{besimp}) one is able to
show that it is equivalent to the expression,
\EQ
\prod_{\stackrel{i,r=p+1}{i < r}}^{t}
\theta(\lambda_{j_r},\lambda_{j_i})
\theta(\lambda_{j_i},\lambda_{j_r}) = 1~~~~\mbox{for}~~~t=2,3.
\label{equivBEA3}
\EN
which is satisfied thanks to the property (\ref{exch}) of function
$\theta(\lambda,\mu)$. As a consequence of the cancellation of all
unwanted terms we find that the three-particle eigenvalue is
given by,
\EQ
\Lambda_{3}(\lambda) = \sum_{a=1}^N
w_a(\lambda) \prod_{i=1}^{3}P_{a}(\lambda,\lambda_{i}).
\label{gama3}
\EN

We conclude by observing that our results for the two-particle
and three-particle states are already capable to unveil us
a common pattern not only for the on-shell properties but
also to the structure of the off-shell Bethe vectors.
It should be emphasized that this fact has become
possible only because we have been working without using
any specific $U(1)$ $R$-matrix.

\subsection{The multi-particle state}
\label{sub44}

The previous solution of the two-particle and three-particle
problems provides us basic guidelines that are useful to construct
multi-particle states. The first lesson we learn is that among the
possible product of creation operators entering the linear
combination there exists a large subset that in practice do not
contribute to the respective state. These terms produce unique easy
unwanted terms that are canceled only by setting the corresponding
linear combination coefficient to zero.  Our results indicate that the surviving product of
creation operators possess the following characteristics. From Eqs.(\ref{ans2simp},\ref{ans3s}) we
observe that the left-hand side of
such products is always governed by the creation fields ${\cal
T}_{1,b}(\lambda_1)$ with $2 \le b \le m\{n+1,N\}$ where $n$ denotes
$n$-particle sector. In addition, each operator ${\cal T}_{1,b}(\lambda_1)$
is then multiplied by the previously determined $(n+1-b)$-particle
vector such that the spin of the $n$-particle state is reproduced.
Finally,  a number of $(b-2)$ diagonal fields ${\cal
T}_{1,1}(\lambda_i)$ are added to the further right to complete the
total number of rapidities $\lambda_1, \cdots,\lambda_n$ present in
the $n$-particle vector. Combining these features together we find
that such building blocks  can be written as, \EQ {\cal T}_{1,
1+\bar{e}}(\lambda_1) \phi_{n-\bar{e}}(\{\lambda_i
\}^{i=2,\dots,n}_{ i \ne j_1, \cdots, j_{\bar{e}-1}} )
\prod_{k_1=1}^{\bar{e}-1} {\cal T}_{1,
1}(\lambda_{j_{k_1}})~~~\mbox{for} ~~ \bar{e}=1,\dots,m \{n,N-1 \}.
\label{build} \EN
where the indices $j_l$ for $l=1,\cdots,\bar{e}-1$
take values on the interval $2 \le j_l \le n$ subjected to the
condition $2 \le j_1 < j_2 <\dots <j_{\bar{e}-1} \le n$.

Considering the above discussion $\phi_n(\lambda_1,\dots,\lambda_n)$
is obtained by taking the most general linear combination of the
terms (\ref{build}), namely \EQ \phi_n(\lambda_1,\dots,\lambda_n) =
\sum_{\bar{e}=1}^{m(n,N-1)}
\phi_n^{(\bar{e})}(\lambda_1,\dots,\lambda_n), \label{ansn1} \EN
where \bear \phi_n^{(\bar{e})}(\lambda_1,\dots,\lambda_n) &=& {\cal
T}_{1, 1+\bar{e}}(\lambda_1) \sideset{}{^*}\sum_{\stackrel{2 \le j_2
<\dots <j_{\bar{e}} \le n}{2 \le j_{\bar{e}+1} < \dots <j_{n} \le
n}} \phi_{n-\bar{e}}(\lambda_{j_{\bar{e}+1}},\dots,\lambda_{j_n} )
\nonumber \\
& \times & \prod_{k_1=2}^{\bar{e}} {\cal T}_{1,1}(\lambda_{j_{k_1}})
g_{\bar{e}}^{( j_2,\dots,
j_{\bar{e}})}(\lambda_{1},\dots,\lambda_{n}), \label{buildn} \ear
where $g_{\bar{e}}^{( j_2,\dots,
j_{\bar{e}})}(\lambda_{1},\dots,\lambda_{n})$ represent the
coefficients of an arbitrary linear combination.

At the present stage the number of unknown coefficients
for given $n$ and $\bar{e}$ is
$\displaystyle{\frac{(n-1)!}{(n-\bar{e})! (\bar{e}-1)!}}$. Here we will choose to normalize
the $n$-particle state by the coefficient proportional to the first term
${\cal T}_{1, 2}(\lambda_1)
\phi_{n-1}(\lambda_2,\dots,\lambda_n)$ of the linear combination (\ref{ansn1}). Therefore, the
total number of coefficients we have to determine is
$\displaystyle{2^{n-1}}-1$ which  at first sight appears as an impracticable task.
This number, however, can be truly reduced by invoking the exchange property of the
$n$-particle vector under the permutation of the rapidities $\lambda_1,\cdots,\lambda_n$.
As emphasized in the previous sections this is a rather important symmetry present in
the explicit construction of the two-particle and three-particle states. Inspired by this
result we demand that the $n$-particle ansatz (\ref{ansn1},\ref{buildn}) should
satisfy similar exchange property. For general $n$  the permutation
$\lambda_j \leftrightarrow \lambda_{j+1}$ for $j=1,\dots, n-1$ reads,
\EQ
\phi_{n}(\lambda_1,\dots,\lambda_j,\lambda_{j+1},\dots,\lambda_n)  = \theta(\lambda_j,\lambda_{j+1})
\phi_{n}(\lambda_1,\dots,\lambda_{j-1},\lambda_{j+1},\lambda_j,\lambda_{j+2},\dots,\lambda_{n}).
\label{symn}
\EN

In our analysis of the three-particle vector
$\phi_{3}(\lambda_1,\lambda_2,\lambda_3)$ we noticed that the
corresponding permutation symmetry  $\lambda_2 \leftrightarrow
\lambda_3$ depends only on the exchange property of the two-particle
vector. It is expected that this feature extends to general
$n$-particle states providing us  a recursive manner to solve the
problem. From now on we assume that the permutation property
(\ref{symn}) of the $n$-particle vector for $j \ge 2$ follows as a
consequence of the exchange symmetry of the previous constructed
$(n-1)$-particle vector. We stress that this assertion has been explicitly verified up
to the four-particle state.
This implies that relation (\ref{symn}) should then
be valid for each term of the linear combination, namely \bear
\phi_n^{(\bar{e})}(\lambda_1,\dots,\lambda_n) =
\theta(\lambda_j,\lambda_{j+1})
\phi_n^{(\bar{e})}(\lambda_1,\dots,\lambda_{j-1},\lambda_{j+1},\lambda_j,\lambda_{j+2},\dots,\lambda_{n}),
\nonumber \\
\mbox{for} ~~ j \ge 2 ~~ \mbox{and} ~~ 1 \le \bar{e} \le n.
\label{symne}
\ear

We now start to explore the consequences of the building blocks permutation
symmetry (\ref{symne}). The first non-trivial case is $\bar{e}=2$  and by substituting
$\phi_n^{(2)}(\lambda_1,\dots,\lambda_n)$ in Eq.(\ref{symne}) we find,
\EQ
g_{2}^{(j+1)}(\lambda_{1},\dots,\lambda_{n}) = \theta(\lambda_j,\lambda_{j+1}) g_{2}^{( j)}(\lambda_1,\dots,\lambda_{j-1},\lambda_{j+1},\lambda_j,\lambda_{j+2},\dots,\lambda_{n}).
\label{recorr2}
\EN
which can recursively be solved and as result we obtain,
\EQ
g_{2}^{(j+1)}(\lambda_{1},\dots,\lambda_{n}) =
\prod_{\stackrel{i=2}{i \ne j+1}}^n \theta_<(\lambda_i,\lambda_{j+1}) g_{2}^{( 2)}(\lambda_1,\lambda_{j+1},
\{ \lambda_i \}^{i=2,\dots,n}_{i \ne j+1}).
\label{recorr21}
\EN

In order to make further progress one needs to find the expression for the
amplitude \newline $g_2^{(2)}(\lambda_1,\cdots,\lambda_n)$. At this point the expressions for the two-particle and three-particle states
obtained in sections \ref{sub42} and \ref{sub43} are of great utility. Direct comparison between Eqs.(\ref{ansn1},\ref{buildn})
and Eqs.(\ref{ans2simp},\ref{ans3s}) for $n=2,3$ reveals us that,
\EQ
g_{2}^{(2)}(\lambda_{1},\lambda_2) = {_1}{\cal F}_{1}^{(2)} (\lambda_1, \lambda_2),~~~~~
g_{2}^{(2)}(\lambda_{1},\lambda_2,\lambda_{3}) =  \frac{R(\lambda_{3},\lambda_{2})_{11}^{11}}{R(\lambda_{3},\lambda_{2})_{21}^{21}}
{_1}{\cal F}_{1}^{(2)} (\lambda_1, \lambda_2).
\label{g22b}
\EN

We next note that the $R$-matrices prefactor in Eq.(\ref{g22b})  is
directly related to the wanted term originated from the commutation rule among
the field ${\cal T}_{1,1}(\lambda_2)$ and the $(n-2)$-particle vector
$\phi_{n-2}(\lambda_3,\dots,\lambda_n)$.
Considering this information
we are able to infer that the expression for
$g_{2}^{( 2)}(\lambda_1,\dots,\lambda_{n})$ is,
\EQ
g_{2}^{( 2)}(\lambda_1,\dots,\lambda_{n})= \prod_{i=3}^n
\frac{R(\lambda_{i},\lambda_{2})_{11}^{11}}{R(\lambda_{i},\lambda_{2})_{21}^{21}}
{_1}{\cal F}_{1}^{(2)} (\lambda_1, \lambda_2).
\label{g22} \EN
and
by substituting Eq.(\ref{g22}) in Eq.(\ref{recorr21}) we finally
find, \EQ g_{2}^{(j)}(\lambda_{1},\dots,\lambda_{n})=
\prod_{\stackrel{i=2}{i \ne j}}^n
\frac{R(\lambda_{i},\lambda_{j})_{11}^{11}}{R(\lambda_{i},\lambda_{j})_{21}^{21}}
\theta_<(\lambda_i,\lambda_{j}) {_1}{\cal F}_{1}^{(2)} (\lambda_1,
\lambda_j). \label{g22geral} \EN

To emphasize the main points of our procedure we shall also consider
explicitly the next case $\bar{e}=3$. By substituting
$\phi_n^{(3)}(\lambda_1,\dots,\lambda_n)$ in Eq.(\ref{symne}) we
found that functions
$g_{3}^{(j_2,j_3)}(\lambda_{1},\dots,\lambda_{n})$ satisfy the
following constraints, \bear
g_{3}^{(j+1,j_3)}(\lambda_{1},\dots,\lambda_{n}) &=&
\theta(\lambda_j,\lambda_{j+1}) g_{3}^{(
j,j_3)}(\lambda_1,\dots,\lambda_{j-1},\lambda_{j+1},\lambda_j,\lambda_{j+2},\dots,\lambda_{n}),
\nonumber \\
&& ~~~~~~~~~~~~~~~~~~~~~~~~~~~~~~~~~~~~\mbox{for} ~~j_2+1 < j_3,
\label{recorr31}
\\
g_{3}^{(j_2,j+1)}(\lambda_{1},\dots,\lambda_{n}) &=&
\theta(\lambda_j,\lambda_{j+1}) g_{3}^{(
j_2,j)}(\lambda_1,\dots,\lambda_{j-1},\lambda_{j+1},\lambda_j,\lambda_{j+2},\dots,\lambda_{n}),
\nonumber \\
&& ~~~~~~~~~~~~~~~~~~~~~~~~~~~~~~~~~~~~\mbox{for} ~~j_2+1 < j_3,
\label{recorr32}
\\
g_{3}^{(j,j+1)}(\lambda_{1},\dots,\lambda_{n}) &=& \theta(\lambda_j,\lambda_{j+1}) g_{3}^{( j,j+1)}(\lambda_1,\dots,\lambda_{j-1},\lambda_{j+1},\lambda_j,\lambda_{j+2},\dots,\lambda_{n}),
\nonumber \\
&& ~~~~~~~~~~~~~~~~~~~~~~~~~~~~~~~~~~~~\mbox{for} ~~j_2+1 = j_3.
\label{recorr33} \ear

Once again it is possible to solve
Eqs.(\ref{recorr31}-\ref{recorr33}) in a recursive manner and the
result is, \EQ g_{3}^{(j_2,j_3)}(\lambda_{1},\dots,\lambda_{n}) =
\prod_{\stackrel{i=2}{i \ne j_2, j_3 }}^n
\theta_<(\lambda_i,\lambda_{j_2}) \theta_<(\lambda_i,\lambda_{j_3})
g_{3}^{(2,3)}(\lambda_1,\lambda_{j_2},\lambda_{j_3},\{ \lambda_i
\}_{\stackrel{i=2,\dots,n}{i \ne j_2,j_3 }}). \label{recorr34} \EN
provided that function $g_3^{(2,3)}(\lambda_1,\dots,\lambda_n)$
satisfies the following exchange property, \EQ
g_{3}^{(2,3)}(\lambda_1,\dots,\lambda_n)=
\theta(\lambda_2,\lambda_3)
g_{3}^{(2,3)}(\lambda_1,\lambda_3,\lambda_2,\lambda_4,\dots,\lambda_n).
\label{recorr33i} \EN

The remaining task consists in the determination of the fundamental amplitude
\newline $g_3^{(2,3)}(\lambda_1,\dots,\lambda_n)$ which is done by considering the same kind of arguments
employed in the case $\bar{e}=2$.  By comparing
Eqs.(\ref{ansn1},\ref{buildn}) for $n=3$ with
Eq.(\ref{ans3s})   tell us that function
$g_{3}^{(2,3)}(\lambda_1,\lambda_2,\lambda_3)={_2}{\cal F}_2^{(2)}(\lambda_1,\lambda_2,\lambda_3)$. The prefactors
for general values of $n$ are now governed by the first term of the commutation rule
between the product ${\cal T}_{1,1}(\lambda_2) {\cal T}_{1,1}(\lambda_3)$ and the creation fields
${\cal T}_{1,2}(\lambda_i)$ present in the leading part of vector $\phi_{n-3}(\lambda_4,\dots,\lambda_n)$. Collecting all
these informations together we find,
\EQ
g_{3}^{(2,3)}(\lambda_1,\dots,\lambda_n)=\prod_{i=4}^n
\frac{R(\lambda_{i},\lambda_{2})_{11}^{11}}{R(\lambda_{i},\lambda_{2})_{21}^{21}}
\frac{R(\lambda_{i},\lambda_{3})_{11}^{11}}{R(\lambda_{i},\lambda_{3})_{21}^{21}}
{_2}{\cal F}_{2}^{(2)} (\lambda_1, \lambda_2,\lambda_3).
\label{g323i}
\EN

We clearly see that Eq.(\ref{g323i}) satisfies the permutation
relation (\ref{recorr33i}) thanks to the exchange symmetry of the
two-particle function ${_2}{\cal
F}_2^{(2)}(\lambda_1,\lambda_2,\lambda_3)$. The coefficients
$g_{3}^{(j_2,j_3)}(\lambda_{1},\dots,\lambda_{n})$ are obtained by
substituting Eq.(\ref{g323i}) in Eq.(\ref{recorr34}) which gives us,
\EQ g_{3}^{(j_2,j_3)}(\lambda_{1},\dots,\lambda_{n}) =
\prod_{\stackrel{i=2}{i \ne j_2, j_3 }}^n
\frac{R(\lambda_{i},\lambda_{j_2})_{11}^{11}}{R(\lambda_{i},\lambda_{j_2})_{21}^{21}}
\frac{R(\lambda_{i},\lambda_{j_3})_{11}^{11}}{R(\lambda_{i},\lambda_{j_3})_{21}^{21}}
\theta_<(\lambda_i,\lambda_{j_2}) \theta_<(\lambda_i,\lambda_{j_3})
{_2}{\cal F}_{2}^{(2)} (\lambda_1, \lambda_{j_2},\lambda_{j_3}).
\label{recorr34geral} \EN

Let us now generalize the above results for arbitrary index
$\bar{e}$. The analysis of the consequences of the permutation
symmetry (\ref{symne}) is made by using  mathematical induction,
leading us to the following general constraint, \EQ g_{\bar{e}}^{(
j_2,\dots, j_{\bar{e}})}(\lambda_{1},\dots,\lambda_{n})=
\prod_{\stackrel{i=2}{i \ne j_2,\dots,j_{\bar{e}}}}^n
\prod_{l=2}^{\bar{e}} \theta_<(\lambda_i,\lambda_{j_l})
g_{\bar{e}}^{(2,\dots,\bar{e})}
(\lambda_1,j_2,\dots,j_{\bar{e}},\{\lambda_{i} \}_{i \ne
\lambda_{j_2},\dots,j_{\bar{e}} }^{i=2,\dots,n}).
\label{recorrgeral} \EN

We are again left to determine the expression of the basic function
$g_{\bar{e}}^{(2,\dots,\bar{e})}(\lambda_1,\dots,\lambda_n)$. To
make progress in this direction we first pause to discuss the
results we have so far obtained. From Eqs.(\ref{g22},\ref{g323i})
we note that functions $g_{2}^{(2)}(\lambda_{1},\dots,\lambda_{n})$
and $g_{3}^{(2,3)}(\lambda_{1},\dots,\lambda_{n})$ are  given in
terms of product among universal prefactors with the off-shell
amplitudes ${_1}{\cal F}_{1}^{(2)} (\lambda_1, \lambda_2)$ and
${_2}{\cal F}_{2}^{(2)} (\lambda_1, \lambda_2,\lambda_3)$. The
former amplitude is a central element of the eigenvalue problem for
the one-particle state while the latter play analogous role for the
two-particle state. It is natural to assume that the structure of
functions
$g_{\bar{e}}^{(2,\dots,\bar{e})}(\lambda_1,\dots,\lambda_n)$ for
arbitrary $\bar{e}$ will follow the same pattern noted above. In
other words, that these elementary functions can be fixed after
the solution of the  eigenvalue problem for the previous
$(\bar{e}-1)$-particle state is completed.  In analogy to sections
\ref{sub41}-\ref{sub43} such solution will provides the expression
for the respective amplitude ${_{\bar{e}-1}}{\cal
F}_{\bar{e}-1}^{(2)} (\lambda_1,
\lambda_{j_2},\dots,\lambda_{j_{\bar{e}}})$. The expression for
$g_{\bar{e}}^{(2,\dots,\bar{e})}(\lambda_1,\dots,\lambda_n)$  is
then obtained by multiplying prefactors involving the ratios
$R(\lambda_{i},\lambda_{j})_{1,1}^{1,1}/R(\lambda_{i},\lambda_{j})_{2,1}^{2,1}$
with the amplitude ${_{\bar{e}-1}}{\cal F}_{\bar{e}-1}^{(2)}
(\lambda_1, \lambda_{j_2},\dots,\lambda_{j_{\bar{e}}})$. As before
the prefactors are easily accounted through the commutation of the
product $ {\cal T}_{1,1}(\lambda_2) \dots {\cal
T}_{1,1}(\lambda_{\bar{e}})$ with operator
$\phi_{n-\bar{e}}(\lambda_{\bar{e}+1}, \dots, \lambda_n)$.
Considering this reasoning we obtain, \EQ g_{\bar{e}}^{(
2,\dots,\bar{e})}(\lambda_1,\dots,\lambda_{n})=
\prod_{i=\bar{e}+1}^n \prod_{j=2}^{\bar{e}}
\frac{R(\lambda_{i},\lambda_{j})_{11}^{11}}{R(\lambda_{i},\lambda_{j})_{21}^{21}}
{_{\bar{e}-1}}{\cal F}_{\bar{e}-1}^{(2)}
(\lambda_1,\dots,\lambda_{\bar{e}}). \label{g2e} \EN and by
substituting Eq.(\ref{g2e}) in Eq.(\ref{recorrgeral}) we found, \EQ
g_{\bar{e}}^{(j_2,\dots,j_{\bar{e}})}(\lambda_1,\dots,\lambda_n)=\prod_{k_1=2}^{\bar{e}}
\prod_{ \stackrel{k_2=2}{k_2 \ne j_2,\dots,j_{\bar{e}} }}^n
\frac{R(\lambda_{k_2},\lambda_{j_{k_1}})_{11}^{11}}{R(\lambda_{k_2},\lambda_{j_{k_1}})_{21}^{21}}
\theta_<(\lambda_{k_2},\lambda_{j_{k_1}}) {_{\bar{e}-1}}{\cal
F}_{\bar{e}-1}^{(2)} (\lambda_1,
\lambda_{j_2},\dots,\lambda_{j_{\bar{e}}}). \label{ggeral} \EN

We have reduced the determination of an arbitrary $n$-particle
vector to the computation of only $N-1$ coefficients
${_{\bar{e}-1}}{\cal F}_{\bar{e}-1}^{(2)} (\lambda_1,
\lambda_{j_2},\dots,\lambda_{j_{\bar{e}}})$ for $\bar{e}=2,\dots,N$.
Collecting together Eqs.(\ref{ansn1}, \ref{buildn}) and
Eq.(\ref{ggeral}) we are able to propose that an educated ansatz for the
$n$-particle vector should be,
\bear
\phi_n(\lambda_1,\dots,\lambda_n)& =& \sum_{\bar{e}=1}^{m(n,N-1)}
{\cal T}_{1, 1+\bar{e}}(\lambda_1) \sideset{}{^*} \sum_{\stackrel{2
\le j_2 <\dots <j_{\bar{e}} \le n}{2 \le j_{\bar{e}+1} < \dots
<j_{n} \le n}} \phi_{n-\bar{e}}(\lambda_{j_{\bar{e}+1}}, \dots,
\lambda_{j_{n}} )~
{_{\bar{e}-1}}{\cal F}_{\bar{e}-1}^{(2)} (\lambda_1,
\lambda_{j_2},\dots,\lambda_{j_{\bar{e}}} )
\nonumber \\
&\times &
\prod_{k_1=2}^{\bar{e}}
{\cal T}_{1,1}(\lambda_{j_{k_1}}) \prod_{ k_2=\bar{e}+1 }^n
\frac{R(\lambda_{j_{k_2}},\lambda_{j_{k_1}})_{11}^{11}}{R(\lambda_{j_{k_2}},\lambda_{j_{k_1}})_{21}^{21}}
\theta_<(\lambda_{j_{k_2}},\lambda_{j_{k_1}}) \label{ansn} \ear
where the normalization ${_0}{\cal F}_{0}^{(2)}(\lambda) =
1$ is assumed.

Having at hand a proposal for the $n$-particle vector we can turn
our attention to the action of the diagonal operators ${\cal
T}_{a,a}(\lambda)$ on the multi-particle state $\ket{\Phi_n}=
\phi_n(\lambda_1,\dots,\lambda_n) \ket{0}$.  This study is
fundamental to find the expressions of the off-shell functions
${_{\bar{e}-1}}{\cal F}_{\bar{e}-1}^{(2)} (\lambda_1,
\lambda_{j_2},\dots,\lambda_{j_{\bar{e}}})$ in terms the $R$-matrix
elements. It should be stressed that the approach put forward here
is clearly self-consistent. After the solution of the eigenvalue
problem for the sector with $(n-1)$ particles we have the basic
ingredients to pursue the next sector having one more particle and
so forth. Considering our previous results for one-particle,
two-particle and three-particle together with the help of
mathematical induction we are able to determine the action of the
diagonal fields on $\ket{\Phi_n}$. The final results are, \bear
{\cal T}_{a, a}(\lambda) \ket{\Phi_n} &=& w_a(\lambda)
\prod_{i=1}^{n}P_{a}(\lambda,\lambda_{i}) \ket{\Phi_n}-
\sum_{t=1}^{n} \sum_{p=M \lbrace 0,a+t-N \rbrace}^{m \lbrace a-1,t
\rbrace} {\cal T}_{a-p, a+t-p}(\lambda)
\nonumber \\
& \times &  \sideset{}{^*}\sum_{\stackrel{1 \le j_1< \dots <
j_{(t-p)} \le n}{1 \le j_{(t-p+1)}  < \dots < j_{t} \le n}}
\phi_{n-t}(\{ \lambda_{i} \}_{i \ne j_1,\dots,j_{t}
}^{i=1,\dots,n})~ {_{t-p}}{\cal F}_{t}^{(a-p)}(\lambda,
\lambda_{j_1},\dots,\lambda_{j_{t}} )
\nonumber \\
& \times & \left(
\prod_{k=1}^{t-p} w_1(\lambda_{j_k})
\prod_{\stackrel{i=1}{i \ne j_1,\dots,j_{t} }}^{n}
\frac{R(\lambda_{i},\lambda_{j_k})_{1, 1}^{1,
1}}{R(\lambda_{i},\lambda_{j_k})_{2, 1}^{2, 1}}
\theta_<(\lambda_{i},\lambda_{j_k}) \right)
\nonumber \\
& \times & \left( \prod_{l=t-p+1}^{t}
w_2(\lambda_{j_l}) \prod_{\stackrel{i=1}{i \ne
j_1,\dots,j_{t} }}^{n}
\frac{R(\lambda_{j_l},\lambda_{i})_{1, 1}^{1,
1}}{R(\lambda_{j_l},\lambda_{i})_{2, 1}^{2, 1}}
\theta_<(\lambda_{j_l},\lambda_{i}) \right)
\prod_{k=1}^{t-p} \prod_{l=t-p+1}^{t}
\theta_<(\lambda_{j_l},\lambda_{j_k}) \ket{0},
\nonumber \\ \label{comAaaPhin}
\ear

The structure of the recurrence relations for the off-shell
amplitudes
${_c}{\cal
F}_{b}^{(a)}( \lambda,\lambda_1,\dots,\lambda_b)$ is as follows. For
$c \neq 0$ and $c \neq b$ our findings for the two-particle and three-particle
states  given in
Eqs.(\ref{f21},\ref{f31i},\ref{f32i})
are sufficient to guide us to the following general structure,
\bear {_c}{\cal
F}_{b}^{(a)}(\lambda,\lambda_1,\dots,\lambda_b) &=& {_0}{\cal
F}_{b-c}^{(a)}(\lambda,\lambda_{(c+1)},\dots,\lambda_{b}) {_c}{\cal
F}_{c}^{(a+b-c)}(\lambda,\lambda_1,\dots,\lambda_c) \prod_{i=c+1}^b
\prod_{j=1}^c \frac{R(\lambda_i,\lambda_j)_{1, 1}^{1, 1}}
{R(\lambda_i,\lambda_j)_{2, 1}^{2, 1}} \nonumber \\
&& \mbox{for}~~b=2,\dots,N-1; ~a=1,\dots,N-b;~c=1,\dots,b-1.
\label{fbc}
\ear

The expressions for $c=0,b$ are, however, more complicated and their multi-particle
extension required us to perform
explicit computations up to the four-particle state. Considering the help of
mathematical induction as well as the expected properties of these functions
under exchange of their rapidities, see for instance Eqs.(\ref{sym2F},\ref{sym3F}), we are able
to infer that,
\bear {_0}{\cal
F}_{b}^{(a)}(\lambda,\lambda_1,\dots,\lambda_b) &=&
\sum_{\bar{e}=1}^b \frac{R(\lambda,\lambda_1)_{a+\bar{e}, 1}^{a,
1+\bar{e}}}{R(\lambda,\lambda_1)_{a+b, 1}^{a+b, 1}} \sideset{}{^*}
\sum_{\stackrel{2 \le j_1< \dots < j_{(b-\bar{e})} \le b}{2 \le
j_{(b-\bar{e}+1)}< \dots < j_{(b-1)} \le b}} {_0}{\cal
F}_{b-\bar{e}}^{(a+\bar{e})}(\lambda,\lambda_{j_1},\dots,
\lambda_{(b-\bar{e})}) \nonumber \\
& \times & {_{\bar{e}-1}}{\cal
F}_{\bar{e}-1}^{(2)}(\lambda_1,\lambda_{j_{(b-\bar{e}+1)}},\dots,\lambda_{j_{(b-1)}})
\prod_{l_1=1}^{b-\bar{e}} \prod_{l_2=b-\bar{e}+1}^{b-1}
\frac{R(\lambda_{j_{l_1}},\lambda_{j_{l_2}})_{1, 1}^{1, 1}}
{R(\lambda_{j_{l_1}},\lambda_{j_{l_2}})_{2, 1}^{2, 1}}
\theta_<(\lambda_{j_{l_1}},\lambda_{j_{l_2}}) \nonumber \\
&\mbox{for}& ~~~~~~~~~~~~~ b=1,\dots,N-1; ~~a=1,\dots,N-b
\label{fb0}
\ear
and
\bear {_b}{\cal
F}_{b}^{(a)}(\lambda,\lambda_1,\dots,\lambda_b) &=& -
\sum_{\bar{f}=0}^{b-1} ~\sum_{1 \le l_1 < l_2 < \dots <
l_{(b-\bar{f})} \le b} {_{\bar{f}}}{\cal F}_{b}^{(a)}(\lambda,\{
\lambda_i \}_{i \ne  l_1,\dots,l_{(b-\bar{f}) }}^{i=1,\dots,b} ,
\lambda_{l_1},\dots,\lambda_{l_{(b-\bar{f})}}) \nonumber \\
&\times & \prod_{s=1}^{b-\bar{f}} \prod_{\stackrel{i=1}{i \ne
l_1,\dots,l_{(b-\bar{f})} }}^{b}
\theta_<(\lambda_{i},\lambda_{l_s})
\frac{R(\lambda_{i},\lambda_{l_s})_{1, 1}^{1,
1}}{R(\lambda_{i},\lambda_{l_s})_{2, 1}^{2, 1}}
\frac{R(\lambda_{l_s},\lambda_{i})_{2, 1}^{2,
1}}{R(\lambda_{l_s},\lambda_{i})_{1, 1}^{1, 1}}
.\nonumber \\
&\mbox{for}& ~~~~~~~~~~~~~ b=1,\dots,N-1; ~~a=1,\dots,N-b.
\label{fbb}
\ear

Here we stress that the recurrence relations
(\ref{fbc},\ref{fb0},\ref{fbb}) have as initial condition the overall normalization
${_0}{\cal F}_{0}^{(a)}(\lambda) =1$ as well as the expressions for the one-particle
off-shell amplitudes
${_0}{\cal F}_{1}^{(a)}(\lambda,\mu)$ and $ {_1}{\cal
F}_{1}^{(a)}(\lambda,\mu)$ defined by Eq.(\ref{psi1}).

The solution of the eigenvalue problem for the multiparticle state is completed
by summing over
Eq.(\ref{comAaaPhin}) for $a=1,\dots,N$. By substituting the
expression (\ref{fbb}) for the off-shell amplitudes
${_b}{\cal
F}_{b}^{(a)}(\lambda,\lambda_1,\dots,\lambda_b)$  we find,
\bear && T(\lambda) \ket{\Phi_n} =
\sum_{a=1}^N w_a(\lambda) \prod_{i=1}^{n}P_{a}(\lambda,\lambda_{i}) \ket{\Phi_n}
- \sum_{t=1}^n \sum_{a=1}^{N-t} {\cal T}_{a, a+t}(\lambda)
\sum_{p=0}^{t-1} \sideset{}{^*}\sum_{\stackrel{1 \le j_1 < \dots <
j_{p} \le n}{1 \le j_{(p+1)} <
 \dots < j_{t} \le n}}
\nonumber \\
&& \times \phi_{n-t}(\{ \lambda_{i} \}_{i \ne
j_1,\dots,j_{t} }^{i=1,\dots,n})~ {_{p}}{\cal
F}_{t}^{(a)}(\lambda,\lambda_{j_1},\dots,\lambda_{t})
\left( \prod_{s=1}^{p} w_1(\lambda_{j_s})
\prod_{\stackrel{i=1}{i \ne  j_1,\dots,j_{t} }}^{n}
\frac{R(\lambda_{i},\lambda_{j_s})_{1, 1}^{1,
1}}{R(\lambda_{i},\lambda_{j_s})_{2, 1}^{2, 1}}
\theta_<(\lambda_{i},\lambda_{j_s}) \right)
\nonumber \\
&& \times \left( \prod_{r=p+1}^{t} \prod_{s=1}^{p}
\theta_<(\lambda_{j_s},\lambda_{j_r}) \prod_{\stackrel{i=1}{i \ne
j_1,\dots,j_{t} }}^{n} \theta_<(\lambda_{i},\lambda_{j_r})
\right)
 \left[ \prod_{r=p+1}^{t} w_2(\lambda_{j_r})
\prod_{\stackrel{i=1}{i \ne j_1,\dots,j_{t} }}^{n}
\frac{R(\lambda_{j_r},\lambda_{i})_{1, 1}^{1,
1}}{R(\lambda_{j_r},\lambda_{i})_{2, 1}^{2, 1}}
\theta(\lambda_{j_r},\lambda_{i}) \right.
\nonumber \\
&& \left. \times \prod_{s=1}^{p} \theta(\lambda_{j_r},\lambda_{j_s})
- \prod_{r=p+1}^{t} w_1(\lambda_{j_r}) \prod_{\stackrel{i=1}{i \ne
j_1,\dots,j_{t} }}^{n} \frac{R(\lambda_{i},\lambda_{j_r})_{1, 1}^{1,
1}}{R(\lambda_{i},\lambda_{j_r})_{2, 1}^{2, 1}} \prod_{s=1}^{p}
\frac{R(\lambda_{j_s},\lambda_{j_r})_{1, 1}^{1,
1}}{R(\lambda_{j_s},\lambda_{j_r})_{2, 1}^{2, 1}}
\frac{R(\lambda_{j_r},\lambda_{j_s})_{2, 1}^{2,
1}}{R(\lambda_{j_r},\lambda_{j_s})_{1, 1}^{1, 1}} \right].
\label{traPhincompactaii} \ear

The unwanted terms are canceled out by requiring that functions
inside the bracket of
Eq.(\ref{traPhincompactaii}) are null for $t=1,\dots, m(n,N-1)$.
The situation here is exactly the same found already for the
two-particle and three-particle state. It is sufficient to consider the constraint
coming from the case $t=1$ which leads us to the following Bethe ansatz equations,
\EQ
\frac{w_1(\lambda_{j_1})}{w_2(\lambda_{j_1})} =
\prod_{\stackrel{i=1}{i \ne j_1}}^{n}
\theta(\lambda_{j_1},\lambda_{i})
\frac{R(\lambda_{j_1},\lambda_{i})_{1, 1}^{1,
1}}{R(\lambda_{j_1},\lambda_{i})_{2, 1}^{2, 1}}
\frac{R(\lambda_{i},\lambda_{j_1})_{2, 1}^{2,
1}}{R(\lambda_{i},\lambda_{j_1})_{1, 1}^{1, 1}}~~~~
\mbox{for}~~j_1=1,\dots,n. \label{beaN}
\EN

To verify  the  cancellations for the remaining values of index $t$ one
just has to substitute Eq.(\ref{beaN}) on the bracket of
Eq.(\ref{traPhincompactaii}) and to consider the exchange property (\ref{exch})
of functions $\theta(\lambda_i,\lambda_j)$. The elimination of
all unwanted terms by Eq.(\ref{beaN}) implies that the $n$-particle eigenstate is,
\EQ \Lambda_{n}(\lambda) =
\sum_{a=1}^N w_a(\lambda) \prod_{i=1}^{n}P_{a}(\lambda,\lambda_{i}).
\label{gaman}
\EN

We close by mentioning that all formulae of this section pass by the
following tests. For $N=2$ we recover the well known algebraic
Bethe ansatz solution of the six-vertex model \cite{TATA}. In this special
case one has to note that both numerator and denominator of Eq.(\ref{thetasym})
vanish and the respective limit gives us $\theta(\lambda,\mu)=
R(\lambda,\mu)_{2,2}^{2,2}/R(\lambda,\mu)_{1,1}^{1,1}$.  For $N=3$
we reproduce the algebraic Bethe ansatz construction proposed
by Tarasov \cite{TA1} for the nineteen-vertex models.  Strictly speaking,
our results should be seen as extensions of the above
mentioned works since no assumption of spectral parameter dependence
for the $R$-matrix has been made. We also verified that the expressions
(\ref{beaN},\ref{gaman}) reproduce the Bethe ansatz equations and eigenvalues
of the higher spin generalization of the six-vertex model \cite{RE}.  We
have also computed explicitly the most complicated
off-shell amplitudes of the four-particle
state  such as ${_{0}}{\cal
F}_{4}^{(a)}(\lambda,\lambda_{1},\dots,\lambda_{4})$  and
${_{4}}{\cal
F}_{4}^{(a)}(\lambda,\lambda_{1},\dots,\lambda_{4})$.
These computations have been checked to be in accordance
with the recurrence relations (\ref{fb0},\ref{fbb}).

\section{Conclusion}

In this paper we have shown how the algebraic Bethe ansatz method
works for arbitrary vertex models whose $R$-matrix commutes with
one $U(1)$ symmetry. This invariance guarantees the existence of
a reference state in which the respective monodromy
matrix acts triangularly. We recall that this is necessary
requirement to start performing an algebraic Bethe ansatz
analysis.
The transfer matrix eigenvalue problem is viewed under the general perspective
that its solution should not depend on a specific functional
form of the $R$-matrix. We argued that the algebraic formulation
of the Bethe states of the transfer matrix can be done
solely on basis of the Yang-Baxter algebra, the Yang-Baxter
equation and the unitarity property satisfied by the $R$-matrix. In fact, we have
presented a method
to obtain the appropriate commutation rules between the
monodromy matrix elements for arbitrary number $N$ of edge states. Moreover,
the necessary identities among the $R$-matrix elements to solve
the eigenvalue problem are derived from the respective Yang-Baxter
and unitary relations.

This approach provided us the expressions for the
on-shell properties such as the transfer matrix eigenvalues
and the Bethe ansatz equations as well as the structure
of the off-shell Bethe vectors in terms of the arbitrary
$R$-matrix weights. The respective off-shell amplitudes
are determined by means of a recurrence formula whose
input are special entries of the $R$-matrix. Note
that the previous understanding of off-shell structure of
$U(1)$ integrable models was restricted to the six-vertex \cite{TATA}
and nineteen-vertex models \cite{TA1}. By contrast, this paper
offers us the basic ingredients to compute such properties
for any $U(1)$ vertex model. In a forthcoming paper \cite{CMM}
we shall indeed exhibit the explicit expressions of
the off-shell amplitudes of some $U(1)$ invariant
vertex models such as the classic higher spin
$XXZ-S$ Heisenberg system and the non-compact
vertex model based on the $SL(2,R)$ algebra.
Interesting enough, we found that all the off-shell
amplitudes ${_c}{\cal F}_b^{(a)}(\lambda,\lambda_1,\dots,\lambda_b)$
factorize in terms of product of elementary functions.

We hope that the framework developed in this paper will also
be relevant to solve other families of integrable models
by the algebraic Bethe ansatz. In particular,
that the universal structure of our expressions for
the eigenvectors and the off-shell amplitudes should play
the role of cornerstones in nested Bethe ansatz solutions
of integrable models with $R$-matrices commuting with
more than one $U(1)$ symmetry. This idea has indeed worked for the
Hubbard model whose algebraic Bethe ansatz solution \cite{HUB}
encoded the main features found in the algebraic Bethe ansatz
formulation of the $N=3$ $U(1)$ vertex model. In general, we
expect that the algebraic solution of a $N$-state integrable model
with $m$ distinct $U(1)$ conserved charges will
be described in terms of a $(N-m+1) \times (N-m+1)$ monodromy
matrix. The partition of the monodromy matrix from $N \times N$
to $(N-m+1) \times (N-m+1)$ are going to be dictated by the form of the
$U(1)$ operators that commute with the $R$-matrix.
This means that the respective length of the recurrence
relations for the eigenvectors and off-shell amplitudes will be
governed by the effective number $\bar{N}=N-m+1$. Now
the off-shell
amplitudes
${_c}{\cal F}_b^{(a)}(\lambda,\lambda_1,\dots,\lambda_b)$   will be
seen as vectors with $m^2$ components while
$\theta(\lambda,\mu)$  will behave as a
$m \times m$ auxiliary factorized $\bar{R}(\lambda,\mu)$
$R$-matrix  satisfying the unitary property. We recall that
this scenario is known to work for specific vertex models
with $N-2$ $U(1)$ symmetries such as those based on the
non-exceptional Lie algebras \cite{MA}.
This has been also verified for the vertex model related to the
fundamental representation of the
$U_q[G_2]$ algebra \cite{KUN}. This system is rather involved
since the size of its R-matrix is $ 7 \times 7$ while its
algebraic Bethe ansatz solution is formulated by a $ 5 \times 5$
monodromy matrix. In all these cases
the corresponding auxiliary $\bar{R}$-matrices
$\bar{R}(\lambda,\mu)$ is derived from the analysis of
the commutation relations following the
general method explained in section  \ref{sec3}. Their
factorization and that they fulfill the unitarity relation were verified using the respective specific
weights. The results of this paper suggest 
that such properties can be derived as a consequence of the
Yang-Baxter equation and the unitary relation
of the original $R$-matrix we have started with. Hopefully, this could be shown 
by generalizing the method discussed in section \ref{secI}. 

Finally, we think that this paper highlights the expectation
that the Bethe ansatz properties of integrable models
should be exhibited in terms of universal formulae depending
only on the $R$-matrix amplitudes. We hope that our results
will inspire the search of a general recipe to solve integrable
systems invariant by many $U(1)$ symmetries through the
algebraic Bethe ansatz method.

\addcontentsline{toc}{section}{Appendix A}
\section*{\bf Appendix A: The two-particle state}
\setcounter{equation}{0}
\renewcommand{\theequation}{A.\arabic{equation}}

In this appendix we provide the technical details entering
the solution of the two-particle eigenvalue problem. Here we shall assume that the easy
unwanted terms mentioned in section \ref{sub42}  have been already canceled out. Therefore, the results
to be described in what follows concern with the
symmetric two-particle ansatz,
\EQ
\ket{\Phi_{2}}= {\cal T}_{1,2}(\lambda_1) {\cal T}_{1, 2}(\lambda_2) \ket{0}
-\frac{R(\lambda_1,\lambda_2)_{3,1}^{2,2}}{R(\lambda_1,\lambda_2)_{3,1}^{3,1}}
w_{1}(\lambda_2)
{\cal T}_{1,3}(\lambda_1) \ket{0}.
\label{ans2simpii}
\EN

The elimination of the easy unwanted terms brings considerable simplifications on the
action of the diagonal operators
${\cal T}_{a,a}(\lambda)$ on $\ket{\Phi_{2}}$. Considering all the steps mentioned
in section \ref{sub42}  we find that
${\cal T}_{a,a}(\lambda)\ket{\Phi_{2}}$ has to be divided in five different parts. We find that
they are given by,
\bear
{\cal T}_{a,a}(\lambda) \ket{\Phi_{2}} & = &
w_{a}(\lambda) P_a(\lambda,\lambda_1)
\left[
P_a(\lambda,\lambda_2) \ket{\Phi_2^{(1)}}
+
\bar{P}_a(\lambda,\lambda_1,\lambda_2) \ket{\Phi_2^{(2)}}
\right]
\nonumber \\
&-&
\bar{\delta}_{a}^N
\sum_{\stackrel{ i,j=1}{i \ne j}}^{2}
w_1(\lambda_i) {_1}{\cal H}_{1}^{(a)}(\lambda,\lambda_1,\lambda_2|i)
{\cal T}_{a,a+1}(\lambda) {\cal T}_{1,2}(\lambda_j) \ket{0}
\nonumber \\
&-&
\bar{\delta}_{a}^1
\sum_{\stackrel{ i,j=1}{i \ne j}}^{2}
w_2(\lambda_i) {_0}{\cal H}_{1}^{(a-1)}(\lambda,\lambda_1,\lambda_2|i)
{\cal T}_{a-1,a}(\lambda) {\cal T}_{1,2}(\lambda_j) \ket{0}
\nonumber \\
&-&
\bar{\delta}_{a}^{N-1,N}
w_1(\lambda_1) w_1(\lambda_2) {_2}{\cal F}_{2}^{(a)}(\lambda,\lambda_1,\lambda_2)
{\cal T}_{a,a+2}(\lambda) \ket{0}
\nonumber \\
&-&
\bar{\delta}_{a}^{1,N}
\sum_{\stackrel{ i,j=1}{i \ne j}}^{2}
w_1(\lambda_i) w_2(\lambda_j) {_1}{\cal H}_{2}^{(a-1)}(\lambda,\lambda_1,\lambda_2|i)
{\cal T}_{a-1,a+1}(\lambda) \ket{0}
\nonumber \\
&-&
\bar{\delta}_{a}^{1,2}
w_2(\lambda_1) w_2(\lambda_2) {_0}{\cal F}_{2}^{(a-2)}(\lambda,\lambda_1,\lambda_2)
{\cal T}_{a-2,a}(\lambda) \ket{0}
\nonumber \\
&&~~~~~~~~~~~~~~~~~~~~~~~~~~~~~~\mathrm{for}~~ 1 \le a \le N,
\label{Aaaphi2}
\ear

Here we recall that functions $P_a(\lambda,\mu)$ have been  defined in the one-particle problem, see Eq.(\ref{pa}).
The dependence of the extra functions $\bar{P}_a(\lambda,\lambda_1,\lambda_2)$
on the $R$-matrix elements are given by,
\EQ
\bar{P}_1(\lambda,\lambda_1,\lambda_2)=
\frac{R(\lambda_1,\lambda_2)_{3,1}^{3,1}}{R(\lambda_1,\lambda_2)_{3,1}^{2,2}}
\left[
\frac{R(\lambda_2,\lambda)_{1,2}^{2,1}}{R(\lambda_2,\lambda)_{2,1}^{2,1}}
\frac{R(\lambda_1,\lambda)^{2,2}_{3,1}}{R(\lambda_1,\lambda)^{3,1}_{3,1}}
+
\frac{R(\lambda_1,\lambda_2)_{3,1}^{2,2}}{R(\lambda_1,\lambda_2)_{3,1}^{3,1}}
\frac{R(\lambda_1,\lambda)_{2,1}^{2,1}}{R(\lambda_1,\lambda)_{3,1}^{3,1}}
\right]
\label{Pbar1}
\EN
\bear
\bar{P}_a(\lambda,\lambda_1,\lambda_2)
&=&
\frac{R(\lambda_1,\lambda_2)_{3,1}^{3,1}}{R(\lambda_1,\lambda_2)_{3,1}^{2,2}}
\left[
\frac{R(\lambda,\lambda_2)_{a+1,1}^{a,2}}{R(\lambda,\lambda_2)_{a+1,1}^{a+1,1}}
\frac{D_{2}^{(a+1,1)}(\lambda,\lambda_1)}{D_{2}^{(a+1,0)}(\lambda,\lambda_1)}
+
\frac{R(\lambda_1,\lambda_2)_{3,1}^{2,2}}{R(\lambda_1,\lambda_2)_{3,1}^{3,1}}
\frac{D_{3}^{(a,0)}(\lambda,\lambda_1)}{D_{2}^{(a,0)}(\lambda,\lambda_1)}
\right.
\nonumber \\
&-&
\left.
\frac{D_{2}^{(a,1)}(\lambda,\lambda_1)}{D_{2}^{(a,0)}(\lambda,\lambda_1)}
\frac{R(\lambda,\lambda_2)_{a,1}^{a-1,2}}{R(\lambda,\lambda_2)_{a,1}^{a,1}}
\right]
~~~~~~~~ \mbox{for} ~~2 \le a \le N-2
\label{Pbar2-N-2}
\ear
\bear
\bar{P}_{N-1}(\lambda,\lambda_1,\lambda_2)
&=&
\frac{R(\lambda_1,\lambda_2)_{3,1}^{3,1}}{R(\lambda_1,\lambda_2)_{3,1}^{2,2}}
\left[
\frac{R(\lambda,\lambda_2)_{N,1}^{N-1,2}}{R(\lambda,\lambda_2)_{N,1}^{N,1}}
\frac{R(\lambda,\lambda_1)_{N-1,3}^{N,2}}{R(\lambda,\lambda_1)_{N,2}^{N,2}}
+
\frac{R(\lambda_1,\lambda_2)_{3,1}^{2,2}}{R(\lambda_1,\lambda_2)_{3,1}^{3,1}}
\frac{R(\lambda,\lambda_1)_{N,1}^{N,1}}{R(\lambda,\lambda_1)_{N,2}^{N,2}}
\right.
\nonumber \\
&\times&
\left.
\frac{\left|
\begin{array}{cc}
R(\lambda,\lambda_1)^{N-1,3}_{N,2} & R(\lambda,\lambda_1)^{N,2}_{N,2} \\
R(\lambda,\lambda_1)^{N-1,3}_{N-1,3} &
R(\lambda,\lambda_1)^{N,2}_{N-1,3}
\end{array} \right|}
{\left| \begin{array}{cc}
R(\lambda,\lambda_1)^{N-1,2}_{N,1} & R(\lambda,\lambda_1)^{N,1}_{N,1} \\
R(\lambda,\lambda_1)^{N-1,2}_{N-1,2} &
R(\lambda,\lambda_1)^{N,1}_{N-1,2}
\end{array} \right|}
-
\frac{D_{2}^{(N-1,1)}(\lambda,\lambda_1)}{D_{2}^{(N-1,0)}(\lambda,\lambda_1)}
\frac{R(\lambda,\lambda_2)_{N-1,1}^{N-2,2}}{R(\lambda,\lambda_2)_{N-1,1}^{N-1,1}}
\right]
\label{PbarN-1}
\ear
\EQ
\bar{P}_N(\lambda,\lambda_1,\lambda_2)
=
\frac{R(\lambda_1,\lambda_2)_{3,1}^{3,1}}{R(\lambda_1,\lambda_2)_{3,1}^{2,2}}
\left[
\frac{R(\lambda_1,\lambda_2)_{3,1}^{2,2}}{R(\lambda_1,\lambda_2)_{3,1}^{3,1}}
\frac{R(\lambda,\lambda_1)_{N,3}^{N,3}}{R(\lambda,\lambda_1)_{N,2}^{N,2}}
-
\frac{R(\lambda,\lambda_1)_{N-1,3}^{N,2}}{R(\lambda,\lambda_1)_{N,2}^{N,2}}
\frac{R(\lambda,\lambda_2)_{N,1}^{N-1,2}}{R(\lambda,\lambda_2)_{N,1}^{N,1}}
\right].
\label{PbarN}
\EN

For sake of clarity we have divided the functions proportional to the unwanted terms
in two distinct
categories. The first class consists of those
that carry an internal discrete index dependence
${_c}{\cal H}_{b}^{(a)}(\lambda,\lambda_1,\lambda_2|1)$  or
${_c}{\cal H}_{b}^{(a)}(\lambda,\lambda_1,\lambda_2|2)$  where
$ 1 \le b \le 2,~~~
1 \le a \le N-b,~~~
b-1 \le c \le 1$.
The expressions for functions
${_c}{\cal H}_{b}^{(a)}(\lambda,\lambda_1,\lambda_2|1)$   are
\bear
\label{H11}
{_1}{\cal H}_{1}^{(a)}(\lambda,\lambda_1,\lambda_2|1)
&=&
\frac{R(\lambda_2,\lambda_1)^{1,1}_{1,1}}{R(\lambda_2,\lambda_1)^{2,1}_{2,1}}
{_1}{\cal F}_{1}^{(a)}(\lambda,\lambda_1)  ~~~~ \mbox{for} ~~ a=1,\dots,N-1
\\
\label{H01}
{_0}{\cal H}_{1}^{(a)}(\lambda,\lambda_1,\lambda_2|1)
&=&
P_2(\lambda_1,\lambda_2)
{_0}{\cal F}_{1}^{(a)}(\lambda,\lambda_1) ~~~~ \mbox{for} ~~ a=1,\dots,N-1
\\
\label{H12}
{_1}{\cal H}_{2}^{(a-1)}(\lambda,\lambda_1,\lambda_2|1)
&=&
{_1}{\cal F}_{1}^{(a)}(\lambda,\lambda_1)
{_0}{\cal H}_{1}^{(a-1)}(\lambda,\lambda_1,\lambda_2|2) ~~~~ \mbox{for} ~~ a=2,\dots,N-1
\ear

The explicit expressions for functions
${_c}{\cal H}_{b}^{(a)}(\lambda,\lambda_1,\lambda_2|2)$  are in general
very cumbersome. Fortunately,
thanks to the exchange symmetry of the two-particle vector (\ref{sym2}) there exists a
direct relationship  between  functions
${_c}{\cal H}_{b}^{(a)}(\lambda,\lambda_1,\lambda_2|2)$
and ${_c}{\cal H}_{b}^{(a)}(\lambda,\lambda_1,\lambda_2|1)$. In fact, by considering
the permutation
$\lambda_1 \leftrightarrow \lambda_2$ on the two-particle state result
(\ref{Aaaphi2}) we are able to  derive the following
consistency relation,
\EQ
{_c}{\cal H}_{b}^{(a)}(\lambda,\lambda_1,\lambda_2|2)
=
\theta(\lambda_1,\lambda_2)
{_c}{\cal H}_{b}^{(a)}(\lambda,\lambda_2,\lambda_1|1)
\label{symH}
\EN

However, in the course of our analysis we find that explicit expressions
for the following special families
${_1}{\cal H}_{1}^{(a)}(\lambda,\lambda_1,\lambda_2|2)$  and
${_0}{\cal H}_{1}^{(a+1)}(\lambda,\lambda_1,\lambda_2|2)$  for $a=1,\cdots,N-1$ are indeed
very relevant.  For this reason it is necessary to quote their expressions, namely
\bear
{_1}{\cal H}_{1}^{(1)}(\lambda,\lambda_1,\lambda_2|2)
&=&
\frac{R(\lambda_2,\lambda)^{2,1}_{1,2}}{R(\lambda_2,\lambda)^{2,1}_{2,1}}
P_2(\lambda_1,\lambda)
-
\frac{R(\lambda_1,\lambda)^{2,1}_{1,2}}{R(\lambda_1,\lambda)^{2,1}_{2,1}}
\frac{R(\lambda_2,\lambda_1)^{2,1}_{1,2}}{R(\lambda_2,\lambda_1)^{2,1}_{2,1}}
-
\frac{R(\lambda_1,\lambda_2)^{2,2}_{3,1}}{R(\lambda_1,\lambda_2)^{3,1}_{3,1}}
\frac{R(\lambda_1,\lambda)_{2,2}^{3,1}}{R(\lambda_1,\lambda)^{3,1}_{3,1}}
\label{H111}
\nonumber \\
\ear
\bear
{_1}{\cal H}_{1}^{(a)}(\lambda,\lambda_1,\lambda_2|2)
&=&
-
\left[
\frac{D_{2}^{(a,0)}(\lambda,\lambda_1)}{D_{2}^{(a+1,0)}(\lambda,\lambda_1)}
\frac{R(\lambda,\lambda_2)_{a+1,1}^{a,2}}{R(\lambda,\lambda_2)_{a+1,1}^{a+1,1}}
\frac{R(\lambda,\lambda_1)_{a,1}^{a,1}}{R(\lambda,\lambda_1)_{a+1,1}^{a+1,1}}
-\frac{R(\lambda,\lambda_1)_{a+1,1}^{a,2}}{R(\lambda,\lambda_1)_{a+1,1}^{a+1,1}}
\frac{R(\lambda_2,\lambda_1)_{1,2}^{2,1}}{R(\lambda_2,\lambda_1)_{2,1}^{2,1}}
\right.
\nonumber \\
&-&
\left.
\frac{R(\lambda_1,\lambda_2)_{3,1}^{2,2}}{R(\lambda_1,\lambda_2)_{3,1}^{3,1}}
\frac{\left|
\begin{array}{cc}
R(\lambda,\lambda_1)_{a+2,1}^{a,3} & R(\lambda,\lambda_1)_{a+1,2}^{a,3} \\
R(\lambda,\lambda_1)_{a+2,1}^{a+2,1} &
R(\lambda,\lambda_1)_{a+1,2}^{a+2,1}
\end{array} \right|}
{\left| \begin{array}{cc}
R(\lambda,\lambda_1)_{a+2,1}^{a+1,2} & R(\lambda,\lambda_1)_{a+1,2}^{a+1,2} \\
R(\lambda,\lambda_1)_{a+2,1}^{a+2,1} &
R(\lambda,\lambda_1)_{a+1,2}^{a+2,1}
\end{array} \right|}
\right]   ~~ \mbox{for} ~ a=2,\dots,N-2
\label{H11|2}
\ear
\bear
\label{H01|2}
{_0}{\cal H}_{1}^{(a)}(\lambda,\lambda_1,\lambda_2|2)
&=&
\left[
\frac{R(\lambda,\lambda_2)_{a+1,1}^{a,2}}{R(\lambda,\lambda_2)_{a+1,1}^{a+1,1}}
\frac{R(\lambda,\lambda_1)_{a,1}^{a,1}}{R(\lambda,\lambda_1)_{a+1,1}^{a+1,1}}
-\frac{R(\lambda,\lambda_1)_{a+1,1}^{a,2}}{R(\lambda,\lambda_1)_{a+1,1}^{a+1,1}}
\frac{R(\lambda_1,\lambda_2)_{2,1}^{1,2}}{R(\lambda_1,\lambda_2)_{2,1}^{2,1}}
\right].
\ear

The remaining functions to be defined are
${_0}{\cal F}{_2^{(a)}}(\lambda,\lambda_1,\lambda_2)$ and
${_2}{\cal F}{_2^{(a)}}(\lambda,\lambda_1,\lambda_2)$. Their explicit
expression in terms of the weights are,
\bear
\label{f20r}
{_0}{\cal F}{_2^{(a)}}(\lambda,\lambda_1,\lambda_2)
=
\frac{R(\lambda,\lambda_1)_{a+1,1}^{a,2}}{R(\lambda,\lambda_1)_{a+2,1}^{a+2,1}}
\frac{R(\lambda,\lambda_2)_{a+2, 1}^{a+1, 2}}{R(\lambda,\lambda_2)_{a+2, 1}^{ a+2, 1}}
-
\frac{R(\lambda,\lambda_1)_{a+2,1}^{a,3}}{R(\lambda,\lambda_1)_{a+2,1}^{a+2,1}}
\frac{R(\lambda_1,\lambda_2)_{3, 1}^{2, 2}}{R(\lambda_1,\lambda_2)_{3, 1}^{3, 1}}
\nonumber \\
\mbox{for} ~~ 1 \le a \le N-2
\ear

\bear
\label{f22a}
{_2}{\cal F}{_2^{(1)}}(\lambda,\lambda_1,\lambda_2)
&=&
\frac{R(\lambda_2,\lambda)_{1,2}^{2,1}}{R(\lambda_2,\lambda)_{2,1}^{2,1}}
D_2^{(2,1)}(\lambda_1,\lambda)
-
\frac{R(\lambda_1,\lambda)_{1,3}^{3,1}}{R(\lambda_1,\lambda)_{3,1}^{3,1}}
\frac{R(\lambda_1,\lambda_2)_{3, 1}^{2, 2}}{R(\lambda_1,\lambda_2)_{3, 1}^{3, 1}}
\\
\label{f22b}
{_2}{\cal F}{_2^{(a)}}(\lambda,\lambda_1,\lambda_2)
&=&
-
\left[
\frac{R(\lambda_1,\lambda_2)_{3,1}^{2,2}}{R(\lambda_1,\lambda_2)_{3,1}^{3,1}}
\frac{\left|
\begin{array}{cc}
R(\lambda,\lambda_1)_{a+2,1}^{a,3} & R(\lambda,\lambda_1)_{a+1,2}^{a,3} \\
R(\lambda,\lambda_1)_{a+2,1}^{a+1,2} &
R(\lambda,\lambda_1)_{a+1,2}^{a+1,2}
\end{array} \right|}
{\left| \begin{array}{cc}
R(\lambda,\lambda_1)_{a+2,1}^{a+1,2} & R(\lambda,\lambda_1)_{a+1,2}^{a+1,2} \\
R(\lambda,\lambda_1)_{a+2,1}^{a+2,1} &
R(\lambda,\lambda_1)_{a+1,2}^{a+2,1}
\end{array} \right|}
\right.
\nonumber \\
&-& \left.
\frac{D_{2}^{(a,0)}(\lambda,\lambda_1)}{D_{2}^{(a+1,0)}(\lambda,\lambda_1)}
\frac{R(\lambda,\lambda_2)_{a+1,1}^{a,2}}{R(\lambda,\lambda_2)_{a+1,1}^{a+1,1}}
\frac{R(\lambda,\lambda_1)_{a,1}^{a,1}}{R(\lambda,\lambda_1)_{a+1,1}^{a+1,1}}
\frac{R(\lambda,\lambda_1)_{a+2,1}^{a+1,2}}{R(\lambda,\lambda_1)_{a+2,1}^{a+2,1}}
\right]
\nonumber \\
&& ~~~~~~~~~~~~~~~~~~~~~~~~~~~~~~~~~~~~~~~~~~~ \mbox{for} ~~ 2 \le a \le N-2
\ear

To make progress on the structure of the two-particle problem we first investigate
the wanted terms. It is essential to assure that the functional form of the products
proportional to either $\ket{\Phi_2^{(1)}}$ or $\ket{\Phi_2^{(2)}}$ are exactly the same.
This means that we have to demonstrate the following relation,
\EQ
\bar{P}_a(\lambda,\lambda_1,\lambda_2)=P_a(\lambda,\lambda_2)
~~~~~\mbox{for} ~~ a=1,\dots, N.
\label{relap}
\EN

The factorization property (\ref{relap}) follows from a series of
identities we have derived in section (\ref{subI2}). The particular
cases $a=1$ and $a=N$  are an immediate consequence of the
Yang-Baxter equation. This is easily seen by comparing the
definitions of the corresponding
$\bar{P}_a(\lambda,\lambda_1,\lambda_2)$ given in
Eqs.(\ref{Pbar1},\ref{PbarN}) with the identities
(\ref{apAid3i},\ref{apAid4i}), respectively. In order to show
property (\ref{relap}) for the remaining cases $2 \le a \le N-1$ we
have to combine together the identities derived from both the
unitarity relation and the Yang-Baxter equation.  In order to show
Eq.(\ref{relap}) for
$2 \le a \le N-2$  we just have to replace the determinants
$D_4^{(i,b)}(\lambda,\lambda_1)$ and $D_5^{(i,2)}(\lambda,\lambda_1)$
in the Yang-Baxter identity (\ref{apAid9i})  by using the relations
(\ref{apAid7},\ref{apAid8}) derived from the unitarity property.
The proof for the case $a=N-1$ requires to consider the extension
of Eqs.(\ref{apAid7},\ref{apAid8}) when the index $i$ goes to
the value $N-1$. As a result of such analytical continuation we found,
\EQ \lim_{i
\rightarrow
N-1}\frac{D_2^{(i+1,1)}(\lambda,\lambda_1)}{D_2^{(i+1,0)}(\lambda,\lambda_1)}=
\frac{R(\lambda,\lambda_1)_{N-1,3}^{N,2} }{
R(\lambda,\lambda_1)_{N,2}^{N,2} }
=-\frac{D_5^{(N+1,2)}(\lambda,\lambda_1)}{D_4^{(N+1,3)}(\lambda,\lambda_1)}
\label{apAid7CAi} \EN
and
\bear \lim_{i \rightarrow N-1}
\frac{D_3^{(i,0)}(\lambda,\lambda_1)}{D_2^{(i,0)}(\lambda,\lambda_1)}&=&
\frac{\left|\begin{array}{cc} R(\lambda,\lambda_1)_{N,2}^{N-1,3} &
R(\lambda,\lambda_1)_{N,2}^{N,2} \\
R(\lambda,\lambda_1)_{N-1,3}^{N-1,3} &
R(\lambda,\lambda_1)_{N-1,3}^{N,2}
\end{array}\right| R(\lambda,\lambda_1)_{N,1}^{N,1}}{\left|\begin{array}{cc}
R(\lambda,\lambda_1)_{N,1}^{N-1,2} &
R(\lambda,\lambda_1)_{N,1}^{N,1} \\
R(\lambda,\lambda_1)_{N-1,2}^{N-1,2} &
R(\lambda,\lambda_1)_{N-1,2}^{N,1}
\end{array} \right|
R(\lambda,\lambda_1)_{N,2}^{N,2}}
\nonumber \\
&=&
\frac{D_4^{(N,2)}(\lambda,\lambda_1)}{D_4^{(N,3)}(\lambda,\lambda_1)}
\frac{D_4^{(N+1,4)}(\lambda,\lambda_1)}{D_4^{(N+1,3)}(\lambda,\lambda_1)}.
\label{apAid8CAi}
\ear

To complete the proof we have to consider the analytical continuation of
Eq.(\ref{apAid9i}) for $a=N-1$ by means of relations
(\ref{D4CAi},\ref{D5CAi}). The  replacement of the determinants
$D_4^{(i,b)}(\lambda,\lambda_1)$ and
$D_5^{(i,2)}(\lambda,\lambda_1)$ in this analytical extension of Eq.(\ref{apAid9i})
with the help of
Eqs.(\ref{apAid7},\ref{apAid7CAi},\ref{apAid8CAi}) leads us to
relation (\ref{relap}) for $a=N-1$.

We now turn our attention to the structure of the unwanted terms. In
order to obtain the result (\ref{comAaaPhi2geral}) presented in
section \ref{sub42} we first identify functions ${_1}{\cal
H}_{2}^{(a)}(\lambda,\lambda_1,\lambda_2|1)$ and ${_1}{\cal
F}_{2}^{(a)}(\lambda,\lambda_1,\lambda_2)$, \EQ {_1}{\cal
H}_{2}^{(a)}(\lambda,\lambda_1,\lambda_2|1)= {_1}{\cal
F}_{2}^{(a)}(\lambda,\lambda_1,\lambda_2) \label{relationFH} \EN

The next step  is to carry out some simplifications on functions
${_c}{\cal F}_{2}^{(a)}(\lambda,\lambda_1,\lambda_2)$ for $c=0,1,2$ proportional
to undesirable terms with spin $s=2$.  It is
possible to express these functions in terms of recurrence relations involving
the one-particle weights
${_0}{\cal F}_{1}^{(a)}(\lambda,\lambda_1,\lambda_2)$,
${_1}{\cal F}_{1}^{(a)}(\lambda,\lambda_1,\lambda_2)$  and certain $R$-matrix amplitudes. For
function
${_0}{\cal F}_{2}^{(a)}(\lambda,\lambda_1,\lambda_2)$ this follows directly by using the definition (\ref{psi1})
in Eq.(\ref{f20r}), namely
\EQ
{_0}{\cal F}{_2^{(a)}}(\lambda,\lambda_1,\lambda_2)=
\frac{R(\lambda,\lambda_1)_{a+1,1}^{a,2}}{R(\lambda,\lambda_1)_{a+2,1}^{a+2,1}}
{_0}{\cal F}_{1}^{(a+1)}(\lambda,\lambda_2)+
\frac{R(\lambda,\lambda_1)_{a+2,1}^{a,3}}{R(\lambda,\lambda_1)_{a+2,1}^{a+2,1}}
{_1}{\cal F}_{1}^{(2)}(\lambda_1,\lambda_2) .
\EN

In the case of
${_1}{\cal F}_{2}^{(a)}(\lambda,\lambda_1,\lambda_2)$ we have to use the identification (\ref{relationFH})
together with the  definitions (\ref{H01},\ref{H12}) as well as the property (\ref{symH}). After few
manipulations we find
\EQ
{_1}{\cal F}_{2}^{(a)}(\lambda,\lambda_1,\lambda_2)
=
{_0}{\cal F}_{1}^{(a)}(\lambda,\lambda_2)
{_1}{\cal F}_{1}^{(a+1)}(\lambda,\lambda_1)
\frac{R(\lambda_2,\lambda_1)_{1,1}^{1,1}}{R(\lambda_2,\lambda_1)_{2,1}^{2,1}}.
\label{simpF12}
\EN
where  we have used the relation
$R(\lambda_2,\lambda_1)_{2,1}^{2,1} P_2(\lambda_2,\lambda_1) \theta(\lambda_1,\lambda_2) =
R(\lambda_2,\lambda_1)_{1,1}^{1,1}$ which can be derived considering Eqs.(\ref{pa},\ref{thetasym},\ref{exch}).

The last simplification concerns to show that the definitions (\ref{f22a}, \ref{f22b}) for functions \newline ${_2}{\cal F}_{2}^{(a)}(\lambda,\lambda_1,\lambda_2)$ are equivalent to the
following expression,
\EQ
\label{F22simp}
{_2}{\cal F}{_2^{(a)}}(\lambda,\lambda_1,\lambda_2)
=
-{_0}{\cal F}{_2^{(a)}}(\lambda,\lambda_1,\lambda_2)
-
\sum_{\stackrel{i,j=1}{j \neq i}}^{2}
{_1}{\cal F}{_2^{(a)}}(\lambda,\lambda_i,\lambda_j) \frac{R(\lambda_j,\lambda_i)_{2,1}^{2,1}}{R(\lambda_j,\lambda_i)_{1,1}^{1,1}}
\frac{R(\lambda_i,\lambda_j)_{1,1}^{1,1}}{R(\lambda_i,\lambda_j)_{2,1}^{2,1}}
\prod_{k=1}^{j-1} \theta(\lambda_k, \lambda_j).
\EN

In order to show the equivalence between
Eqs.(\ref{f22a},\ref{f22b}) and
(\ref{F22simp}) it is necessary to use particular relations
coming from the symmetry relation
Eq.(\ref{symH}), namely
\bear
{_0}{\cal
H}_1^{(a+1)}(\lambda,\lambda_1,\lambda_2|2) =
\theta(\lambda_1,\lambda_2) {_0}{\cal
H}_1^{(a+1)}(\lambda,\lambda_2,\lambda_1|1) ~~~~\mbox{for} ~~1 \le
a \le N-2 \label{h01a+1}
\\
%\mbox{and}
\nonumber \\
{_1}{\cal H}_1^{(a)}(\lambda,\lambda_1,\lambda_2|2) =
\theta(\lambda_1,\lambda_2) {_1}{\cal
H}_1^{(a)}(\lambda,\lambda_2,\lambda_1|1) ~~~~\mbox{for} ~~1 \le a
\le N-2. \label{h11a} \ear

We first substitute the expressions for these functions given by
Eqs.(\ref{H11},\ref{H111},\ref{H11|2}) in the relation (\ref{h11a}).
We then reorder the corresponding amplitude ratios of type
$\frac{R(\lambda,\mu)_{1,2}^{2,1}}{R(\lambda,\mu)_{2,1}^{2,1}}$ for
$(\lambda,\mu)=(\lambda_2,\lambda_1)$ and
$(\lambda,\mu)=(\lambda_1,\lambda)$ with the help of identity
(\ref{apAid1}). As a result we obtain,\bear
\frac{R(\lambda_2,\lambda)_{1,2}^{2,1}}{R(\lambda_2,\lambda)_{2,1}^{2,1}}
P_2(\lambda_1,\lambda) -
\frac{R(\lambda,\lambda_1)_{2,1}^{1,2}}{R(\lambda,\lambda_1)_{2,1}^{2,1}}
\frac{R(\lambda_1,\lambda_2)_{2,1}^{1,2}}{R(\lambda_1,\lambda_2)_{2,1}^{2,1}}
-
\frac{R(\lambda_1,\lambda_2)_{3,1}^{2,2}}{R(\lambda_1,\lambda_2)_{3,1}^{3,1}}
\frac{R(\lambda_1,\lambda)_{2,2}^{3,1}}{R(\lambda_1,\lambda)_{3,1}^{3,1}}
\nonumber \\
= \theta(\lambda_1,\lambda_2)
\frac{R(\lambda_1,\lambda_2)_{1,1}^{1,1}}{R(\lambda_1,\lambda_2)_{2,1}^{2,1}}
%\times
{_1}{\cal F}_1^{(1)}(\lambda,\lambda_2) ~~~~\mbox{for}~~a=1
\nonumber \\
\label{h111i} \ear \bear -
\frac{D_{2}^{(a,0)}(\lambda,\lambda_1)}{D_{2}^{(a+1,0)}(\lambda,\lambda_1)}
{_0}{\cal F}_1^{(a)}(\lambda,\lambda_2)
\frac{R(\lambda,\lambda_1)_{a,1}^{a,1}}{R(\lambda,\lambda_1)_{a+1,1}^{a+1,1}}
- {_0}{\cal F}_1^{(a)}(\lambda,\lambda_1)
\frac{R(\lambda_1,\lambda_2)_{2,1}^{1,2}}{R(\lambda_1,\lambda_2)_{2,1}^{2,1}}
\nonumber \\
+
\frac{R(\lambda_1,\lambda_2)_{3,1}^{2,2}}{R(\lambda_1,\lambda_2)_{3,1}^{3,1}}
\frac{\left|
\begin{array}{cc}
R(\lambda,\lambda_1)_{a+2,1}^{a,3} & R(\lambda,\lambda_1)_{a+1,2}^{a,3} \\
R(\lambda,\lambda_1)_{a+2,1}^{a+2,1} &
R(\lambda,\lambda_1)_{a+1,2}^{a+2,1}
\end{array} \right|}
{\left| \begin{array}{cc}
R(\lambda,\lambda_1)_{a+2,1}^{a+1,2} & R(\lambda,\lambda_1)_{a+1,2}^{a+1,2} \\
R(\lambda,\lambda_1)_{a+2,1}^{a+2,1} &
R(\lambda,\lambda_1)_{a+1,2}^{a+2,1}
\end{array} \right|}
= \theta(\lambda_1,\lambda_2)
\frac{R(\lambda_1,\lambda_2)_{1,1}^{1,1}}{R(\lambda_1,\lambda_2)_{2,1}^{2,1}}
{_1}{\cal F}_1^{(a)}(\lambda,\lambda_2)
\nonumber \\
~~~~~~~~\mbox{for}~~ 2 \le a \le N-2.
\label{h11ai}
\ear

We next multiply Eq.(\ref{h11ai}) by function ${_0}{\cal
F}_1^{(a+1)}(\lambda,\lambda_1)$ and Eq.(\ref{h01a+1}) by function $
{_0}{\cal F}_1^{(a)}(\lambda,\lambda_1)$.  By subtracting the former
equation from the latter equation and by considering the explicit
expressions for ${_0}{\cal
H}_1^{(a+1)}(\lambda,\lambda_1,\lambda_2|2)$ given in
Eq.(\ref{H01|2})  we found,
\bear
\lefteqn{{_0}{\cal
F}_1^{(a)}(\lambda,\lambda_1) {_0}{\cal
F}_1^{(a+1)}(\lambda,\lambda_2)
\frac{R(\lambda,\lambda_1)_{a+1,1}^{a+1,1}}{R(\lambda,\lambda_1)_{a+2,1}^{a+2,1}}
+
\frac{D_{2}^{(a,0)}(\lambda,\lambda_1)}{D_{2}^{(a+1,0)}(\lambda,\lambda_1)}
{_0}{\cal F}_1^{(a)}(\lambda,\lambda_2) {_0}{\cal
F}_1^{(a+1)}(\lambda,\lambda_1)}
\nonumber \\
&\times&
\frac{R(\lambda,\lambda_1)_{a,1}^{a,1}}{R(\lambda,\lambda_1)_{a+1,1}^{a+1,1}}
-
\frac{R(\lambda_1,\lambda_2)_{3,1}^{2,2}}{R(\lambda_1,\lambda_2)_{3,1}^{3,1}}
{_0}{\cal F}_1^{(a+1)}(\lambda,\lambda_1) \frac{\left|
\begin{array}{cc}
R(\lambda,\lambda_1)_{a+2,1}^{a,3} & R(\lambda,\lambda_1)_{a+1,2}^{a,3} \\
R(\lambda,\lambda_1)_{a+2,1}^{a+2,1} &
R(\lambda,\lambda_1)_{a+1,2}^{a+2,1}
\end{array} \right|}
{\left| \begin{array}{cc}
R(\lambda,\lambda_1)_{a+2,1}^{a+1,2} & R(\lambda,\lambda_1)_{a+1,2}^{a+1,2} \\
R(\lambda,\lambda_1)_{a+2,1}^{a+2,1} &
R(\lambda,\lambda_1)_{a+1,2}^{a+2,1}
\end{array} \right|}
= {_0}{\cal F}_1^{(a)}(\lambda,\lambda_1)
\nonumber \\
&\times&
{_0}{\cal F}_1^{(a+1)}(\lambda,\lambda_2)
\frac{R(\lambda_2,\lambda_1)_{1,1}^{1,1}}{R(\lambda_2,\lambda_1)_{2,1}^{2,1}}
- \theta(\lambda_1,\lambda_2)
\frac{R(\lambda_1,\lambda_2)_{1,1}^{1,1}}{R(\lambda_1,\lambda_2)_{2,1}^{2,1}}
{_1}{\cal F}_1^{(a)}(\lambda,\lambda_2) {_0}{\cal
F}_1^{(a+1)}(\lambda,\lambda_1). \label{fb22demons}
\ear

After few manipulations in Eq.(\ref{fb22demons}) we are able to write it as,
\bear
&&\frac{D_{2}^{(a,0)}(\lambda,\lambda_1)}{D_{2}^{(a+1,0)}(\lambda,\lambda_1)}
{_0}{\cal F}_1^{(a)}(\lambda,\lambda_2) {_0}{\cal
F}_1^{(a+1)}(\lambda,\lambda_1)
\frac{R(\lambda,\lambda_1)_{a,1}^{a,1}}{R(\lambda,\lambda_1)_{a+1,1}^{a+1,1}}
- \frac{\left|
\begin{array}{cc}
R(\lambda,\lambda_1)_{a+2,1}^{a,3} & R(\lambda,\lambda_1)_{a+1,2}^{a,3} \\
R(\lambda,\lambda_1)_{a+2,1}^{a+1,2} &
R(\lambda,\lambda_1)_{a+1,2}^{a+1,2}
\end{array} \right|}
{\left| \begin{array}{cc}
R(\lambda,\lambda_1)_{a+2,1}^{a+1,2} & R(\lambda,\lambda_1)_{a+1,2}^{a+1,2} \\
R(\lambda,\lambda_1)_{a+2,1}^{a+2,1} &
R(\lambda,\lambda_1)_{a+1,2}^{a+2,1}
\end{array} \right|}
\nonumber \\
&& \times
\frac{R(\lambda_1,\lambda_2)_{3,1}^{2,2}}{R(\lambda_1,\lambda_2)_{3,1}^{3,1}}
=
\frac{R(\lambda_1,\lambda_2)_{3,1}^{2,2}}{R(\lambda_1,\lambda_2)_{3,1}^{3,1}}
\frac{R(\lambda,\lambda_1)_{a+2,1}^{a,3}}{R(\lambda,\lambda_1)_{a+2,1}^{a+2,1}}
- {_0}{\cal F}_1^{(a+1)}(\lambda,\lambda_2)
\frac{R(\lambda,\lambda_1)_{a+1,1}^{a,2}}{R(\lambda,\lambda_1)_{a+2,1}^{a+2,1}}
- {_0}{\cal F}_1^{(a)}(\lambda,\lambda_1)
\nonumber \\
&& \times {_1}{\cal F}_1^{(a+1)}(\lambda,\lambda_2)
\frac{R(\lambda_2,\lambda_1)_{1,1}^{1,1}}{R(\lambda_2,\lambda_1)_{2,1}^{2,1}}
- \theta(\lambda_1,\lambda_2)
\frac{R(\lambda_1,\lambda_2)_{1,1}^{1,1}}{R(\lambda_1,\lambda_2)_{2,1}^{2,1}}
{_0}{\cal F}_1^{(a)}(\lambda,\lambda_2) {_1}{\cal
F}_1^{(a+1)}(\lambda,\lambda_1).
\label{fb22demonsi}
\ear
which in fact shows that
Eqs.(\ref{f22b},\ref{F22simp}) are equivalent once we consider Eq.(\ref{simpF12}).

Finally, it remains to show that
Eq.(\ref{F22simp}) for $a=1$ is equivalent to
Eq.(\ref{f22a}). In this case, we multiply
Eq.(\ref{h01a+1}) for a=1  by
${_0}{\cal F}_1^{(1)}(\lambda,\lambda_1)$  and
Eq.(\ref{h111i}) by
${_0}{\cal
F}_1^{(2)}(\lambda,\lambda_1)$. By subtracting these equations we found,
\bear
\lefteqn{{_0}{\cal
F}_1^{(2)}(\lambda,\lambda_2) {_0}{\cal
F}_1^{(1)}(\lambda,\lambda_1)
\frac{R(\lambda,\lambda_1)_{2,1}^{2,1}}{R(\lambda,\lambda_1)_{3,1}^{3,1}}
-
\frac{R(\lambda_2,\lambda)_{1,2}^{2,1}}{R(\lambda_2,\lambda)_{2,1}^{2,1}}
P_2(\lambda_1,\lambda) {_0}{\cal F}_1^{(2)}(\lambda,\lambda_1)}
\nonumber \\
&+&
{_0}{\cal F}_1^{(2)}(\lambda_1,\lambda_2)
{_0}{\cal F}_1^{(2)}(\lambda,\lambda_1)
\frac{R(\lambda_1,\lambda)_{2,2}^{3,1}}{R(\lambda_1,\lambda)_{3,1}^{3,1}}
=
\frac{R(\lambda_2,\lambda_1)_{1,1}^{1,1}}{R(\lambda_2,\lambda_1)_{2,1}^{2,1}}
{_0}{\cal F}_1^{(2)}(\lambda,\lambda_2) {_0}{\cal
F}_1^{(1)}(\lambda,\lambda_1)
\nonumber \\
&-& \theta(\lambda_1,\lambda_2)
\frac{R(\lambda_1,\lambda_2)_{1,1}^{1,1}}{R(\lambda_1,\lambda_2)_{2,1}^{2,1}}
{_1}{\cal F}_1^{(1)}(\lambda,\lambda_2) {_0}{\cal
F}_1^{(2)}(\lambda,\lambda_1).
\label{fb22demonsii}
\ear

We have now to eliminate the terms $P_2(\lambda_1,\lambda)$ and
${_0}{\cal F}_1^{(2)}(\lambda,\lambda_1)
\frac{R(\lambda_1,\lambda)_{2,2}^{3,1}}{R(\lambda_1,\lambda)_{3,1}^{3,1}}$
from Eq.(\ref{fb22demonsii}). This step is implemented with the help
of two identities derived from the unitarity relation.  More
specifically, these are the identity (\ref{uni1ans2}) and that
coming from $U[1,3]_1^3$ of Eq.(\ref{uniai}), namely
\EQ
R(\lambda,\mu)_{3,1}^{1,3} R(\mu,\lambda)_{3,1}^{3,1} +
R(\lambda,\mu)_{3,1}^{2,2} R(\mu,\lambda)_{2,2}^{3,1} +
R(\lambda,\mu)_{3,1}^{3,1} R(\mu,\lambda)_{1,3}^{3,1} = 0.
\label{uniq2}
\EN and as result we obtain,
\bear
\lefteqn{\frac{R(\lambda_2,\lambda)_{1,2}^{2,1}}{R(\lambda_2,\lambda)_{2,1}^{2,1}}
D_2^{(2,1)}(\lambda_1,\lambda) + {_1}{\cal
F}_1^{(2)}(\lambda_1,\lambda_2)
\frac{R(\lambda_1,\lambda)_{1,3}^{3,1}}{R(\lambda_1,\lambda)_{3,1}^{3,1}}
= -
\frac{R(\lambda_2,\lambda_1)_{1,1}^{1,1}}{R(\lambda_2,\lambda_1)_{2,1}^{2,1}}
{_0}{\cal F}_1^{(2)}(\lambda,\lambda_2) {_1}{\cal
F}_1^{(1)}(\lambda,\lambda_1)}
\nonumber \\
&-& \theta(\lambda_1,\lambda_2)
\frac{R(\lambda_1,\lambda_2)_{1,1}^{1,1}}{R(\lambda_1,\lambda_2)_{2,1}^{2,1}}
{_1}{\cal F}_1^{(1)}(\lambda,\lambda_2) {_0}{\cal
F}_1^{(2)}(\lambda,\lambda_1) - {_0}{\cal
F}_1^{(2)}(\lambda,\lambda_2)
\frac{R(\lambda,\lambda_1)_{2,1}^{1,2}}{R(\lambda,\lambda_1)_{3,1}^{3,1}}
\nonumber \\
&+& {_0}{\cal F}_1^{(2)}(\lambda_1,\lambda_2)
\frac{R(\lambda,\lambda_1)_{3,1}^{1,3}}{R(\lambda,\lambda_1)_{3,1}^{3,1}}.
\label{fb22demonsiii}
\ear
which shows the equivalence of
Eqs.(\ref{f22a},\ref{F22simp}) for $a=1$.

Putting all these results together we will find the final results
(\ref{comAaaPhi2geral}) presented in section (\ref{sub42}).

\addcontentsline{toc}{section}{Appendix B}
\section*{\bf Appendix B: The three-particle state}
\setcounter{equation}{0}
\renewcommand{\theequation}{B.\arabic{equation}}

In this appendix we describe some of the technical details  used to solve the
three-particle problem. The steps necessary for the last two terms of the
three-particle vector (\ref{ans3s}) have already been discussed in the text of
section \ref{sub43}. Therefore we shall concentrate our attention on the most
intricate part of the calculations which is associated to the first term
${\cal T}_{1, 2}(\lambda_1)\phi_{2}(\lambda_2,\lambda_3)$. In this case we have
to carry the operators ${\cal T}_{a,a}(\lambda)$ through the state
$\ket{\Phi_3^{(1)}} = {\cal T}_{1, 2}(\lambda_1)\phi_{2}(\lambda_2,\lambda_3)\ket{0}$.
The first step is done
with the help of
Eqs.(\ref{1part1}-\ref{1partN}) and as result we obtain,
\bear
{\cal T}_{1,1}(\lambda) \ket{\Phi_3^{(1)}}
& = &
\left[
\frac{R(\lambda_1,\lambda)^{1,1}_{1,1}}
{R(\lambda_1,\lambda)^{2,1}_{2,1}}
{\cal T}_{1,2}(\lambda_1) {\cal T}_{1,1}(\lambda)
-\frac{R(\lambda_1,\lambda)_{1, 2}^{2, 1}}{R(\lambda_1,\lambda)_{2, 1}^{2, 1}}
{\cal T}_{1,2}(\lambda) {\cal T}_{1,1}(\lambda_1)
\right]
\phi_2(\lambda_2,\lambda_3) \ket{0},
\nonumber \\
&&
\label{3part112}
\ear
\bear
\lefteqn{{\cal T}_{a,a}(\lambda) \ket{\Phi_3^{(1)}}
=
\left[
D_{2}^{(a,0)}(\lambda,\lambda_1){\cal T}_{1,2}(\lambda_1) {\cal T}_{a,a}(\lambda)
+
\sum_{\bar{e}=3}^{a+1}
D_{2}^{(a,\bar{e}-2)}(\lambda,\lambda_1)
{\cal T}_{1,\bar{e}}(\lambda_1) {\cal T}_{a,a+2-\bar{e}}(\lambda)
\right.}
\nonumber \\
&+& \sum_{\bar{e}=1}^{a}
\frac{R(\lambda,\lambda_1)_{a+1,1}^{a,2}}{R(\lambda,\lambda_1)_{a+1,1}^{a+1,1}}
\frac{R(\lambda,\lambda_1)_{a, 1}^{\bar{e}, a-\bar{e}+1}}{R(\lambda,\lambda_1)_{a, 1}^{a, 1}}
{\cal T}_{\bar{e},a+1}(\lambda) {\cal T}_{a-\bar{e}+1,1}(\lambda_1)
\nonumber \\
&-&
\left.
\sum_{\bar{e}=1}^{a-1}
\frac{R(\lambda,\lambda_1)_{a, 1}^{\bar{e}, a-\bar{e}+1}}{R(\lambda,\lambda_1)_{a,1}^{a,1}}
{\cal T}_{\bar{e},a}(\lambda) {\cal T}_{a-\bar{e}+1,2}(\lambda_1)
\right]
\phi_2(\lambda_2,\lambda_3) \ket{0},
~~~~~\mathrm{for}~~~
2 \le a \le N-1,
\label{3parta12}
\ear
\bear
{\cal T}_{N, N}(\lambda) \ket{\Phi_3^{(1)}}
&=&
\left[
\frac{R(\lambda,\lambda_1)_{N,2}^{N, 2}}{R(\lambda,\lambda_1)_{N, 1}^{N, 1}}
{\cal T}_{1, 2}(\lambda_1)
{\cal T}_{N, N}(\lambda)
+ \sum_{\bar{e}=3}^{N}
\frac{R(\lambda,\lambda_1)_{N+2-\bar{e}, \bar{e}}^{N, 2}}{R(\lambda,\lambda_1)_{N, 1}^{N, 1}}
{\cal T}_{1, \bar{e}}(\lambda_1)
{\cal T}_{N, N+2-\bar{e}}(\lambda)
\right.
\nonumber \\
&-&
\left.
\sum_{\bar{e}=1}^{N-1}
\frac{R(\lambda,\lambda_1)_{N, 1}^{\bar{e}, N-\bar{e}+1}}{R(\lambda,\lambda_1)_{N, 1}^{N, 1}}
{\cal T}_{\bar{e}, N}(\lambda)
{\cal T}_{N-\bar{e}+1, 2}(\lambda_1)
\right]
\phi_2(\lambda_2,\lambda_3) \ket{0}.
\label{3partN12}
\ear

By examining Eqs.(\ref{3part112}-\ref{3partN12}) we conclude that
our next problem is to compute the action of operators ${\cal
T}_{d+a, a}(\lambda)$ with $d=0,\dots,N-a$ on the two-particle state
$\phi_2(\lambda_2,\lambda_3) \ket{0}$. The main stages to
disentangle  this problem is as follows. For the diagonal fields
$d=0$ this task is implemented by using the results
(\ref{comAaaPhi2geral}) for the two-particle state. This operation,
however, is able to produce products of creation operators of the
form ${\cal T}_{1, 2}(\lambda_1) {\cal T}_{a_1, a_1+1}(\lambda)
{\cal T}_{1, 2}(\lambda_j)\ket{0}$ with $j=2,3$ and ${\cal T}_{1,
2}(\lambda_1) {\cal T}_{a_2, a_2+2}(\lambda)\ket{0}$ for
$a_i=a-i,\dots,a$ that need to be reordered as far as the rapidity
$\lambda$ is concerned. The product of the creation fields is sorted
out by commuting the operators ${\cal T}_{1, 2}(\lambda_1)$ and
${\cal T}_{a_1, a_1+1}(\lambda)$ by using
Eqs.(\ref{b12ba-1a},\ref{b12bN-1N}). After that we carry on the
diagonal and annihilation operators through the field ${\cal T}_{1,
2}(\lambda_j)$ with the help of Eq.(\ref{aniq0},\ref{1partsimp}).
These two steps together are able to generate the following product
of creation fields ${\cal T}_{1, 3}(\lambda_1) {\cal T}_{a_1,
a_1+1}(\lambda)$ which also has a wrong order on the rapidities
$\lambda$ and $\lambda_1$. This term as far as ${\cal T}_{1,
2}(\lambda_1) {\cal T}_{a_2, a_2+2}(\lambda)\ket{0}$ are fortunately
fixed by another set of commutation rules defined by
Eq.(\ref{eq37}) as well as by the linear system of equations (\ref{creation1},\ref{creation2}) with $d_1=0$ and $b_1=3$.
Putting together all the above steps we are finally able to reorder the products of
creation fields in a suitable manner.

We now turn to the computations involving ${\cal T}_{d+a,
a}(\lambda) \phi_2(\lambda_2,\lambda_3) \ket{0}$ for $d=1,\cdots,
N-a$. To this end we shall first discuss the cases where $d \ge 3$.
In this situation the computations are somehow simpler since the
azimuthal spin of  the operator ${\cal T}_{d+a, a}(\lambda)$ exceeds
that associated to the two-particle state
$\phi_2(\lambda_2,\lambda_3) \ket{0}$. In fact, by using the
commutation rules
(\ref{aniq0},\ref{saniq131},\ref{comT11Taa+1},\ref{1partsimp}) we
can reduce the corresponding calculations to the action of several
annihilators on the reference state $\ket{0}$ and consequently we
find, \EQ \label{comTa+daPhi2} {\cal T}_{d+a, a}(\lambda)
\phi_2(\lambda_2,\lambda_3)\ket{0}=0 ~~\mbox{for}~~~d \ge 3, \EN

The results for $d=1$ and $d=2$ are obtained after a number of extra steps are
performed. First we have to commute the operators
${\cal T}_{a+1, a}(\lambda)$  and
${\cal T}_{a+2, a}(\lambda)$  with the fields
${\cal T}_{1,2}(\lambda_j)$   for $j=2,3$ present on the first part of
the state
$\phi_2(\lambda_2,\lambda_3) \ket{0}$. This task is accomplished with the help of
Eqs.(\ref{aniq0},\ref{1partsimp}) which generates the following type
of products
${\cal T}_{1, 1}(\lambda_2) {\cal T}_{a, a+1}(\lambda) \ket{0}$,
${\cal T}_{1, 1}(\lambda_2) {\cal T}_{a+1, a+2}(\lambda) \ket{0}$,
${\cal T}_{a, a}(\lambda) {\cal T}_{2, 3}(\lambda_2) \ket{0}$ and
${\cal T}_{a+1, a}(\lambda) {\cal T}_{2, 3}(\lambda_2) \ket{0}$.
We note that these terms possess either a diagonal or an annihilation
operator that still need to be carried out to the right-hand side. This
operation for the first two terms is done by using
Eq.(\ref{comT11Taa+1}). For the last two terms we have to employ
Cramer's rule in  Eq.(\ref{aniq132}) to compute the commutation
rules among the fields
${\cal T}_{a, a}(\lambda)$  and
${\cal T}_{a+1, a}(\lambda)$  with
${\cal T}_{2, 3}(\mu) $. The last step concerns with the commutations
among the operators
${\cal T}_{a+1, a}(\lambda)$,
${\cal T}_{a+2, a}(\lambda)$   and
${\cal T}_{1,3}(\lambda_2)$. This is easily performed with the help of
of the commutation rules described at end of section \ref{sub33} by Eqs.(\ref{saniq131}-\ref{sist2x2T21TN-1N+1}).
Collecting together
all the above mentioned steps we find that,
\bear
\label{comTa+2aPhi2}
\lefteqn{{\cal T}_{a+2, a}(\lambda) \phi_2(\lambda_2,\lambda_3)\ket{0}
=
w_{a+2}(\lambda) w_{1}(\lambda_2) w_{1}(\lambda_3)
{_0}{\cal F}_2^{(a)}(\lambda,\lambda_2,\lambda_3) \ket{0}}
\nonumber \\
&+&
w_{a+1}(\lambda) w_{2}(\lambda_2) w_{1}(\lambda_3)
{_1}{\cal F}_2^{(a)}(\lambda,\lambda_2,\lambda_3)
\frac{R(\lambda_2,\lambda_3)_{1, 1}^{1,1}}{R(\lambda_2,\lambda_3)_{2, 1}^{2, 1}}
\frac{R(\lambda_3,\lambda_2)_{2, 1}^{2, 1}}{R(\lambda_3,\lambda_2)_{1, 1}^{1,1}}
\theta(\lambda_2,\lambda_3) \ket{0}
\nonumber \\
&-&
w_{a+1}(\lambda) w_{1}(\lambda_2) w_{2}(\lambda_3)
{_0}{\cal F}_1^{(a)}(\lambda,\lambda_2)
{_0}{\cal H}_1^{(a+1)}(\lambda,\lambda_2,\lambda_3|2) \ket{0}
\nonumber \\
&+& w_{a}(\lambda) w_{2}(\lambda_2) w_{2}(\lambda_3) {_2}{\cal
F}_2^{(a)}(\lambda,\lambda_2,\lambda_3|1) \ket{0}, ~~\mbox{for}~~ a
\le N-2 \ear

\bear \label{comTa+1aPhi2} \lefteqn{{\cal T}_{a+1, a}(\lambda)
\phi_2(\lambda_2,\lambda_3)\ket{0} = {\cal T}_{1, 2}(\lambda_3)
{_0}{\cal F}_1^{(a)}(\lambda,\lambda_2) \left[ w_{a+1}(\lambda)
w_1(\lambda_2) P_1(\lambda_2,\lambda_3) P_{a+1}(\lambda,\lambda_3)
\right.}
\nonumber \\
&-&
\left.
w_{a}(\lambda)
w_2(\lambda_2)
P_2(\lambda_2,\lambda_3)
P_{a}(\lambda,\lambda_3)
\right] \ket{0}
\nonumber \\
&+&
{\cal T}_{1, 2}(\lambda_2)
\left[
w_{a+1}(\lambda)
w_1(\lambda_3)
P_{1,a+1}(\lambda,\lambda_2,\lambda_3)
\right.
\nonumber \\
&-&
\left.
w_{a}(\lambda)
w_2(\lambda_3)
P_{2,a}(\lambda,\lambda_2,\lambda_3)
\right] \ket{0}
\nonumber \\
&+&
{\cal T}_{a+1, a+2}(\lambda)
w_1(\lambda_2)
w_1(\lambda_3)
{_0}{\cal F}_2^{(a)}(\lambda,\lambda_2,\lambda_3) \ket{0}
\nonumber \\
&-&
{\cal T}_{a, a+1}(\lambda)
{_0}{\cal F}_1^{(a)}(\lambda,\lambda_2)
\left[
w_1(\lambda_2)
w_2(\lambda_3)
{_0}{\cal H}_1^{(a)}(\lambda,\lambda_2,\lambda_3|2)
\right.
\nonumber \\
&+&
\left.
w_2(\lambda_2)
w_1(\lambda_3)
{_0}{\cal F}_1^{(a)}(\lambda,\lambda_3)
\frac{R(\lambda_2,\lambda_3)_{1, 1}^{1,1}}{R(\lambda_2,\lambda_3)_{2, 1}^{2, 1}}
\theta(\lambda_2,\lambda_3)
\right] \ket{0}
\nonumber \\
&+& \bar{\delta}_{a}^{1} {\cal T}_{a-1, a}(\lambda) w_2(\lambda_2)
w_2(\lambda_3) {_2}{\cal F}_2^{(a-1)}(\lambda,\lambda_2,\lambda_3)
\ket{0}, ~~\mbox{for}~~ a \le N-1 \ear where functions ${_2}{\cal
F}_2^{(a)}(\lambda,\lambda_2,\lambda_3|1)$,
$P_{1,a+1}(\lambda,\lambda_2,\lambda_3)$ and
$P_{2,a+2}(\lambda,\lambda_2,\lambda_3)$  are given by, \bear
{_2}{\cal F}_2^{(a)}(\lambda,\lambda_2,\lambda_3|1) &=&
\frac{R(\lambda,\lambda_2)_{a+2,
1}^{a,3}}{R(\lambda,\lambda_2)_{a+2, 1}^{a+2, 1}} {_0}{\cal
F}_1^{(2)}(\lambda_2,\lambda_3) + P_2(\lambda_2,\lambda_3) {_0}{\cal
F}_1^{(a+1)}(\lambda,\lambda_2) {_0}{\cal
F}_1^{(a)}(\lambda,\lambda_3)
\nonumber \\
&-& {_0}{\cal F}_1^{(a+1)}(\lambda,\lambda_2) {_0}{\cal
F}_1^{(a)}(\lambda,\lambda_2) {_0}{\cal
F}_1^{(1)}(\lambda_2,\lambda_3),~~ \mbox{for} ~~ a \le N-2 \ear
\bear P_{1,a+1}(\lambda,\lambda_2,\lambda_3) =
P_{a+1}(\lambda,\lambda_3) {_0}{\cal F}_1^{(a)}(\lambda,\lambda_2)
{_0}{\cal F}_1^{(1)}(\lambda_2,\lambda_3) - {_0}{\cal
F}_1^{(a+1)}(\lambda,\lambda_3) {_0}{\cal
F}_1^{(a)}(\lambda,\lambda_2)
\nonumber \\
\times \frac{R(\lambda,\lambda_2)_{a+1, 2}^{a+2,1}}{R(\lambda,\lambda_2)_{a+2, 1}^{a+2, 1}}
+
\frac{R(\lambda,\lambda_2)_{a, 2}^{a, 2}}{R(\lambda,\lambda_2)_{a+1, 1}^{a+1, 1}}
{_0}{\cal F}_1^{(a)}(\lambda,\lambda_3)
-
{_1}{\cal F}_1^{(2)}(\lambda_2,\lambda_3)
\frac{\left|
\begin{array}{cc}
R(\lambda,\lambda_2)_{a+2,1}^{a,3} &
R(\lambda,\lambda_2)_{a+1,2}^{a,3} \\
R(\lambda,\lambda_2)_{a+2,1}^{a+2,1} &
R(\lambda,\lambda_2)_{a+1,2}^{a+2,1}
\end{array} \right|}
{R(\lambda,\lambda_2)_{a+2,1}^{a+2,1} R(\lambda,\lambda_2)_{a+1,1}^{a+1,1}}
\ear \bear
P_{2,a}(\lambda,\lambda_2,\lambda_3)
&=&
P_{a}(\lambda,\lambda_2)
{_0}{\cal F}_1^{(a)}(\lambda,\lambda_2)
{_0}{\cal F}_1^{(1)}(\lambda_2,\lambda_3)
+
{_0}{\cal F}_1^{(a)}(\lambda,\lambda_3)
{_0}{\cal F}_1^{(a)}(\lambda,\lambda_2)
\frac{R(\lambda,\lambda_2)_{a, 2}^{a+1,1}}{R(\lambda,\lambda_2)_{a+1, 1}^{a+1, 1}}
\nonumber \\
&-& \frac{R(\lambda,\lambda_2)_{a, 2}^{a,
2}}{R(\lambda,\lambda_2)_{a+1, 1}^{a+1, 1}} {_0}{\cal
F}_1^{(a)}(\lambda,\lambda_3),~~ \mbox{for} ~~ a \le N-1. \ear

It turns out that the above functions can be further simplified as follows. Considering
Eqs. (\ref{H01|2}, \ref{f20r}, \ref{simpF12}) we are able to rewrite ${_2}{\cal F}_2^{(a)}(\lambda,\lambda_2,\lambda_3|1)$ as,
\bear
{_2}{\cal F}_2^{(a)}(\lambda,\lambda_2,\lambda_3|1)
&=&
-
{_0}{\cal F}_2^{(a)}(\lambda,\lambda_2,\lambda_3)
-
{_1}{\cal F}_2^{(a)}(\lambda,\lambda_2,\lambda_3)
\frac{R(\lambda_2,\lambda_3)_{1, 1}^{1,1}}{R(\lambda_2,\lambda_3)_{2, 1}^{2, 1}}
\frac{R(\lambda_3,\lambda_2)_{2, 1}^{2, 1}}{R(\lambda_3,\lambda_2)_{1, 1}^{1,1}}
\theta(\lambda_2,\lambda_3)
\nonumber \\
&+& {_0}{\cal F}_1^{(a)}(\lambda,\lambda_2) {_0}{\cal
H}_1^{(a+1)}(\lambda,\lambda_2,\lambda_3|2). \label{F22|1} \ear

By using the Eq.(\ref{H01},\ref{symH}) and then the direct comparison between
Eq.(\ref{F22simp}) and Eq.(\ref{F22|1}) leads us to the identity,
\EQ {_2}{\cal F}_2^{(a)}(\lambda,\lambda_2,\lambda_3|1)={_2}{\cal
F}_2^{(a)}(\lambda,\lambda_2,\lambda_3). \label{idFF} \EN

Next, by exploring the permutation property of the two-particle
vector $\phi_2(\lambda_2,\lambda_3)$ (\ref{sym2}) it is not
difficult to conclude that the right-hand side of
Eqs.(\ref{comTa+2aPhi2},\ref{comTa+1aPhi2}) must be  symmetrical
under the rapidity exchange $\lambda_2 \leftrightarrow \lambda_3$
and therefore we find that, \bear
P_{1,a+1}(\lambda,\lambda_2,\lambda_3)=\theta(\lambda_2,\lambda_3)
P_1(\lambda_3,\lambda_2) P_{a+1}(\lambda,\lambda_2) {_0}{\cal
F}_1^{(a)}(\lambda,\lambda_3) \label{P1a+1}
\\
P_{2,a}(\lambda,\lambda_2,\lambda_3)=\theta(\lambda_2,\lambda_3) P_2(\lambda_3,\lambda_2) P_{a}(\lambda,\lambda_2) {_0}{\cal F}_1^{(a)}(\lambda,\lambda_3).
\label{P2a}
\ear

Taking into account the identities (\ref{idFF},
\ref{P1a+1},\ref{P2a}) we are able to bring
Eqs.(\ref{comTa+2aPhi2},\ref{comTa+1aPhi2}) into their more
symmetrical form, namely
\bear
\label{comTa+2aPhi2s}
\lefteqn{{\cal T}_{a+2, a}(\lambda) \phi_2(\lambda_2,\lambda_3)\ket{0}
=
w_{a+2}(\lambda)
w_{1}(\lambda_2) w_{1}(\lambda_3) {_0}{\cal
F}_2^{(a)}(\lambda,\lambda_2,\lambda_3) \ket{0}}
\nonumber \\
&+&
w_{a+1}(\lambda) w_{2}(\lambda_2) w_{1}(\lambda_3)
{_1}{\cal F}_2^{(a)}(\lambda,\lambda_2,\lambda_3)
\frac{R(\lambda_2,\lambda_3)_{1, 1}^{1,1}}{R(\lambda_2,\lambda_3)_{2, 1}^{2, 1}}
\frac{R(\lambda_3,\lambda_2)_{2, 1}^{2, 1}}{R(\lambda_3,\lambda_2)_{1, 1}^{1,1}}
\theta(\lambda_2,\lambda_3) \ket{0}
\nonumber \\
&+&
w_{a+1}(\lambda) w_{1}(\lambda_2) w_{2}(\lambda_3)
{_1}{\cal F}_2^{(a)}(\lambda,\lambda_3,\lambda_2)
\frac{R(\lambda_3,\lambda_2)_{1, 1}^{1,1}}{R(\lambda_3,\lambda_2)_{2, 1}^{2, 1}}
\frac{R(\lambda_2,\lambda_3)_{2, 1}^{2, 1}}{R(\lambda_2,\lambda_3)_{1, 1}^{1,1}} \ket{0}
\nonumber \\
&+& w_{a}(\lambda) w_{2}(\lambda_2) w_{2}(\lambda_3) {_2}{\cal
F}_2^{(a)}(\lambda,\lambda_2,\lambda_3) \ket{0}, ~~ \mbox{for} ~~ a
\le N-2 \ear

\bear
\label{comTa+1aPhi2s}
{\cal T}_{a+1, a}(\lambda) \phi_2(\lambda_2,\lambda_3)\ket{0}
&=&
{\cal T}_{1, 2}(\lambda_3)
{_0}{\cal F}_1^{(a)}(\lambda,\lambda_2)
\left[
w_{a+1}(\lambda)
w_1(\lambda_2)
\frac{R(\lambda_3,\lambda_2)_{1, 1}^{1,1}}{R(\lambda_3,\lambda_2)_{2, 1}^{2, 1}}
P_{a+1}(\lambda,\lambda_3)
\right.
\nonumber \\
&-&
\left.
w_{a}(\lambda)
w_2(\lambda_2)
\theta(\lambda_2,\lambda_3)
\frac{R(\lambda_2,\lambda_3)_{1, 1}^{1,1}}{R(\lambda_2,\lambda_3)_{2, 1}^{2, 1}}
P_{a}(\lambda,\lambda_3)
\right] \ket{0}
\nonumber \\
&+&
{\cal T}_{1, 2}(\lambda_2)
{_0}{\cal F}_1^{(a)}(\lambda,\lambda_3)
\left[
w_{a+1}(\lambda)
w_1(\lambda_3)
\theta(\lambda_2,\lambda_3)
\frac{R(\lambda_2,\lambda_3)_{1, 1}^{1,1}}{R(\lambda_2,\lambda_3)_{2, 1}^{2, 1}}
P_{a+1}(\lambda,\lambda_2)
\right.
\nonumber \\
&-&
\left.
w_{a}(\lambda)
w_2(\lambda_3)
\frac{R(\lambda_3,\lambda_2)_{1, 1}^{1,1}}{R(\lambda_3,\lambda_2)_{2, 1}^{2, 1}}
P_{a}(\lambda,\lambda_2)
\right] \ket{0}
\nonumber \\
&+&
{\cal T}_{a+1, a+2}(\lambda)
w_1(\lambda_2)
w_1(\lambda_3)
{_0}{\cal F}_2^{(a)}(\lambda,\lambda_2,\lambda_3) \ket{0}
\nonumber \\
&-&
{\cal T}_{a, a+1}(\lambda)
{_0}{\cal F}_1^{(a)}(\lambda,\lambda_2)
{_0}{\cal F}_1^{(a)}(\lambda,\lambda_3)
\left[
w_1(\lambda_2)
w_2(\lambda_3)
\frac{R(\lambda_3,\lambda_2)_{1, 1}^{1,1}}{R(\lambda_3,\lambda_2)_{2, 1}^{2, 1}}
\right.
\nonumber \\
&+&
\left.
w_2(\lambda_2)
w_1(\lambda_3)
\frac{R(\lambda_2,\lambda_3)_{1, 1}^{1,1}}{R(\lambda_2,\lambda_3)_{2, 1}^{2, 1}}
\theta(\lambda_2,\lambda_3)
\right] \ket{0}
\nonumber \\
&+&  \bar{\delta}_{a}^{1} {\cal T}_{a-1, a}(\lambda) w_2(\lambda_2)
w_2(\lambda_3) {_2}{\cal F}_2^{(a-1)}(\lambda,\lambda_2,\lambda_3)
\ket{0}, ~~ \mbox{for} ~~ a \le N-1. \ear

The commutation of the diagonal field with the last two terms ${\cal T}_{1, 3}(\lambda_i)
\phi_{1}(\lambda_j)$ and ${\cal T}_{1, 4}(\lambda_1)$ have been already discussed in the main text of subsection \ref{sub43}.
Therefore there is no need to repeat the procedure used for such two terms once again.
Considering the above results together with the steps explained for the operators ${\cal T}_{1, 3}(\lambda_i)
\phi_{1}(\lambda_j)$ and ${\cal T}_{1, 4}(\lambda_1)$ we find out the action of the
operator ${\cal T}_{a, a}(\lambda)$ on the three particle ansatz $\ket{\Phi_3}$ is given by
\bear
\lefteqn{ {\cal
T}_{a, a}(\lambda) \ket{\Phi_3} = w_a(\lambda)
\prod_{i=1}^{3}P_{a}(\lambda,\lambda_{i}) \ket{\Phi_3}}
\nonumber \\
&&
-\bar{\delta}_{a}^{N}
{\cal T}_{a, a+1}(\lambda)
\sum_{1 \le i_1<i_2 \le 3}
\sum_{\stackrel{j_1=1}{i_1 \neq j_1 \neq i_2}}^{3}
\phi_2(\lambda_{i_1},\lambda_{i_2})
w_1(\lambda_{j_1})
\prod_{k=1}^{2}
\frac{R(\lambda_{i_k},\lambda_{j_1})_{1, 1}^{1, 1}}{R(\lambda_{i_k},\lambda_{j_1})_{2, 1}^{2, 1}}
\theta_<(\lambda_{i_k},\lambda_{j_1})
{_1}{\cal F}_{1}^{(a)}(\lambda,\lambda_{j_1})  \ket{0}
\nonumber \\
&&
-\bar{\delta}_{a}^{1}
{\cal T}_{a-1, a}(\lambda)
\sum_{1 \le i_1<i_2 \le 3}
\sum_{\stackrel{l_1=1}{i_1 \neq l_1 \neq i_2}}^{3}
\phi_2(\lambda_{i_1},\lambda_{i_2})
w_2(\lambda_{l_1})
\prod_{k=1}^{2}
\frac{R(\lambda_{l_1},\lambda_{i_k})_{1, 1}^{1, 1}}{R(\lambda_{l_1},\lambda_{i_k})_{2, 1}^{2, 1}}
\theta_<(\lambda_{l_1},\lambda_{i_k})
{_0}{\cal F}_{1}^{(a-1)}(\lambda,\lambda_{l_1})  \ket{0}
\nonumber \\
&&
-\bar{\delta}_{a}^{N-1,N}
{\cal T}_{a, a+2}(\lambda)
\sum_{i_1=1}^{3}
\sum_{\stackrel{1 \le j_1<j_2 \le 3}{j_1 \neq i_1 \neq j_2}}
\phi_1(\lambda_{i_1})
w_1(\lambda_{j_1})
w_1(\lambda_{j_2})
\prod_{k=1}^{2}
\frac{R(\lambda_{i_1},\lambda_{j_k})_{1, 1}^{1, 1}}{R(\lambda_{i_1},\lambda_{j_k})_{2, 1}^{2, 1}}
\theta_<(\lambda_{i_1},\lambda_{j_k})
\nonumber \\
&&
\times {_2}{\cal F}_{2}^{(a)}(\lambda,\lambda_{j_1},\lambda_{j_2}) \ket{0}
-\bar{\delta}_{a}^{1,N}
{\cal T}_{a-1, a+1}(\lambda)
\sum_{i_1=1}^{3}
\sum_{\stackrel{j_1=1}{j_1 \neq i_1}}^{3}
\sum_{\stackrel{l_1=1}{i_1 \neq l_1 \neq j_1}}^{3}
\phi_{1}(\lambda_{i_1})
w_1(\lambda_{j_1})
w_2(\lambda_{l_1})
\frac{R(\lambda_{i_1},\lambda_{j_1})_{1, 1}^{1, 1}}{R(\lambda_{i_1},\lambda_{j_1})_{2, 1}^{2, 1}}
\nonumber \\
&& \times
\theta_<(\lambda_{i_1},\lambda_{j_1}) \frac{R(\lambda_{l_1},\lambda_{i_1})_{1, 1}^{1, 1}}{R(\lambda_{l_1},\lambda_{i_1})_{2, 1}^{2, 1}}
\theta_<(\lambda_{l_1},\lambda_{i_1})
\theta_<(\lambda_{l_1},\lambda_{j_1})
{_1}{\cal F}_{2}^{(a-1)}(\lambda,\lambda_{j_1},\lambda_{l_1}) \ket{0}
-\bar{\delta}_{a}^{1,2}
{\cal T}_{a-2, a}(\lambda)
\nonumber \\
&&
\times \sum_{i_1=1}^{3}
\sum_{\stackrel{1 \le l_1 < l_2 \le 3}{l_1 \neq i_1 \neq l_2}}
\phi_{1}(\lambda_{i_1})
w_2(\lambda_{l_1})
w_2(\lambda_{l_2})
\prod_{k=1}^{2}
\frac{R(\lambda_{l_k},\lambda_{i_1})_{1, 1}^{1, 1}}{R(\lambda_{l_k},\lambda_{i_1})_{2, 1}^{2, 1}}
\theta_<(\lambda_{l_k},\lambda_{i_1})
{_0}{\cal F}_{2}^{(a-2)}(\lambda,\lambda_{l_1},\lambda_{l_2}) \ket{0}
\nonumber \\
&&
-\bar{\delta}_{a}^{N-2,N-1,N}
{\cal T}_{a, a+3}(\lambda) w_1(\lambda_1) w_1(\lambda_2) w_1(\lambda_3)
{_3}{\cal F}_{3}^{(a)}(\lambda,\lambda_1,\lambda_2,\lambda_3) \ket{0}
-\bar{\delta}_{a}^{1,N-1,N}
{\cal T}_{a-1, a+2}(\lambda)
\nonumber \\
&&
\times \sum_{1 \le j_1 < j_2 \le 3}
\sum_{\stackrel{l_1=1}{j_1 \neq l_1 \neq j_2}}^{3}
w_1(\lambda_{j_1})
w_1(\lambda_{j_2})
w_2(\lambda_{l_1})
\prod_{k=1}^{2} \theta_<(\lambda_{l_1},\lambda_{j_k})
{_2}{\cal F}_{3}^{(a-1)}(\lambda,\lambda_{j_1},\lambda_{j_2},\lambda_{l_1}) \ket{0}
\nonumber \\
&&
-\bar{\delta}_{a}^{1,2,N}
{\cal T}_{a-2, a+1}(\lambda)
\sum_{j_1=1}^{3}
\sum_{\stackrel{1 \le l_1 < l_2 \le 3}{l_1 \neq j_1 \neq l_2}}
w_1(\lambda_{j_1})
w_2(\lambda_{l_1})
w_2(\lambda_{l_2})
\prod_{k=1}^{2} \theta_<(\lambda_{l_k},\lambda_{j_1})
{_1}{\cal F}_{3}^{(a-2)}(\lambda,\lambda_{j_1},\lambda_{l_1},\lambda_{l_2}) \ket{0}
\nonumber \\
&&
-\bar{\delta}_{a}^{1,2,3}
{\cal T}_{a-3, a}(\lambda)
w_2(\lambda_1) w_2(\lambda_2) w_2(\lambda_3)
{_0}{\cal F}_{3}^{(a-3)}(\lambda,\lambda_1,\lambda_2,\lambda_3) \ket{0},
\label{comAaaPhi3ss}
\ear
where functions ${_c}{\cal
F}_{3}^{(a)}(\lambda,\lambda_1,\lambda_2,\lambda_3)$ for $c=0,1,2,3$
have been summarized in Eqs(\ref{f30i}-\ref{f33i}).

\section*{Acknowledgments}
The authors thank the Brazilian Research Agencies FAPESP and CNPq for financial support.

\end{document}